\documentclass[aps,prc,floatfix,showkeys,nofootinbib,twocolumn]{revtex4-1}
\usepackage[T1]{fontenc}
\usepackage{amsmath}
\usepackage{amssymb}
\usepackage{mathtools}
\usepackage{booktabs}
\usepackage{siunitx}
\usepackage[referable]{threeparttablex}
\usepackage{longtable}
\usepackage{hyperref}
\usepackage{graphicx}
\usepackage{color}

\begin{document}

\title{Gamma-ray and conversion-electron spectroscopy of the high-spin isomer in \texorpdfstring{$^{145}$}{145}Sm}

\author{M.~S.~M.~Gerathy}
\email[Corresponding author: ]{matthew.gerathy@anu.edu.au}

\author{G.~J.~Lane}
\author{A.~E.~Stuchbery}
\author{G.~D.~Dracoulis}
\thanks{Deceased}
\author{T.~Kib\'edi}
\author{A.~Akber}
\author{L.~J.~Bignell}
\author{B.~J.~Coombes}
\author{J.~T.~H.~Dowie}
\author{T.~J.~Gray}
\author{B.~Q.~Lee}
\author{B.~P.~McCormick}
\author{A.~J.~Mitchell}
\author{N.~Palalani}

\affiliation{Department of Nuclear Physics, Research School of Physics, The Australian National University, Canberra, ACT, 2601, Australia}

\date{\today}

\begin{abstract}
\begin{description}
\item[Background] High-spin isomers at $\approx$9-MeV excitation energies have been reported in several $N=83$ isotones near $Z=64$.  Spin and parity assignments of $J^{\pi}=49/2^{+}$ remain tentative for a number of these states in the odd-A nuclei.
\item[Purpose] To study the decay of the (49/2$^{+}$) high-spin isomer in $^{145}$Sm, make firm spin and parity assignments to the isomer and states populated in its decay, and investigate the structure of the nucleus.
\item[Methods] The $^{145}$Sm isomer was populated in the $^{124}$Sn($^{26}$Mg,5n) reaction.  Gamma-ray and conversion-electron data were collected using the Solenogam array.
\item[Results] A revised lifetime of $t_{1/2}=3.52(16)$~$\mu$s was measured for the high-spin isomer.  Several new states have been added to the level scheme, and a new state at 8815~keV is proposed as the isomer, based on decay-property systematics, transition strengths, and spin and parity assignments.  Firm spin and parity assignments have been made to states up to and including the isomer and the new level scheme is interpreted using shell-model calculations performed with the KShell program.
\item[Conclusions] The interpretation of the 49/2$^{+}$ isomer as a deformed excitation of the core neutrons remains unchanged, although there has been a significant revision of the level scheme below the isomer, and hence significant reinterpretations of the lower-lying states.
\end{description}
\end{abstract}


\maketitle
\sloppy


\section{Introduction}

Long-lived nuclear states can provide a powerful probe of nuclear structure and the underlying behavior of the nucleons.  In general, transition rates depend on changes in energy, angular momentum and parity, as well as the character of the parent and daughter states.  In particular, transitions between states with very different character are hindered, leading to long state lifetimes.

\begin{figure*}[htb]
\centering
\includegraphics{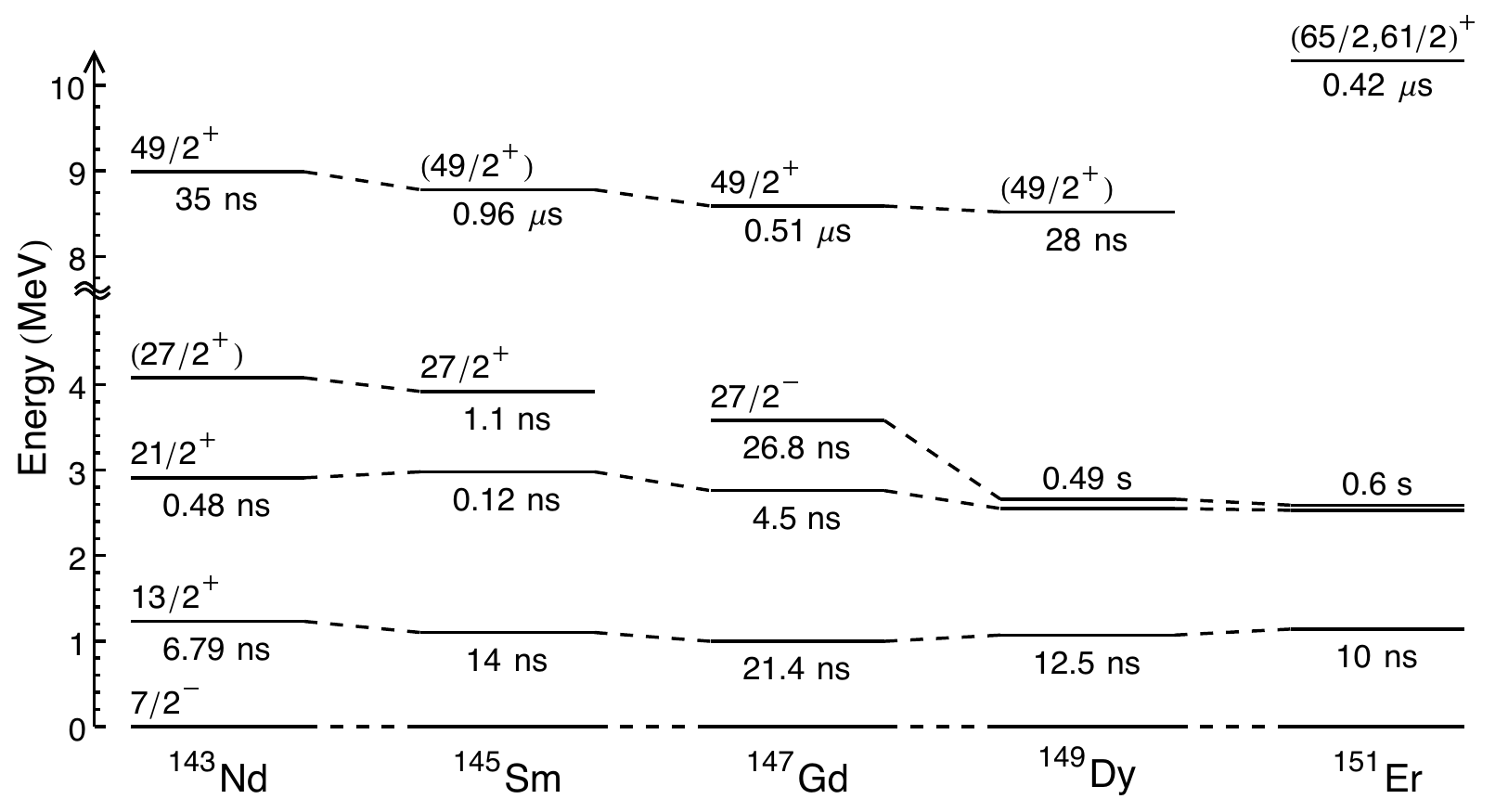}
\caption{Energy systematics of isomers in the odd-mass, $N=83$ isotones.  Experimental values were taken from Refs.~\cite{Zhou1999,Odahara1997,Broda2020,Gono2002,Foin2000}; results from the present work are not included.}
\label{fig:145Sm-N=83_Isomers}
\end{figure*}

Among the $N=83$ isotones, high-spin isomers have been reported at $\approx$9-MeV excitation energy in $^{143}$Nd~\cite{Zhou1999}, $^{144}$Pm~\cite{Murakami1993}, $^{145}$Sm~\cite{Odahara1997}, $^{146}$Eu~\cite{Ideguchi1995}, $^{147}$Gd~\cite{Broda2020}, $^{148}$Tb~\cite{Ideguchi1999}, $^{149}$Dy~\cite{Gono2002}, $^{150}$Ho~\cite{Fukuchi2006}, and $^{151}$Er~\cite{Foin2000}; the systematics of the odd-mass isomers are shown in Figure~\ref{fig:145Sm-N=83_Isomers}.  The  deformed independent particle model (DIPM)~\cite{Neergard1981} has been successful in predicting the measured electromagnetic moments of the 49/2$^{+}$ isomer in $^{147}$Gd~\cite{Hausser1979,Hausser1982}.  This has led to the association of the 49/2$^{+}$ isomeric state with an oblate-deformed, core-neutron excitation; $[\pi(h^{2}_{11/2})\nu(f_{7/2}h_{9/2}i_{13/2})]_{49/2^{+}}$.  This interpretation has been extended to all of the $N=83$ isotones, based on the agreement between experimentally observed energy systematics and those predicted by the DIPM (see Figures~8 and 9 in Ref.~\cite{Odahara1997}).   The $[\pi(h^{2}_{11/2}d^{-1}_{5/2})\nu(f_{7/2}h_{9/2}i_{13/2})]_{27^{+}}$ configuration was assigned for the even-mass systems.  For many of these isomers, however, the spin and parity assignments are based only on the predictions of the DIPM and lack experimental confirmation.  Firm experimental spin and parity assignments are needed along the entire isotone chain.

In $^{145}$Sm, DIPM calculations predict a 49/2$^{+}$ isomer at 9540~keV~\cite{Odahara1997}.  Previous work by Ferragut \textit{et al.}~\cite{Ferragut1993} reported a high-spin isomer in $^{145}$Sm with a half-life of 0.96~$\mu$s, and later work by  Odahara \textit{et al.}~\cite{Odahara1997} reported the decay scheme of this state and placed it at 8785~keV.  While a spin and parity of $J^{\pi}=49/2^{+}$ was tentatively assigned based on energy systematics and DIPM predictions, above $\approx$4~MeV the spin and parity assignments are either tentative or missing.

In the present work, the decay of the high-spin isomer in $^{145}$Sm has been studied using the Solenogam array~\cite{Gerathy2020}.  Solenogam is a system of $\gamma$-ray and electron detectors coupled to the recoil separator SOLITAIRE at the Australian National University Heavy-Ion Accelerator Facility (ANU-HIAF).  The high-spin isomer in $^{145}$Sm was isolated in a low-background environment at the focal plane of Solenogam and the subsequent isomeric decay observed.  A revised level scheme has been deduced using $\gamma$-ray and e$^{-}$ singles data, as well as $\gamma$-$\gamma$, $\gamma$-e$^{-}$, and $\gamma$-X coincidence data\footnote{`X' refers to $\gamma$ rays observed in the LEPS detector while `$\gamma$' refers to $\gamma$ rays observed in the HPGe detectors.}.  The new level scheme is compared with nuclear shell-model predictions and discussed within the broader context of the $N=83$ isotones.

\section{Experimental details}

The high-spin isomer in $^{145}$Sm was populated via the $^{124}$Sn($^{26}$Mg,5n)$^{145}$Sm reaction using a 1.5-mg/cm$^{2}$ target of enriched $^{124}$Sn.  Evaporation residues were transported through the SOLITAIRE separator using a 5.5-T magnetic field and 0.25 Torr of He gas.  After implantation in a catcher tape at the focal point of the Solenogam array, $\gamma$ rays and electrons were detected using five Ortec Gamma-X HPGe detectors (four with Compton suppression), one Ortec LEPS detector, and six Si(Li) detectors; see Refs.~\cite{Gerathy2020} and~\cite{Rodriguez2010} for more complete descriptions of SOLITAIRE and Solenogam.

Initial measurements were obtained at a beam energy of 110 MeV.  In order to determine the lifetime of the isomer, a chopped beam of 9.63~$\mu$s on, 48.15~$\mu$s off was used.  While PACE2~\cite{Gavron1980} calculations predicted a maximum $^{145}$Sm yield at 110~MeV, a second measurement was performed with the beam energy increased to 115~MeV.  This increase in beam energy brought in more angular momentum, leading to an increased population of the isomeric state, outweighing the slight decrease in absolute production cross-section.  The beam chopping was also changed to 4.17/4.17~$\mu$s (on/off), as the isomer production yield and data cleanliness were optimised jointly by having irradiation and measurement times approximately equal to the half-life of the isomeric state (see \S\ref{sec:lifetimes} for details of the revised half-life measured in this work).  The $^{145}$Sm level scheme was constructed using the out-of-beam data from this second run.

The data were collected using a system of XIA Pixie-16 digitizers~\cite{XIA2018}.  The pre-amplifier signals for each each HPGe and Si(Li) detector were directly digitized along with signals from each of the Compton suppressors.  An RF pulse train synced with the chopped beam was also digitized so the event times could be measured relative to the beam burst.  Data were collected with a singles trigger and later sorted into coincidence events using a $\pm$192-ns coincidence window.  Compton suppression was carried out at this stage and events were removed from the data stream if a HPGe detector was in coincidence with its associated suppressor.

\section{Results}

\begin{figure*}[htb]
\centering
\includegraphics[width=0.85\textwidth]{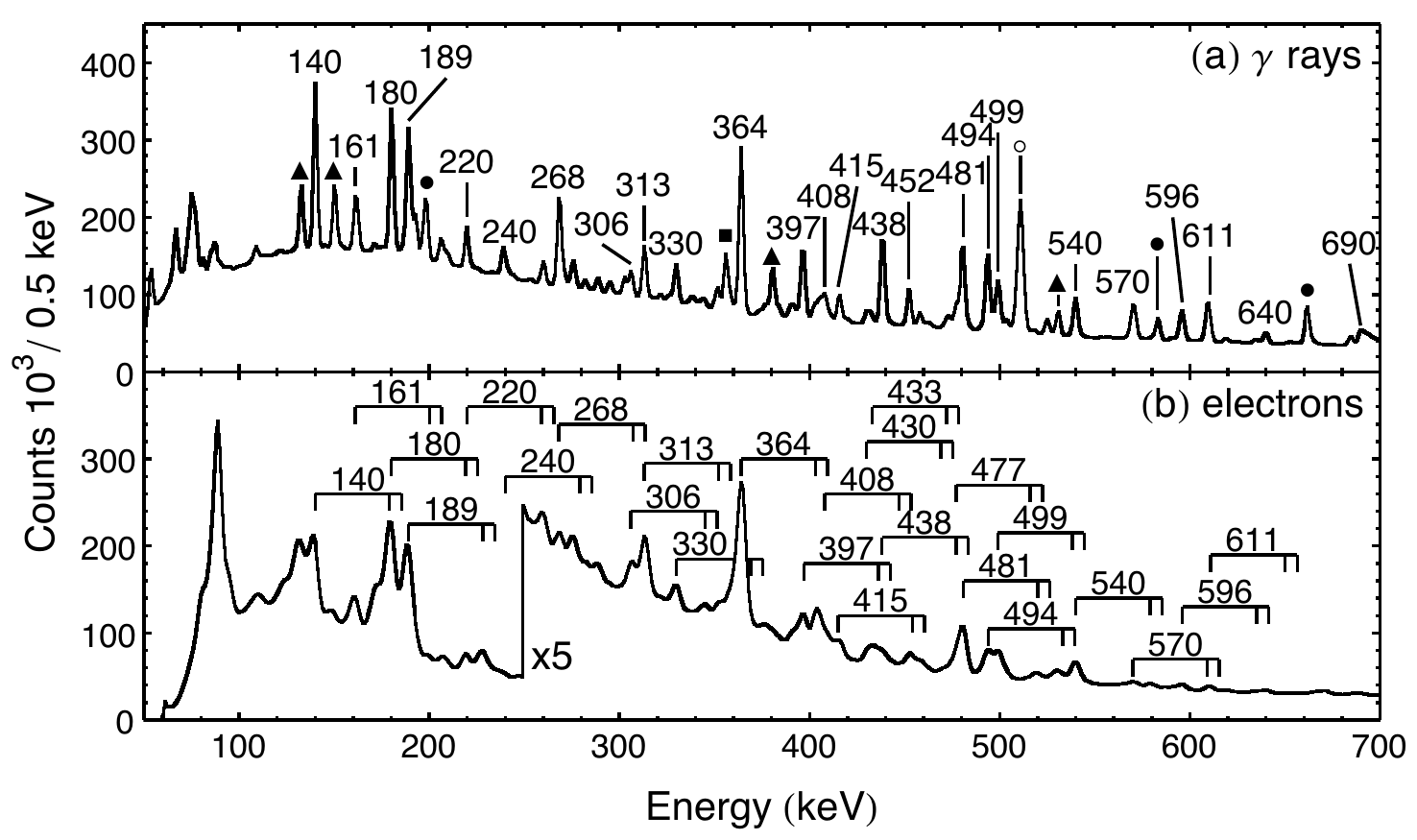}
\caption{Out-of-beam singles spectra for (a) $\gamma$ rays and (b) electrons between 0 and 700~keV.  The electron energies have been shifted to align the $\gamma$-ray and $K$-conversion peaks.  Electron conversion lines for the $K$, $L$, and $M$ shells are indicated for selected peaks.  Contaminants from $^{144}$Sm ($\blacktriangle$), $^{133}$Cs ({\small$\blacksquare$}, see \S\ref{sec:HSIsomer}), the 511-keV positron annihilation peak ($\circ$), and the room background ($\bullet$) have been marked.}
\label{fig:145Sm-Singles_Lo}

\vspace{\floatsep}

\includegraphics[width=0.85\textwidth]{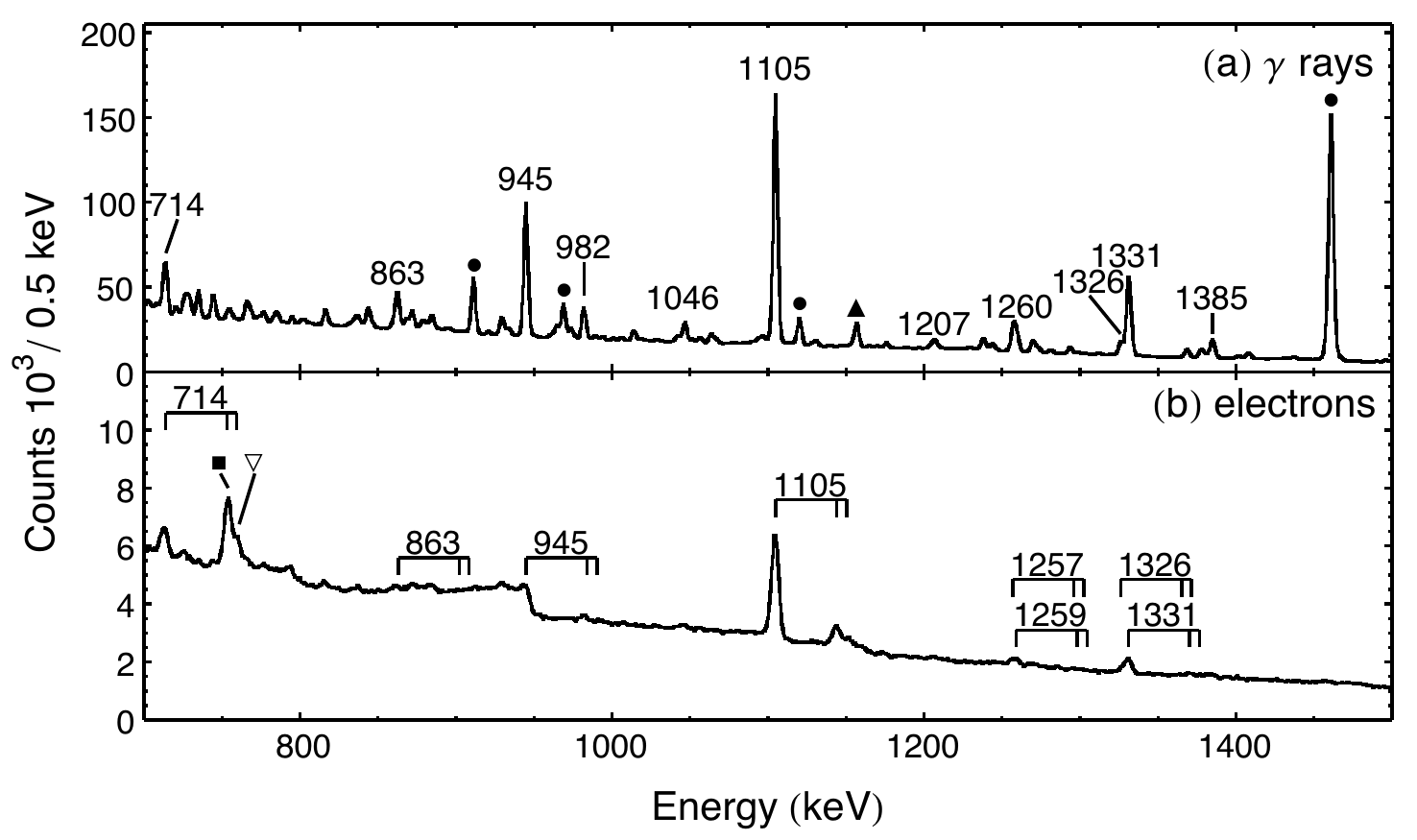}
\centering
\caption{Out-of-beam singles spectra for (a) $\gamma$ rays and (b) electrons between 700 and 1500~keV.  The electron energies have been shifted to align the $\gamma$-ray and $K$-conversion peaks.  The step at 945~keV (shifted) in the electron spectrum is from the Compton edge of the 1105-keV $\gamma$ ray.  Electron conversion lines for the $K$, $L$, and $M$ shells are indicated for selected peaks.  Contaminants from $^{143}$Sm ({\small$\blacksquare$}), $^{141}$Nd ($\triangledown$), $^{44}$Ca ($\blacktriangle$, from the $\beta$ decay of $^{44}$Sc produced via the $^{27}$Al($^{26}$Mg,2$\alpha$n) reaction on the aluminium wall of the target chamber), and the room background ($\bullet$) have been marked.}
\label{fig:145Sm-Singles_Hi}
\end{figure*}

\begin{figure*}[htb]
\centering
\includegraphics[width=0.85\textwidth]{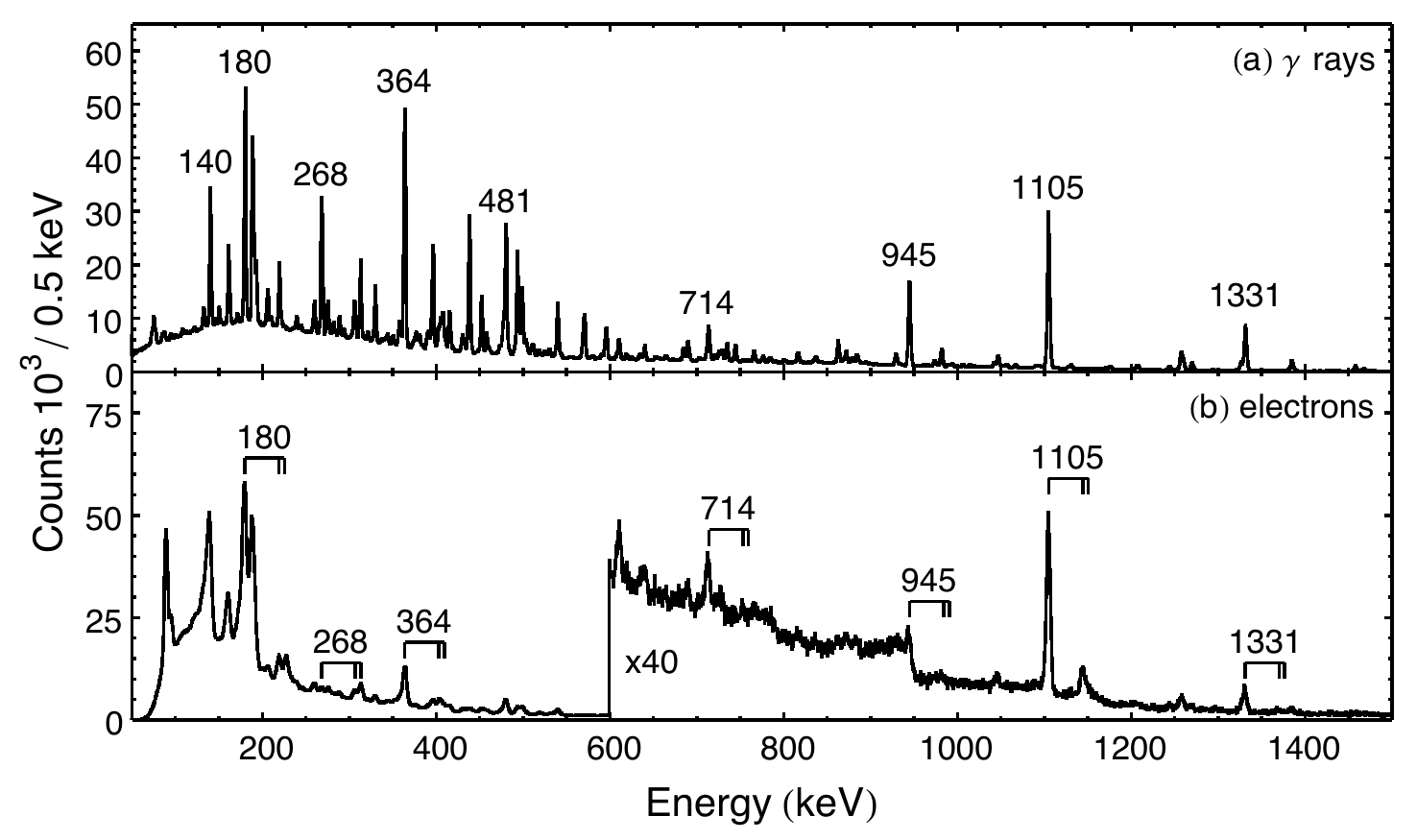}
\centering
\caption{Total projection of out-of-beam (a)~$\gamma$ rays and (b)~electrons, from $\gamma$-$\gamma$ and $\gamma$-e$^{-}$ coincidence matrices in $^{145}$Sm.  The electron energies have been shifted to align the $\gamma$-ray and $K$-conversion peaks.  Conversion lines for the $K$, $L$, and $M$ shells are indicated for selected peaks.}
\label{fig:145Sm-Full_Projection}
\end{figure*}

The out-of-beam $\gamma$-ray and $e^{-}$ singles data from the present work are shown in Figures~\ref{fig:145Sm-Singles_Lo} and~\ref{fig:145Sm-Singles_Hi}, with the total projections of $\gamma$-rays and electrons from $\gamma$-$\gamma$ and $\gamma$-e$^{-}$ coincidences given in Figure~\ref{fig:145Sm-Full_Projection}.  The presence of the 364-, 945-, 1105-, and 1331-keV transitions from the known $^{145}$Sm level scheme~\cite{Odahara1997} confirmed the production of $^{145}$Sm.  Other transition were assigned to $^{145}$Sm based on coincidences with these known transitions.  The list of transition properties measured in the present work is given in Table~\ref{tab:transitions}.  Details on states above the isomer can be found in Ref.~\cite{Odahara1997}.  However, due to the long flight time through the solenoid, no data could be obtained for these states in the present work.  A number of contaminants in the singles spectra were identified as known transitions from long-lived isomers in isotopes produced alongside $^{145}$Sm.

\begingroup

\begin{ThreePartTable}
\setlength{\LTcapwidth}{\textwidth}
\renewcommand\TPTminimum{\textwidth}
\renewcommand{\baselinestretch}{1}
\renewcommand{\arraystretch}{1}
\centering
\begin{TableNotes}[para]
\footnotesize
\item[a]{$\alpha_{\text{tot}}$ value.  Experimental value deduced from intensity balance (see text).}
\item[b]{$\alpha_{K}$ deduced using coincidence spectra.}
\item[c]{New transition added in the current work.}
\item[d]{Unplaced transition in coincidence with transitions assigned to $^{145}$Sm.}
\item[e]{$\alpha_{K}(708.1; M2)=2.12\times 10^{-2}$}
\item[f]{$\alpha_{K}(1042.4; M2)=0.75\times 10^{-2}$}
\item[g]{$\alpha_{K}(1104.6; E3)=0.35\times 10^{-2}$}
\item[h]{$\alpha_{K}(1175.9; M2)=0.55\times 10^{-2}$}
\item[i]{$\alpha_{K}(1457.9; E3)=0.19\times 10^{-2}$}
\end{TableNotes}
\begin{longtable*}{@{\extracolsep{\fill}}c c c c c c c c c c c@{}}
\caption{Properties of transitions in $^{145}$Sm measured in the current work.  Gamma-ray intensities are normalized to the 364-keV transition.  Experimental conversion coefficients are mostly from singles data with some determined from coincidence measurements  (see Table~\ref{tab:doubleICCs}).  Theoretical conversion coefficients for the pure multipolarities are from BrIcc~\cite{Kibedi2008}.}  \label{tab:transitions}\\		
\toprule
\toprule
                   &              & \multicolumn{4}{c}{$100 \alpha_{K}$} &    &         &         &         &         \\
\cmidrule{3-6}
$E_{\gamma}$ (keV) & $I_{\gamma}$ (rel) & expt & $E1$ & $M1$ & $E2$                       & $XL$ & $E_{i}$(keV) & $E_{f}$(keV) & $J_{i}$ & $J_{f}$ \\
\midrule

\endfirsthead

\caption[]{(Continued).}\\
\toprule
\toprule
                   &              & \multicolumn{4}{c}{$100 \alpha_{K}$} &    &         &         &         &         \\
\cmidrule{3-6}
$E_{\gamma}$ (keV) & $I_{\gamma}$ (rel) & expt & $E1$ & $M1$ & $E2$                      & $XL$ & $E_{i}$(keV) & $E_{f}$(keV) & $J_{i}$ & $J_{f}$ \\
\midrule
\endhead

\bottomrule
\bottomrule
\insertTableNotes
\endfoot
(15)                                                           &           &                                    &                               &                               &
& ($M1$)                           & 2979 & 2964 & $21/2^{+}$ & $19/2^{+}$ \\
30.2(8)                                                        & 61(3)     & 127\tnote{a}     & 127\tnote{a} & 860\tnote{a} & 36020\tnote{a} & $E1$                             & 8815 & 8785 & $49/2^{+}$   & $47/2^{-}$   \\
140.3(3)                                                       & 32(3) & 47.8(26)\tnote{b}    & 9.30                          & 54.8                          & 44.5                            & $M1/E2$                          & 3119 & 2979 & $23/2^{+}$ & $21/2^{+}$   \\
161.3(4)                                                       & 20(2)    & 37.4(34)\tnote{b} & 6.39                          & 37.1                          & 29.1                            & $M1/E2$                          & 5029 & 4868 & $31/2^{+}$   & $29/2^{+}$   \\
161.4(9)                                                       & 5(2)     & 31.0(51)\tnote{b} & 6.38                          & 37.0                          & 29.0                            & $M1/E2$                          & 3483 & 3322 & $25/2^{+}$   & $23/2^{+}$   \\
180.2(3)                                                       & 55(4)    & 26(2)                              & 4.74                          & 27.3                          & 20.7                            & $M1$                             & 2230 & 2049 & $17/2^{-}$   & $15/2^{-}$   \\
186.6(8)                                                       & 7.4(8) &                                    & 4.32                          & 24.8                          & 18.6                            & $(M1)$                           & 8377 & 8190 & $45/2^{-}$   & $43/2^{-}$   \\
189.3(3)                                                       & 36(3)    & 23(2)                              & 4.16                          & 23.8                          & 17.8                            & $M1$                             & 3119 & 2930 & $23/2^{+}$   & $21/2^{+}$   \\
192.7(3)                                                       & 16(1)    & 21(1)                              & 3.97                          & 22.7                          & 16.8                            & $M1$                             & 4420 & 4228 & $29/2^{-}$   & $27/2^{-}$   \\
206.2(4)                                                       & 10.9(8)  & 4.3(4)                             & 3.31                          & 18.5                          & 13.7                            & $E1$                             & 8785 & 8579 & $47/2^{-}$   & $45/2^{+}$   \\
209.9(10)                                                      & 5.0(4)   &                                    & 3.16                          & 17.9                          & 12.9                            &                                & 3140 & 2930 &              & $21/2^{+}$   \\
219.7(3)                                                       & 17(1)    & 2.6(4)\tnote{b}   & 2.80                          & 15.8                          & 11.2                            & $E1$                             & 2930 & 2710 & $21/2^{+}$   & $19/2^{-}$   \\
221.8(10)                                                       & 1.5(1)   &                                    & 2.73                          & 15.4                          & 10.9                            &                                & 3032 & 2810 &              & $15/2^{~}$   \\
(236)                                                          & <0.96    &                                    & 2.33                          & 13.1                          & 9.06                            & $(E2)$                           & 8815 & 8579 & $49/2^{+}$   & $45/2^{+}$   \\
240.7(9)                                                       & 9.5(7)   & 13.1(9)                            & 2.21                          & 12.4                          & 8.50                            & $M1$                             & 6361 & 6121 & $37/2^{-}$   & $35/2^{-}$   \\
260.1(4)                                                       & 10.7(8)  & 9.4(11)\tnote{b}     & 1.81                          & 10.1                          & 6.72                            & $M1$                             & 6216 & 5956 & $37/2^{-}$   & $35/2^{-}$   \\
268.4(3)                                                       & 44(3)    & 2.2(2)                             & 1.67                          & 9.24                          & 6.11                            & $E1$                             & 2979 & 2710 & $21/2^{+}$   & $19/2^{-}$   \\
275.6(4)                                                       & 12.7(9)  & 7.2(5)                             & 1.56                          & 8.61                          & 5.64                            & $M1/E2$                          & 5956 & 5680 & $35/2^{-}$   & $33/2^{-}$   \\
282.1(6)                                                       & 3.5(3)   & 8.6(7)                             & 1.46                          & 8.07                          & 5.24                            & $M1$                             & 4868 & 4586 & $29/2^{+}$   & $27/2^{+}$   \\
288.7(5)                                                       & 6.3(5)   & 7.1(5)                             & 1.38                          & 7.61                          & 4.91                            & $M1$                             & 5029 & 4741 & $31/2^{+}$   & $29/2^{+}$   \\
294.1(10)\tnote{c}                             & 5.9(6)   & <1.23                              & 1.32                          & 7.25                          & 4.64                            & $E1$                             & 7742 & 7449 & $43/2^{+}$   & $41/2^{-}$   \\
306.2(4)                                                       & 12.1(9)  & 6.1(4)                             & 1.19                          & 6.51                          & 4.12                            & $M1/E2$                       & 4228 & 3921 & $27/2^{-}$   & $27/2^{-}$   \\
313.2(3)                                                       & 26(2)    & 5.9(4)                             & 1.13                          & 6.14                          & 3.86                            & $M1$                             & 6216 & 5903 & $37/2^{-}$   & $35/2^{-}$   \\
330.0(4)                                                       & 20.7(15) & 3.2(2)                             & 0.99                          & 5.35                          & 3.31                            & $M1/E2$                          & 8072 & 7742 & $45/2^{+}$   & $43/2^{+}$   \\
344.0(3)\tnote{c,d}  & 6.7(6)   & 3.9(3)                             & 0.89                          & 4.80                          & 2.93                            & $M1/E2$                          &      &      &              &              \\
350.5(2)\tnote{c,d}  & 1.2(3)   & 3.1(2)                             & 0.85                          & 4.57                          & 2.78                            & $M1/E2$                          &      &      &              &              \\
358.2(5)                                                       & 7.1(5)   & 5.3(13)\tnote{b}   & 0.81                          & 4.32                          & 2.61                            & $M1$                             & 6719 & 6361 & $39/2^{-}$   & $37/2^{-}$   \\
364.0(3)                                                       & 100      & 3.9(3)                             & 0.78                          & 4.14                          & 2.50                            & $M1$                             & 3483 & 3119 & $25/2^{+}$   & $23/2^{+}$   \\
377.0(4)\tnote{c}                             & 6.9(5)   & 2.8(3)                             & 0.71                          & 3.78                          & 2.26                            & $M1/E2$                          & 5903 & 5526 & $35/2^{-}$   & $33/2^{-}$   \\
391.4(6)                                                       & 5.5(4)   & 3.6(3)                             & 0.65                          & 3.43                          & 2.04                            & $M1$                             & 3322 & 2930 & $23/2^{+}$   & $21/2^{+}$   \\
396.1(4)                                                       & 22.2(16) & 4.1(13)\tnote{b}  & 0.63                          & 3.33                          & 1.97                            & $M1$                             & 7327 & 6931 & $41/2^{+}$   & $39/2^{+}$   \\
396.7(4)                                                       & 25.1(18) & <1\tnote{b}       & 0.63                          & 3.31                          & 1.96                            & $E1$                             & 5903 & 5507 & $35/2^{-}$   & $33/2^{+}$   \\
402.8(6)                                                       & 8.2(6)   &                                    & 0.61                          & 3.18                          & 1.88                            & $(M1)$                           & 6121 & 5719 & $35/2^{-}$   & $33/2^{-}$   \\
405.3(4)                                                       & 11.3(9)  & 3(1)\tnote{b}     & 0.60                          & 3.13                          & 1.85                            & $M1$                             & 6361 & 5956 & $37/2^{-}$   & $35/2^{-}$   \\
407.2(7)                                                       & 11.3(8)  & 3(1)\tnote{b}     & 0.59                          & 3.10                          & 1.82                            & $M1$                             & 8333 & 7926 & $43/2^{-}$   & $41/2^{-}$   \\
408.3(4)                                                       & 11.6(8)  & 4(1)\tnote{b}     & 0.59                          & 3.08                          & 1.81                            & $M1$                             & 8785 & 8377 & $47/2^{-}$   & $45/2^{-}$   \\
415.6(4)                                                       & 16.3(12) & 2.2(2)                             & 0.57                          & 2.94                          & 1.72                            & $M1/E2$                          & 7742 & 7327 & $43/2^{+}$   & $41/2^{+}$   \\
430.3(5)                                                       & 8.4(6)   & 3.0(3)                             & 0.52                          & 2.69                          & 1.57                            & $M1$                             & 5956 & 5526 & $35/2^{-}$   & $33/2^{-}$   \\
432.8(9)                                                       & 5.7(4)   & 2.7(9)\tnote{b}   & 0.51                          & 2.65                          & 1.54                            & $M1$                             & 5680 & 5248 & $33/2^{-}$   & $31/2^{-}$   \\
438.4(3)                                                       & 63.7(45) & 0.35(6)                            & 0.50                          & 2.56                          & 1.49                            & $E1$                             & 3921 & 3483 & $27/2^{-}$   & $25/2^{+}$   \\
452.4(5)                                                       & 26.7(2)  & 1.3(1)                             & 0.46                          & 2.37                          & 1.37                            & $E2$                             & 8785 & 8333 & $47/2^{-}$   & $43/2^{-}$   \\
458.2(5)                                                       & 9.3(7)   & 2.2(3)                             & 0.45                          & 2.29                          & 1.32                            & $M1$                             & 6361 & 5903 & $37/2^{-}$   & $35/2^{-}$   \\
473.8(10)\tnote{c}                             & 9.2(7)   & <0.96\tnote{b}    & 0.42                          & 2.10                          & 1.21                            & $E1$                             & 7404 & 6931 & $41/2^{-}$   & $39/2^{+}$   \\
477.4(4)                                                       & 18.4(13) & 1.8(1)                             & 0.41                          & 2.06                          & 1.19                            & $M1/E2$                          & 5507 & 5029 & $33/2^{+}$   & $31/2^{+}$   \\
480.6(3)                                                       & 66.3(5)  & 1.5(1)                             & 0.40                          & 2.03                          & 1.17                            & $M1/E2$                          & 2710 & 2230 & $19/2^{-}$   & $17/2^{-}$   \\
493.6(4)                                                       & 56(4)    & 1.1(8)                             & 0.38                          & 1.90                          & 1.09                            & $E2$                             & 2930 & 2436 & $21/2^{+}$   & $17/2^{+}$   \\
496.7(4)                                                       & 5.0(4)   &                                    & 0.37                          & 1.87                          & 1.07                            & $(E2)$                           & 6216 & 5719 & $37/2^{-}$   & $33/2^{-}$   \\
499.0(3)                                                       & 35.0(25) & 1.6(1)                             & 0.37                          & 1.84                          & 1.06                            & $M1/E2$                          & 4420 & 3921 & $29/2^{-}$   & $27/2^{-}$   \\
503.7(6)                                                       & 5.3(4)   & 1.6(2)                             & 0.36                          & 1.80                          & 1.03                            & $M1/E2$                          & 6719 & 6216 & $39/2^{-}$   & $37/2^{-}$   \\
511.8(10)                                                      & 2.5(2)   &                                    & 0.35                          & 1.73                          &      
0.99                            & $(E2)$                           & 2049 & 1538 & $15/2^{-}$   & ($11/2^{-}$) \\
518.1(9)\tnote{c}                                              & 2.5(2)   &                                    & 0.34                          & 1.68                          & 0.96                            & $(E1)$                           & 7449 & 6931 & $41/2^{-}$   & $39/2^{+}$   \\
540.0(3)                                                       & 31(2)    & 1.5(1)                             & 0.31                          & 1.51                          & 0.86                            & $M1$                             & 6756 & 6216 & $39/2^{-}$   & $37/2^{-}$   \\
543.4(10)                                                       & 1.60(15) &                                    & 0.31                          & 1.48                          & 0.85                            & $(E2)$                           & 2979 & 2436 & $21/2^{+}$   & $17/2^{+}$   \\
569.6(3)\tnote{c}                             & 13.7(11) & 0.3(1)                             & 0.28                          & 1.32                          & 0.76                            & $E1$                             & 6931 & 6361 & $39/2^{+}$   & $37/2^{-}$   \\
571.0(3)                                                       & 21.0(16) & 0.28(6)                            & 0.28                          & 1.32                          & 0.75                            & $E1$                             & 7327 & 6756 & $41/2^{+}$   & $39/2^{-}$   \\
590.9(8)\tnote{c,d}  & 2.8(2)   & 1.1(2)                             & 0.25                          & 1.21                          & 0.69                            & $M1$                             &      &      &              &              \\
595.6(4)                                                       & 29(2)    & 0.37(3)                            & 0.25                          & 1.18                          & 0.68                            & $E2$                             & 8785 & 8190 & $47/2^{-}$   & $43/2^{-}$   \\
609.7(5)                                                       & 1.9(2)   &                                    & 0.24                          & 1.12                          & 0.64                            & $(E1)$                           & 5029 & 4420 & $31/2^{+}$   & $29/2^{-}$   \\
610.7(5)                                                       & 14(1)    & 1.3(3)\tnote{b}   & 0.24                          & 1.11                          & 0.64                            & $M1$                             & 5031 & 4420 & $31/2^{-}$   & $29/2^{-}$   \\
640.1(6)                                                       & 11.2(8)  & 0.65(9)                            & 0.21                          & 0.99                          & 0.57                            & $E2$                             & 5029 & 4389 & $31/2^{+}$   & $27/2^{+}$   \\
684.8(6)                                                       & 9.7(7)   & 0.33(5)                            & 0.19                          & 0.84                          & 0.49                            & $(M1/E2)$                        & 7404 & 6719 & $41/2^{-}$   & $39/2^{-}$   \\
689.8(4)                                                       & 17(1)    & 0.17(3)                            & 0.18                          & 0.82                          & 0.48                            & $E1$                             & 5719 & 5029 & $33/2^{-}$   & $31/2^{+}$   \\
708.1(5)                                                       & 6.3(5)   & 2.3(17)\tnote{b}     & 0.17                          & 0.77                          & 0.45                            & $M2$\tnote{e}   & 6216 & 5507 & $37/2^{-}$   & $33/2^{+}$   \\
712.8(5)                                                       & 23(2)    & 0.14(5)                            & 0.17                          & 0.76                          & 0.44                            & $E1$                             & 8785 & 8072 & $47/2^{-}$   & $45/2^{+}$   \\
714.3(4)\tnote{c}                             & 11.5(8)  & 0.16(4)                            & 0.17                          & 0.76                          & 0.44                            & $E1$                             & 6931 & 6216 & $39/2^{+}$   & $37/2^{-}$   \\
720.7(3)\tnote{c,d}  & 4.6(4)   & 0.36(9)                            & 0.17                          & 0.74                          & 0.43                            & $M1/E2$                          &      &      &              &              \\
725.7(7)                                                       & 6.5(9) & 0.9(2)\tnote{b}   & 0.16                          & 0.73                          & 0.42                            & $M1$                             & 4647 & 3921 & $29/2^{-}$   & $27/2^{-}$   \\
729.3(6)                                                       & 8.5(8) & 0.31(8)                            & 0.16                          & 0.72                          & 0.42                            & $(M1/E2)$                        & 7449 & 6719 & $41/2^{-}$   & $39/2^{-}$   \\
734.8(5)                                                       & 14(1)    & 0.13(5)                            & 0.16                          & 0.71                          & 0.41                            & $E1$                             & 2964 & 2230 & $19/2^{+}$   & $17/2^{-}$   \\
(743)                                                       & <0.52    &                                    & 0.16                          & 0.69                          & 0.40                            & $(E2)$                           & 8815 & 8072 & $49/2^{+}$   & $45/2^{+}$   \\
744.5(5)                                                       & 11.0(8)  & 0.18(5)                            & 0.16                          & 0.68                          & 0.40                            & $E1$                             & 4228 & 3483 & $27/2^{-}$   & $25/2^{+}$   \\
766.0(5)                                                       & 9.3(7)   & 0.35(6)                            & 0.15                          & 0.64                          & 0.38                            & $E2$                             & 5507 & 4741 & $33/2^{+}$   & $29/2^{+}$   \\
776.7(6)                                                       & 5.2(4)   & 0.2(1)                             & 0.14                          & 0.62                          & 0.36                            & $E1$                             & 8579 & 7803 & $45/2^{+}$   & $43/2^{-}$   \\
816.5(6)                                                       & 8.5(6)   & 0.3(1)                             & 0.13                          & 0.55                          & 0.33                            & $E2$                             & 6719 & 5903 & $39/2^{-}$   & $35/2^{-}$   \\
837.1(5)                                                       & 6.6(6)   & 0.34(9)                            & 0.12                          & 0.52                          & 0.31                            & $M1/E2$                          & 8579 & 7742 & $45/2^{+}$   & $43/2^{+}$   \\
862.7(4)                                                       & 22.3(16) & 0.09(4)                            & 0.12                          & 0.48                          & 0.29                            & $E1$                             & 8190 & 7327 & $43/2^{-}$   & $41/2^{+}$   \\
871.8(5)                                                       & 10.7(8)  & 0.29(7)                            & 0.12                          & 0.47                          & 0.28                            & $E2$                             & 5903 & 5031 & $35/2^{-}$   & $31/2^{-}$   \\
879.0(8)                                                       & 5.1(1)   & 0.31(8)                            & 0.11                          & 0.46                          & 0.28                            & $E2$                             & 5526 & 4647 & $33/2^{-}$   & $29/2^{-}$   \\
884.6(7)                                                       & 8.7(6)   & 0.31(6)                            & 0.11                          & 0.45                          & 0.27                            & $(M1/E2)$                        & 8333 & 7449 & $43/2^{-}$   & $41/2^{-}$   \\
929.5(6)                                                       & 11.5(8)  & 0.26(6)                            & 0.10                          & 0.40                          & 0.25                            & $(M1/E2)$                        & 8333 & 7404 & $43/2^{-}$   & $41/2^{-}$   \\
944.8(3)                                                       & 81(6)    & 0.14(3)\tnote{b}  & 0.10                          & 0.39                          & 0.24                            & $E1$                             & 2049 & 1105 & $15/2^{-}$   & $13/2^{+}$   \\
973.7(3)\tnote{c,d}  & 6.34(55) & 0.11(6)                            & 0.09                          & 0.36                          & 0.22                            & $E1$                             &      &      &              &              \\ 
981.9(5)                                                       & 20(1)    & 0.10(2)                            & 0.09                          & 0.35                          & 0.22                            & $(E2)$                           & 8785 & 7803 & $47/2^{-}$   & $43/2^{-}$   \\
1006.2(10)\tnote{c}                            & 2.1(2)   &                                    & 0.09                          & 0.33                          & 0.21                            & $(E1)$                           & 8333 & 7327 & $43/2^{-}$   & $41/2^{+}$   \\
1042.4(6)                                                      & 4.1(3)   & 0.3(1)                             & 0.08                          & 0.31                          & 0.19                            & $M2$\tnote{f}   & 8785 & 7742 & $47/2^{-}$   & $43/2^{+}$   \\
1046.7(5)                                                      & 13(1)    & 0.12(4)                            & 0.08                          & 0.30                          & 0.19                            & $(E2)$                           & 7803 & 6756 & $43/2^{-}$   & $39/2^{-}$   \\
1051.6(7)\tnote{c,d} & 2.2(2)   &                                    & 0.08                          & 0.30                          & 0.19                            &                                &      &      &              &              \\
1056.6(4)\tnote{c,d} & 3.6(3)   &                                    & 0.08                          & 0.29                          & 0.19                            &                                &      &      &              &              \\
1066.7(10)                                                      & 2.8(2)   &                                    & 0.08                          & 0.29                          & 0.18                            & $(E2)$                           & 4389 & 3322 & $27/2^{+}$   & $23/2^{+}$   \\
1104.6(3)                                                      & 137(9)   & 0.37(8)\tnote{b}  & 0.07                          & 0.27                          & 0.17                            & $E3$\tnote{g}   & 1105 & 0    & $13/2^{+}$   & $ 7/2^{-}$   \\
1105.9(5)                                                      & 36(3)    &                                    & 0.07                          & 0.27                          & 0.17                            & $(E2)$                           & 5526 & 4420 & $33/2^{-}$   & $29/2^{-}$   \\
1127.9(10)\tnote{c,d}   & 3.3(3)   &                                    & 0.07                          & 0.25                          & 0.16                            &                                &      &      &              &              \\
1175.9(10)\tnote{c,d} & 2.6(2)   & 0.7(1)                             & 0.07                          & 0.23                          & 0.15                            & $M2$\tnote{h}   &      &      &              &              \\
1207.1(9)                                                      & 6.6(6)   & 0.17(6)                            & 0.06                          & 0.22                          & 0.14                            & $(M1/E2)$                        & 7926 & 6719 & $41/2^{-}$   & $39/2^{-}$   \\
1243.8(6)\tnote{c,d} & 5.0(4)   &                                    & 0.06                          & 0.20                          & 0.13                            &                                &      &      &              &              \\
1257.3(4)                                                      & 20.7(15) & 0.10(1)                            & 0.06                          & 0.20                          & 0.13                            & $E2$                             & 4741 & 3483 & $29/2^{+}$   & $25/2^{+}$   \\
1259.6(5)                                                      & 11.8(9)  & 0.12(2)                            & 0.06                          & 0.20                          & 0.13                            & $E2$                             & 5680 & 4420 & $33/2^{-}$   & $29/2^{-}$   \\
1270.1(6)                                                      & 9.0(7) & 0.11(2)                            & 0.06                          & 0.19                          & 0.13                            & $E2$                             & 4389 & 3119 & $27/2^{+}$   & $23/2^{+}$   \\
1326.4(6)                                                      & 11.6(8)  & 0.12(2)                            & 0.05                          & 0.17                          & 0.12                            & $E2$                             & 5248 & 3921 & $31/2^{-}$   & $27/2^{-}$   \\
1331.4(3)                                                      & 69(5)    & 0.13(3)\tnote{b}  & 0.05                          & 0.17                          & 0.12                            & $E2$                             & 2436 & 1105 & $17/2^{+}$   & $13/2^{+}$   \\
1385.0(4)                                                      & 17(1)    &                                    & 0.05                          & 0.16                          & 0.11                            & $(E2)$                           & 4868 & 3483 & $29/2^{+}$   & $25/2^{+}$   \\
1457.9(8)                                                      & 10.3(8)  &                                    & 0.05                          & 0.14                          & 0.10                            & $(E3)$\tnote{i} & 8785 & 7327 & $47/2^{-}$   & $41/2^{+}$   \\
1467.5(7)                                                      & 3.6(3)   &                                    & 0.04                          & 0.14                          & 0.10                            & $(E2)$                           & 4586 & 3119 & $27/2^{+}$   & $23/2^{+}$   \\
1538.0(10)                                                        & 2.8(2)   &                                    & 0.04                          & 0.12                          & 0.09                            & $(E2)$                           & 1538 & 0    & ($11/2^{-}$) & $ 7/2^{-}$   \\
1705.8(10)                                                        & 0.6(3)   &                                    & 0.03                          & 0.98                          & 0.07                            & $(E1)$                           & 2810 & 1105 & $15/2^{~}$   & $13/2^{+}$   \\
\end{longtable*}
\end{ThreePartTable}

\endgroup

\subsection{Conversion coefficients}

\begin{figure}[htb]
\centering
\includegraphics[width=\columnwidth]{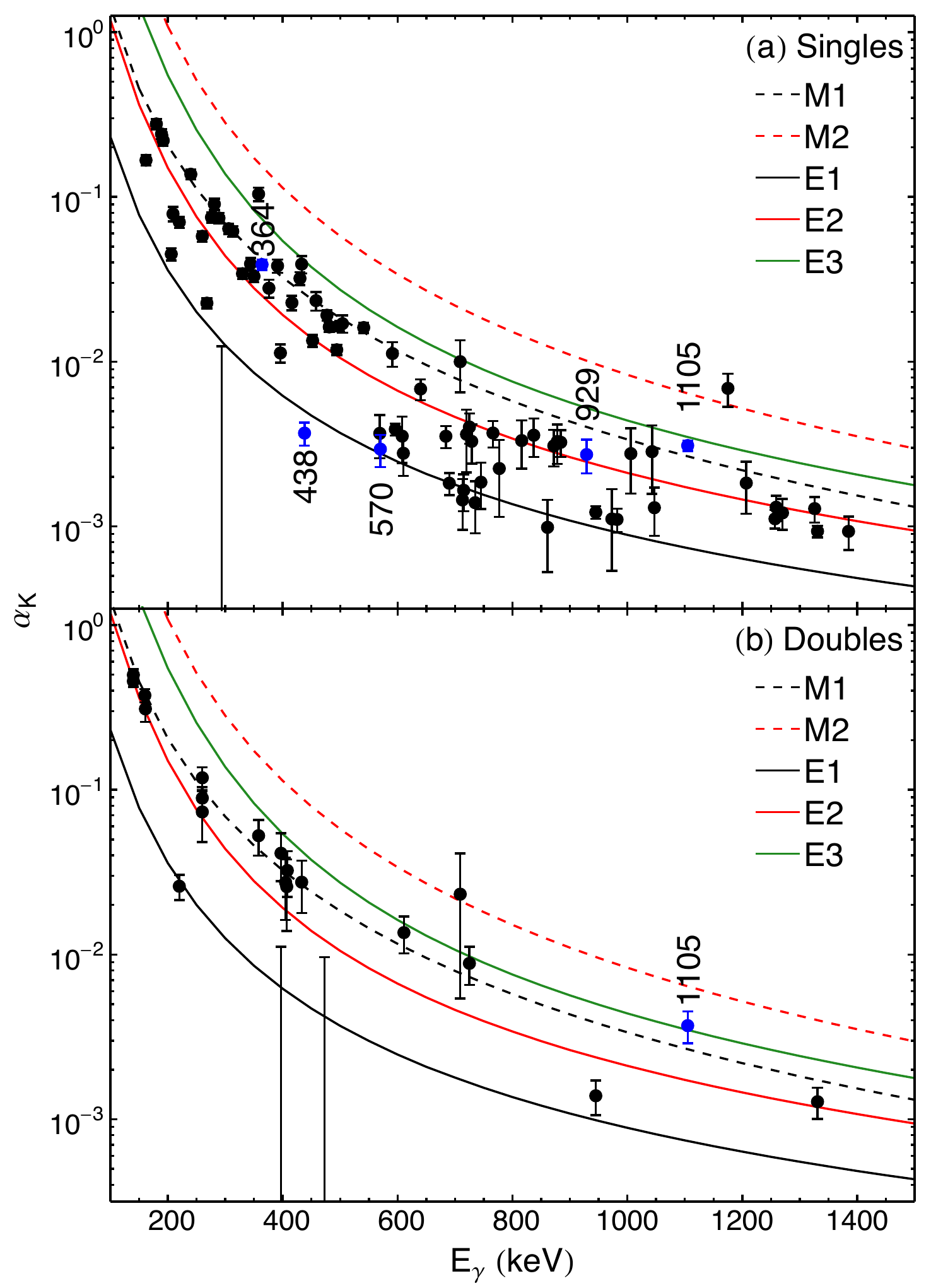}
\caption{Internal conversion coefficients measured for transitions in $^{145}$Sm from (a) $\gamma$ and $e^{-}$ singles, and (b) $\gamma$-$\gamma$ and $\gamma$-$e^{-}$ coincidences (doubles).  Theoretical curves were calculated using BrIcc~\cite{Kibedi2008}.  Some transitions relevant in the discussion of the $^{145}$Sm level scheme have been labelled with the data points highlighted in blue.}
\label{fig:145Sm-ICCs}
\end{figure}

Conversion coefficients were deduced for $\approx$80 transitions assigned to $^{145}$Sm.  These values are listed in Table~\ref{tab:transitions} and shown in Figure~\ref{fig:145Sm-ICCs}.  Conversion coefficients were determined using the $\gamma$-ray and $e^{-}$ singles data, as well as the $\gamma$-$\gamma$ and $\gamma$-$e^{-}$ coincidence data.  $K$-shell conversion coefficients were extracted directly from the singles data by comparing the efficiency-corrected intensities of the $K$-electron peak with those for the equivalent $\gamma$~ray.

In cases where the electron peaks were contaminated, a $\gamma$-ray gate was placed on the $\gamma$-$\gamma$ and $\gamma$-$e^{-}$ coincidence matrices to isolate the transitions of interest (for example, the 1105-keV, $1105 \rightarrow 0$~keV transition was isolated from the 1106-keV, $5526 \rightarrow 4420$~keV transition by gating on the 1260-keV, $5680 \rightarrow 4420$~keV transition).  The conversion coefficient was then determined from the efficiency-corrected intensities in the resulting projections.  Conversion coefficients extracted using this method are listed in Table~\ref{tab:doubleICCs}. 

Transition multipolarities (see Table~\ref{tab:transitions}) were assigned by comparing the measured conversion coefficients with theoretical values calculated using the BrIcc program~\cite{Kibedi2008}.  Due to the highly fragmented decay through the $^{145}$Sm level scheme, in many cases transition multipolarities were restricted by assignments made to parallel decay paths.  For the most part the assignments made to parallel branches were in agreement; however, in cases where the conversion coefficients disagreed (e.g.~the 929-keV transition, discussed further in \S\ref{sec:SpinsandParities}), or where they could not be measured (e.g.~the 1006-keV transition), multipolarity assignments have been labelled as tentative.

\begin{table*}[htb]
\begin{threeparttable}
\centering
\caption{Conversion coefficients in $^{145}$Sm extracted from $\gamma$-$\gamma$ and $\gamma$-$e^{-}$ coincidence spectra.  Theoretical conversion coefficients for the pure multipolarities are from BrIcc~\cite{Kibedi2008}.}
\label{tab:doubleICCs}
\begin{tabular*}{\textwidth}{@{\extracolsep{\fill}}c c c c c c c c c c c c}

\toprule
\toprule
                  &                        & \multicolumn{4}{c}{$100 \alpha_{K}$} &    &         &         &         &         \\
\cmidrule{3-6}
$E_{\gamma}$(keV) & $E_{\text{gate}}$(keV) & expt & $E1$ & $M1$ & $E2$                       & $XL$ & $E_{i}$(keV) & $E_{f}$(keV) & $J_{i}$ & $J_{f}$ \\
\midrule
140.3(3) & 364.0(3)  & 45.7(35)&      &      &      &         &      &      &            &            \\
         & 1104.6(3) & 49.9(39)&      &      &      &         &      &      &            &            \\    
         & mean      & 47.8(26)& 9.30 & 54.8 & 44.5 & $M1/E2$ & 3119 & 2979 & $23/2^{+}$ & $21/2^{+}$ \\
\midrule
161.3(4) & 364.0(3)  & 37.4(34)& 6.39 & 37.1 & 29.1 & $M1/E2$ & 5029 & 4868 & $31/2^{+}$ & $29/2^{+}$ \\
\midrule
161.4(9) & 438.4(3)  & 31.0(51)& 6.38 & 37.0 & 29.0 & $M1/E2$ & 3483 & 3322 & $25/2^{+}$ & $23/2^{+}$ \\
\midrule
219.7(3) & 180.2(3)  & 2.6(4)  & 2.80 & 15.8 & 11.2 & $E1$    & 2930 & 2710 & $21/2^{+}$ & $19/2^{-}$ \\
\midrule
260.1(4) & 364.0(3)  & 8.9(15) &      &      &      &         &      &      &            &            \\
         & 540.0(3)  & 11.8(18)&      &      &      &         &      &      &            &            \\
         & 1326.4(6) & 7.3(25) &      &      &      &         &      &      &            &            \\
         & mean      & 9.4(11) & 1.81 & 10.1 & 6.72 & $M1$    & 6216 & 5956 & $37/2^{-}$ & $35/2^{-}$ \\
\midrule
358.2(5) & 364.0(3)  & 5.3(13) & 0.81 & 4.32 & 2.61 & $M1$    & 6719 & 6361 & $39/2^{-}$ & $37/2^{-}$ \\
396.1(4) & 1326.4(6) & 4.1(13) & 0.63 & 3.33 & 1.97 & $M1$    & 7327 & 6931 & $41/2^{+}$ & $39/2^{+}$ \\
396.7(4) & 540.0(3)  & <1      & 0.63 & 3.31 & 1.96 & $E1$    & 5903 & 5507 & $35/2^{-}$ & $33/2^{+}$ \\
405.3(4) & 1326.4(6) & 3(1)    & 0.60 & 3.13 & 1.85 & $M1$    & 6361 & 5956 & $37/2^{-}$ & $35/2^{-}$ \\
407.2(7) & 452.4(5)  & 3(1)    & 0.59 & 3.10 & 1.82 & $M1$    & 8333 & 7926 & $43/2^{-}$ & $41/2^{-}$ \\
408.3(4) & 540.0(3)  & 4(1)    & 0.59 & 3.08 & 1.81 & $M1$    & 8785 & 8377 & $47/2^{-}$ & $45/52^{-}$ \\
432.8(9) & 1326.4(6) & 2.7(9)  & 0.51 & 2.65 & 1.54 & $M1$    & 5680 & 5248 & $33/2^{-}$ & $31/2^{-}$ \\
473.8(10) & 1257.3(4) & <0.96   & 0.42 & 2.10 & 1.21 & $E1$    & 7404 & 6931 & $41/2^{-}$ & $39/2^{+}$ \\
610.7(5) & 871.8(5)  & 1.3(3)  & 0.24 & 1.11 & 0.64 & $M1$    & 5031 & 4420 & $31/2^{-}$ & $29/2^{-}$ \\
708.1(5) & 1104.6(3) & 2.3(17) & 0.17 & 0.77 & 0.45 & $M2$\tnote{a}    & 6216 & 5507 & $37/2^{-}$ & $33/2^{+}$ \\
725.7(7) & 438.4(3)  & 0.9(2)  & 0.16 & 0.73 & 0.42 & $M1$    & 4647 & 3921 & $29/2^{-}$ & $27/2^{-}$ \\
944.8(3) & 1104.6(3) & 0.14(3) & 0.10 & 0.39 & 0.24 & $E1$    & 2049 & 1105 & $15/2^{-}$ & $13/2^{+}$ \\
1104.6(3)& 1259.6(5) & 0.37(8) & 0.07 & 0.27 & 0.17 & $E3$\tnote{b}    & 1105 & 0    & $13/2^{+}$ & $ 7/2^{-}$ \\
1331.4(3)& 1104.6(3) & 0.13(3) & 0.05 & 0.17 & 0.12 & $E2$    & 2436 & 1105 & $17/2^{+}$ & $13/2^{+}$ \\
\bottomrule
\bottomrule
\end{tabular*}
\begin{tablenotes}\footnotesize
\item[a]{$\alpha_{K}(708.1; M2)=2.12\times 10^{-2}$}
\item[b]{$\alpha_{K}(1104.6; E3)=0.35\times 10^{-2}$}
\end{tablenotes}
\end{threeparttable}
\end{table*}

\subsection{Lifetimes}
\label{sec:lifetimes}

Lifetimes were measured for both the high-spin isomer and the 1105-keV state.  Other previously reported lifetimes in the level scheme were too short to measure in the current experiment, and no evidence of other long-lived states was found.
  
\subsubsection{High-spin isomer}

\begin{figure}[htb]
\centering
\includegraphics[width=\columnwidth]{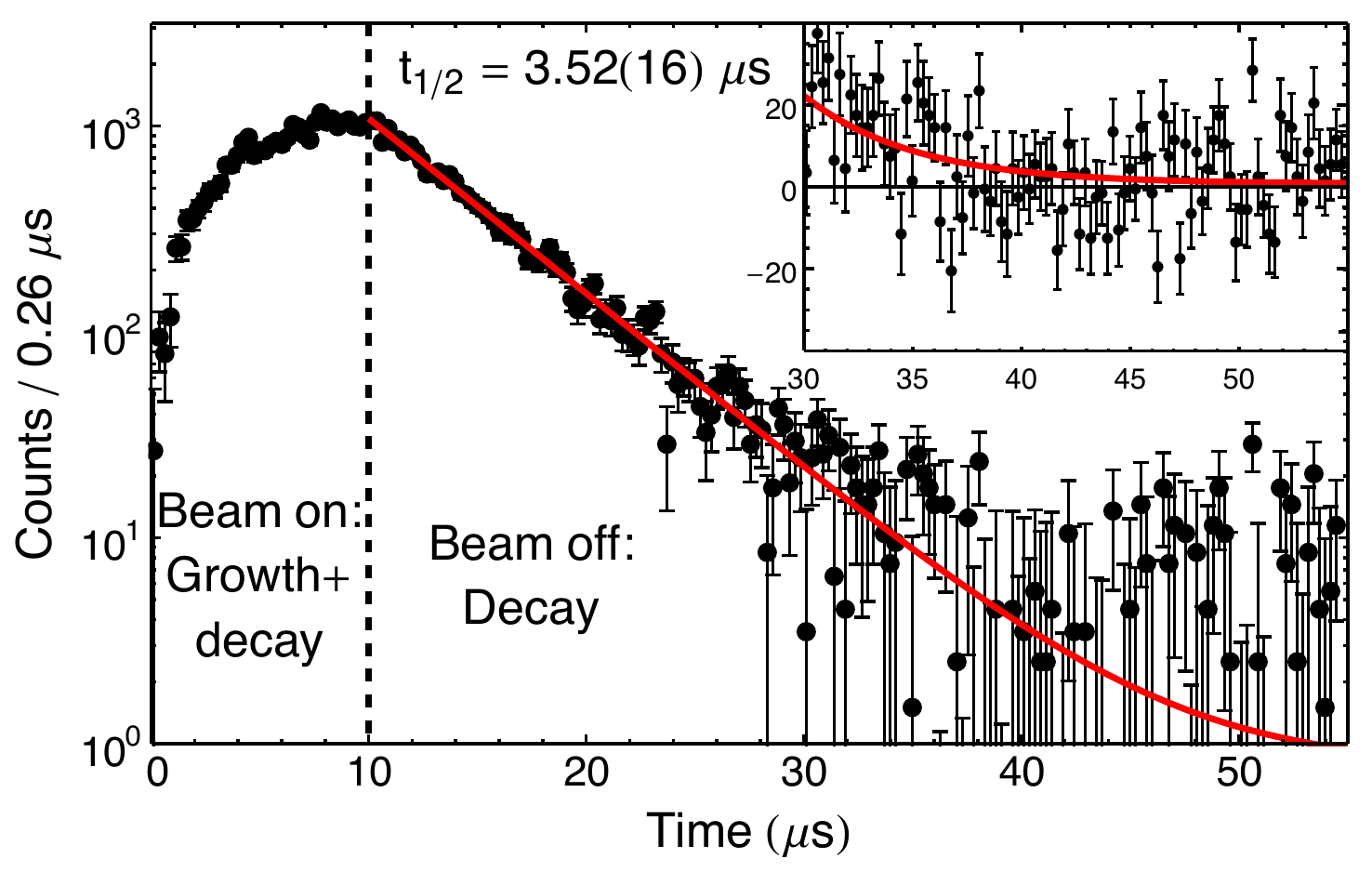}
\caption{Decay curve of the high-spin isomer (see text for details).  The inset shows the background region on a linear scale.}
\label{fig:145Sm-Lifetime_8815}
\end{figure}

The lifetime of the high-spin isomer in $^{145}$Sm was determined using data from the chopped-beam measurement with 9.63/48.15~$\mu$s on/off cycles.  A sum of gates was placed on known $^{145}$Sm $\gamma$~rays (the 140-, 161-, 180-, 193-, 268-, 276-, 313-, 364-, 396-, 397-, 438-, 481-, 494-, 499-, 570-, 571-, 611-, 863-, 945-, 1105-, and 1331-keV transitions) and the projected time-of-arrival for these transitions relative to the beam burst is shown in Figure~\ref{fig:145Sm-Lifetime_8815}.  The out-of-beam portion of this distribution shows the decay of the isomer and, when fitted with a decaying exponential above a flat background (consistent with zero within uncertainty), gives a half life of 3.52(16)~$\mu$s, significantly longer than the value of 0.96$^{+0.19}_{-0.15}~\mu$s reported by Ferragut \textit{et~al.}~\cite{Ferragut1993}.

In their work, Ferragut \textit{et~al.}~\cite{Ferragut1993} populated the $^{145}$Sm isomer using the $^{16}$O($^{136}$Xe,7n) reaction.  Evaporation residues were implanted in a plastic scintillator and particle-$\gamma$ events were recorded.  The lifetime of the isomer was then measured by projecting the time difference for all particle-$\gamma$ coincidences.  This lifetime was measured without applying any $\gamma$-ray gates, with Ferragut \textit{et~al.} attributing most of the observed $\gamma$~rays to the decay of the $^{145}$Sm isomer.  Figure~1 of Ref.~\cite{Ferragut1993} shows the $\gamma$-ray projection of particle-$\gamma$ coincidences measured in that work.  A significant continuous background can be seen at low energies.  Considering that the detectors in their measurement were Compton suppressed~\cite{Ferragut1993}, this background could not have be caused by Compton scattering.  Therefore, the continuum is attributed to room background, a conclusion supported by the presence of a dominant 1461-keV peak from the $^{40}$K decay.  Considering this contamination, the background subtraction required to produce the particle-$\gamma$ time distribution for the lifetime measurement would have been non-trivial without the use of $\gamma$-ray gates.  These difficulties may account for the marked difference in their measured lifetime compared with the present work.  Further analysis in the present work will use the new lifetime value of ${t_{1/2}=3.52~\mu}$s.

\subsubsection{1105-keV state}

\begin{figure}[htb]
\includegraphics[width=\columnwidth]{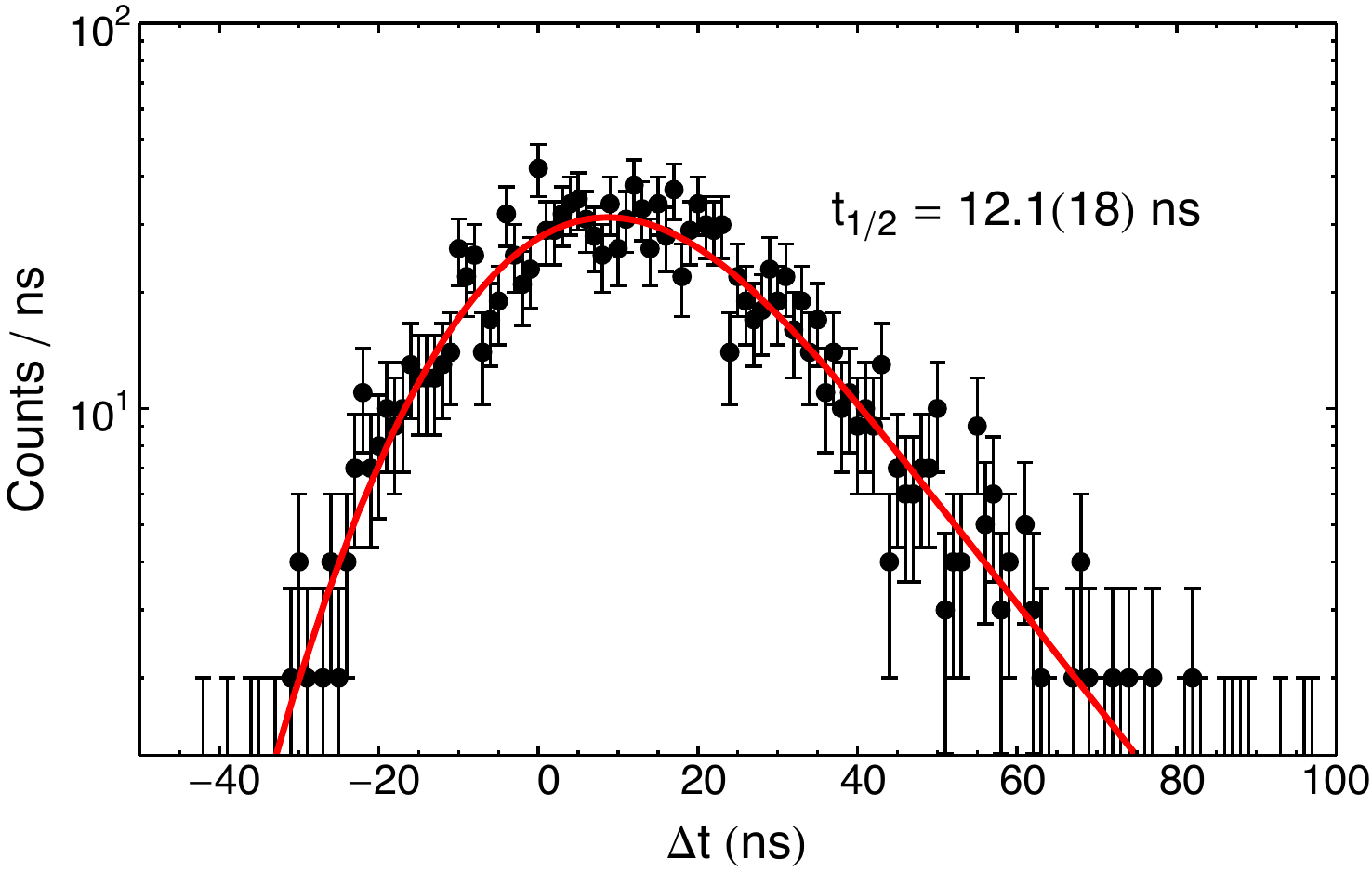}
\centering
\caption{$\gamma$-$\gamma$ time difference spectrum showing the lifetime of the 1105-keV 13/2$^{+}$ state (see text for details).}
\label{fig:145Sm-Lifetime_1105}
\end{figure}

The lifetime of the 1105-keV state was measured using $\gamma$-$\gamma$ data from the 4.17-/4.17-$\mu$s (on/off) measurement.  Projecting the time difference between the depopulating 1105-keV transition and a sum of gates on the 945- and 1331-keV transitions feeding the state produces a typical time-difference spectrum with an exponential component (corresponding to the lifetime of the state) convoluted with a prompt Gaussian (due to the time response of the detectors); this is shown in Figure~\ref{fig:145Sm-Lifetime_1105}.  Fitting this distribution yields a half life of 12.1(18)~ns, in agreement with the literature value of 13.5(15)~ns~\cite{Browne2001}.  A weighted average of these two values gives a lifetime of $t_{1/2}=12.9(12)$~ns, which is adopted in subsequent analysis and interpretation.

\begin{turnpage}
\begin{figure*}
\centering
\includegraphics[width=\textheight]{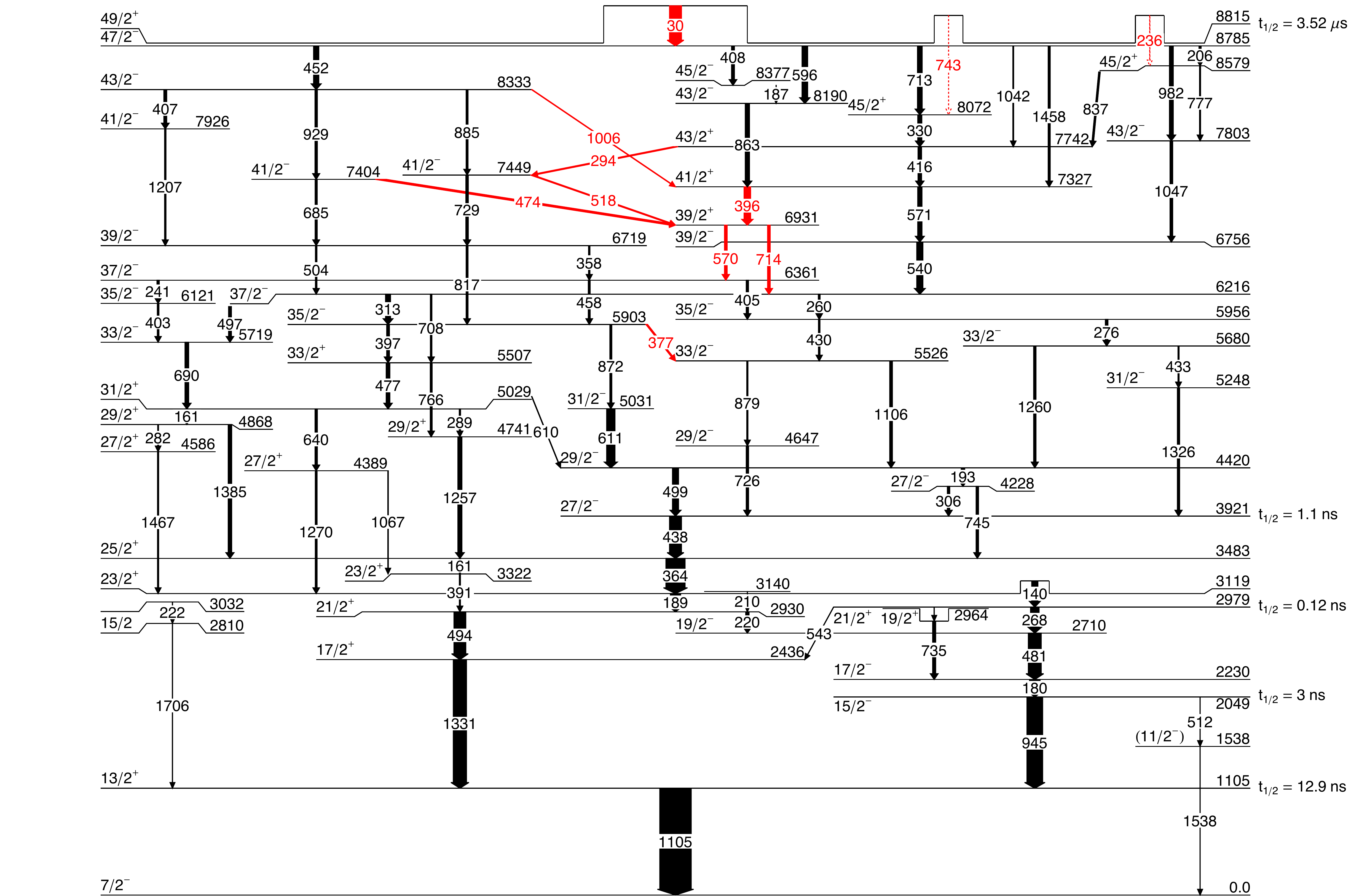}
\caption{Proposed level scheme for $^{145}$Sm.  Newly added transitions are highlighted in red.  The lifetime of the 49/2$^{+}$ isomer is from the present work, while the lifetime for the 1105-keV state is a weighted average of the values from the present work and Ref.~\cite{Browne2001}. The remaining lifetimes are from Ref.~\cite{Browne2001}.  The dashed 743- and 236-keV transitions depopulating the 8815-keV state are unobserved $E2$ transitions that are discussed further in the text.}
\label{fig:145Sm-levelscheme}
\end{figure*}
\end{turnpage}

\subsection{Level scheme}

The proposed level scheme for $^{145}$Sm is shown in Figure~\ref{fig:145Sm-levelscheme}.  For the most part, the level scheme presented here agrees with that of Odahara \textit{et al.}~\cite{Odahara1997}; however, there are a number of significant changes that will be discussed.

\subsubsection{Arrangement of \texorpdfstring{$\gamma$}{gamma} rays}

The $^{145}$Sm level scheme was deduced using $\gamma$-$\gamma$ coincidences.  Several new transitions have been added to the decay scheme and some of the previously assigned $^{145}$Sm transitions have been reassigned.  Throughout the following discussion, intensities are reported relative to the 364-keV, $3483 \rightarrow 3119$~keV transition in the relevant gate or spectrum.

\begin{figure}[htb]
    \centering
    \includegraphics[width=0.8\columnwidth]{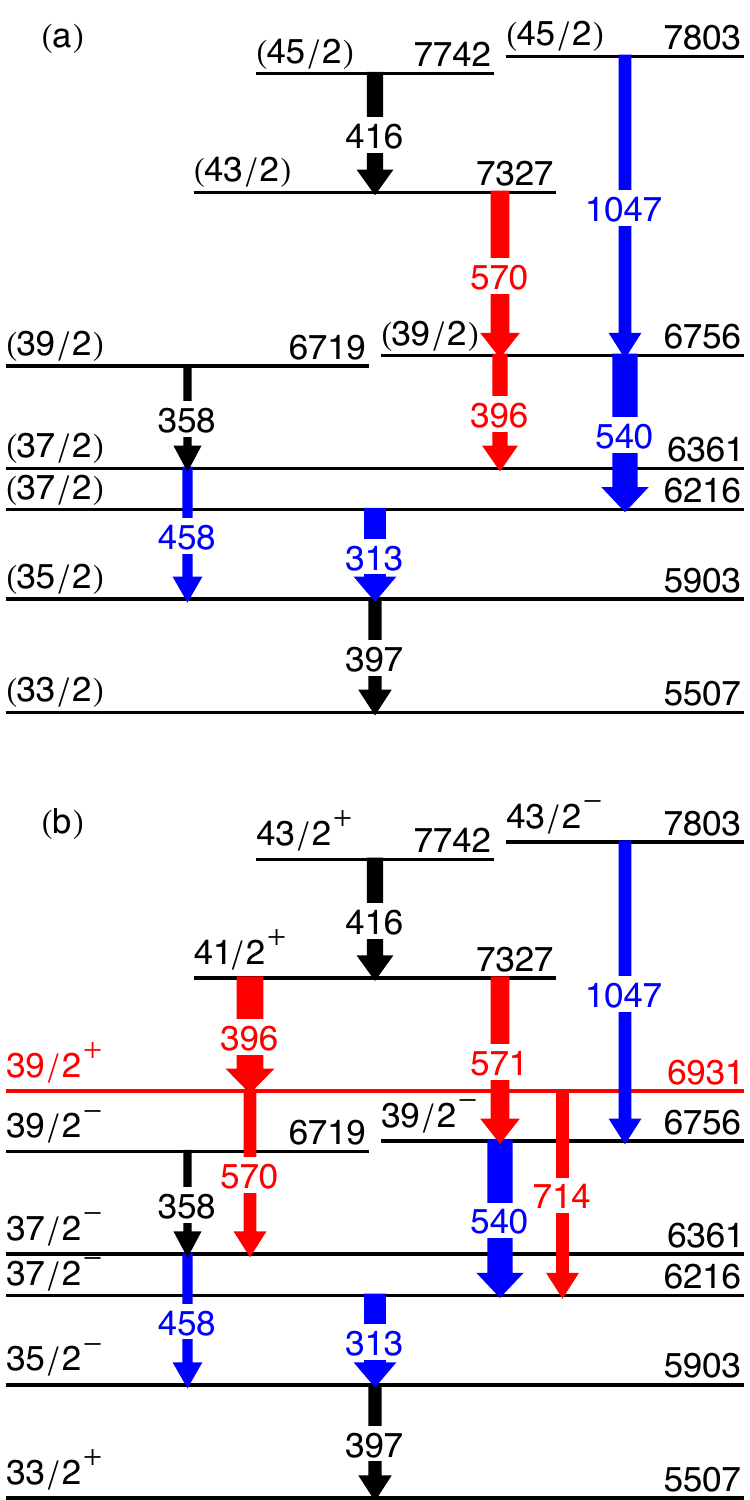}
\caption{Subset of the $^{145}$Sm level scheme proposed in (a) Ref.~\cite{Odahara1997} and (b) the present work.  Some key transitions and states have been highlighted in red, and gating transitions are highlighted in blue (see text).}
    \label{fig:145Sm-levelscheme_396}
\end{figure}

Odahara \textit{et~al.}~\cite{Odahara1997} identified the presence of a 396/397-keV doublet in the $^{145}$Sm $\gamma$-ray spectrum.  The 397-keV $\gamma$ ray was assigned to the $5903 \rightarrow 5507$~keV transition and the 396-keV $\gamma$ ray was assigned to depopulate the state at 6756~keV, feeding the 6361-keV state  (see Figure~\hyperref[{fig:145Sm-levelscheme_396}]{\ref*{fig:145Sm-levelscheme_396}a}).  This second assignment was made based on a comparison of the intensity of the 396-keV peak in gates on the 313-keV, $6216 \rightarrow 5903$~keV, 458-keV, $6361 \rightarrow 5903$~keV, and 540-keV, $6756 \rightarrow 6216$~keV transitions.  In particular, the 396-keV $\gamma$~ray was much weaker in gates on the 313- and 540-keV transitions than in the gate on the 458-keV transition.

A similar comparison of gates in the present work shows the relative intensity of the 396-keV $\gamma$~ray to be comparable in both the 458- and 313-keV gates, as shown in Figure~\ref{fig:145Sm-Gates_458,313,540,1047}.  Also shown is a gate on the 1047-keV transition, populating the state at 6756~keV, which shows a distinct lack of intensity in the 396-keV peak, similar to the gate on the 540-keV transition.  If the Odahara level scheme was correct, the 1047-keV transition should be in direct coincidence with the 396-keV transition and the associated gate should have significant intensity in the 396-keV peak.  The present data, therefore, strongly suggest that the placement of the 396-keV transition by Odahara \textit{et~al.}~\cite{Odahara1997} is incorrect.  Following a careful analysis of the $\gamma$-$\gamma$ coincidences in this region, the 396-keV transition has been reassigned to depopulate the 7327-keV state, feeding a newly proposed state at 6931~keV; this is discussed in more detail below.

\begin{figure}[htb]
\centering
\includegraphics[width=\columnwidth]{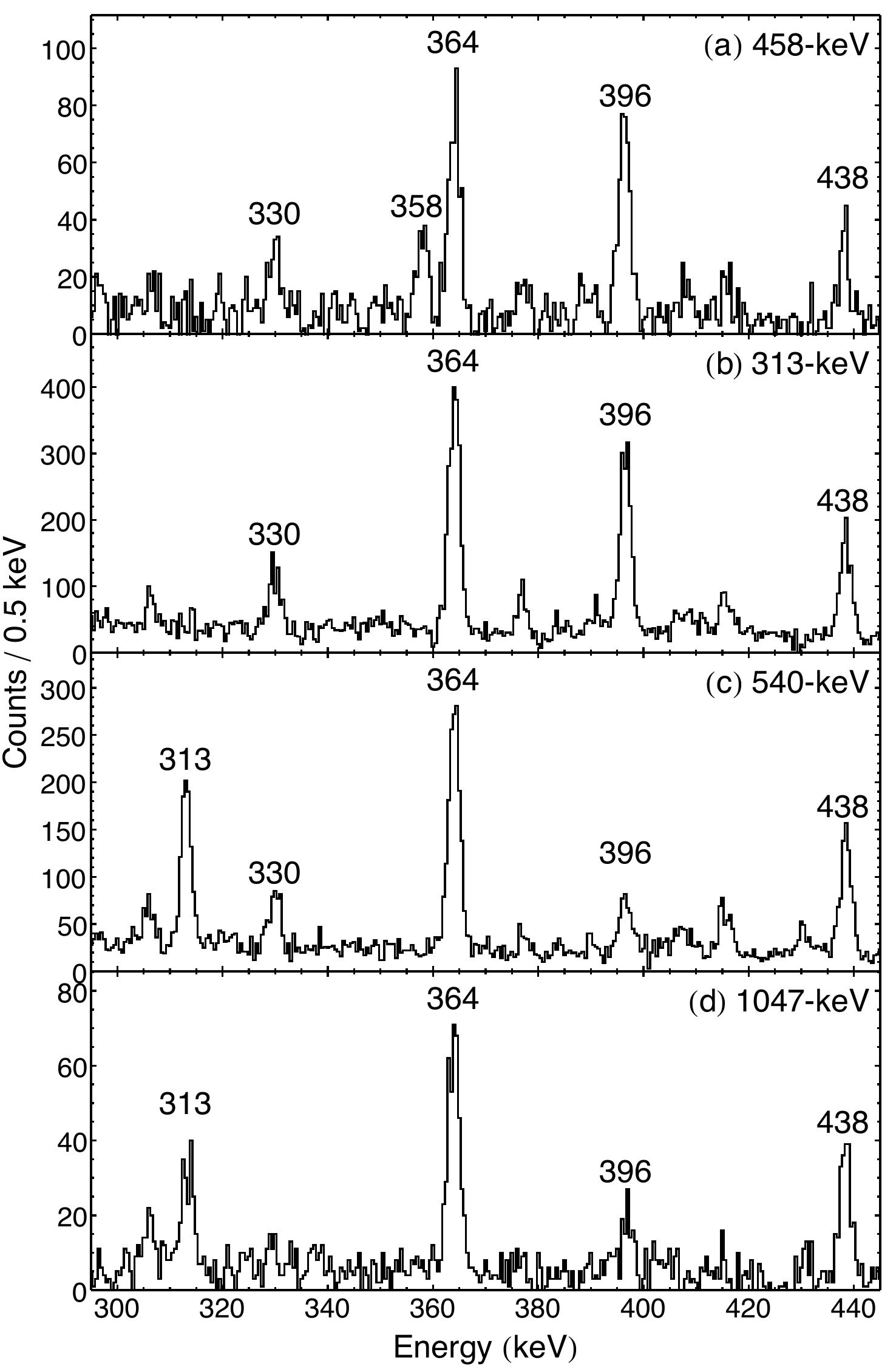}
\caption{Coincidence spectra showing the peak at 396~keV in gates on the (a)~458-, (b)~313-, (c)~540- and (d)~1047-keV transitions.  The intensity of the 396-keV peak relative to the 364-keV peak in each gate is 1.02(8), 0.79(4), 0.24(6), and 0.23(3), respectively.  The low intensity in the 540- and 1047-keV gates is associated with the 397-keV, $5903 \rightarrow 5507$~keV transition.}
\label{fig:145Sm-Gates_458,313,540,1047}
\end{figure}

The comparison of gates on the 458-, 313-, 396- and 540-keV $\gamma$~rays shown in Figure~\ref{fig:145Sm-Gates_458,313,396,540} indicates that the 570-keV $\gamma$~ray is, in fact, a 570/571-keV doublet.  This conclusion is further supported by gates on each half of the combined peak (see Figure~\ref{fig:145Sm-Gates_570,571}) which show the 571-keV $\gamma$~ray in coincidence with the 540-keV $\gamma$~ray (but not the 396-keV $\gamma$~ray), and the 570-keV transition in coincidence with the 396-keV $\gamma$~ray (but not the 540-keV $\gamma$~ray).  It should also be noted that the 570- and 571-keV transitions are not in self-coincidence, placing the 396-570-keV cascade parallel to the 571-540-keV cascade.  This indicates that both cascades depopulate the state at 7327~keV and connect to the 6361- and 6216-keV states, respectively.

\begin{figure}[htb]
\includegraphics[width=\columnwidth]{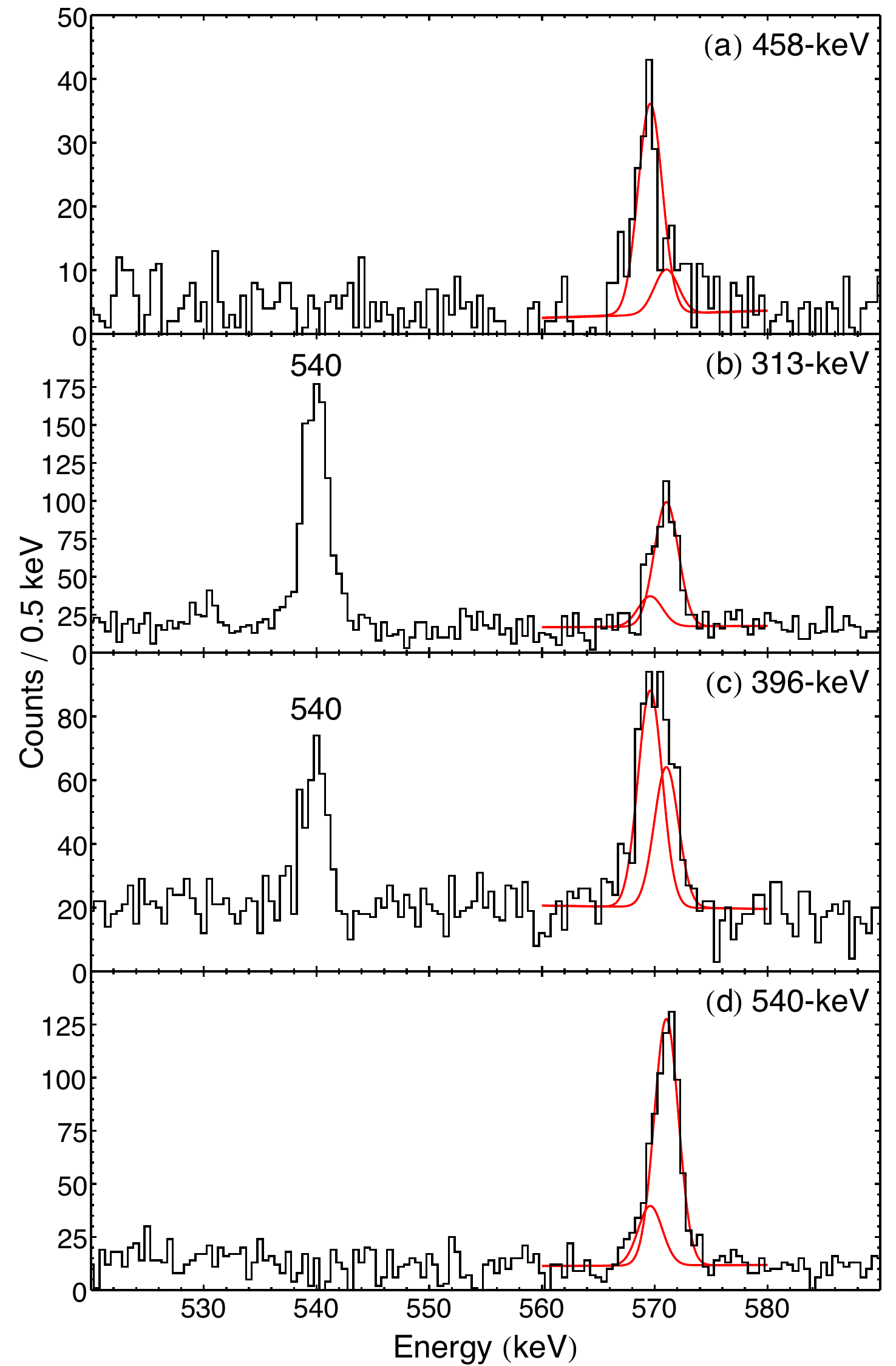}
\centering
\caption{Coincidence spectra showing the doublet at 570/571~keV in gates on the (a)~458-, (b)~313-, (c)~396- and (d)~540-keV transitions.  The fitted peaks at 569.6 and 571.0~keV are shown in red.  The peak positions were fixed in each of these fits.}
\label{fig:145Sm-Gates_458,313,396,540}
\end{figure}

\begin{figure}[htb]
\centering
\includegraphics[width=\columnwidth]{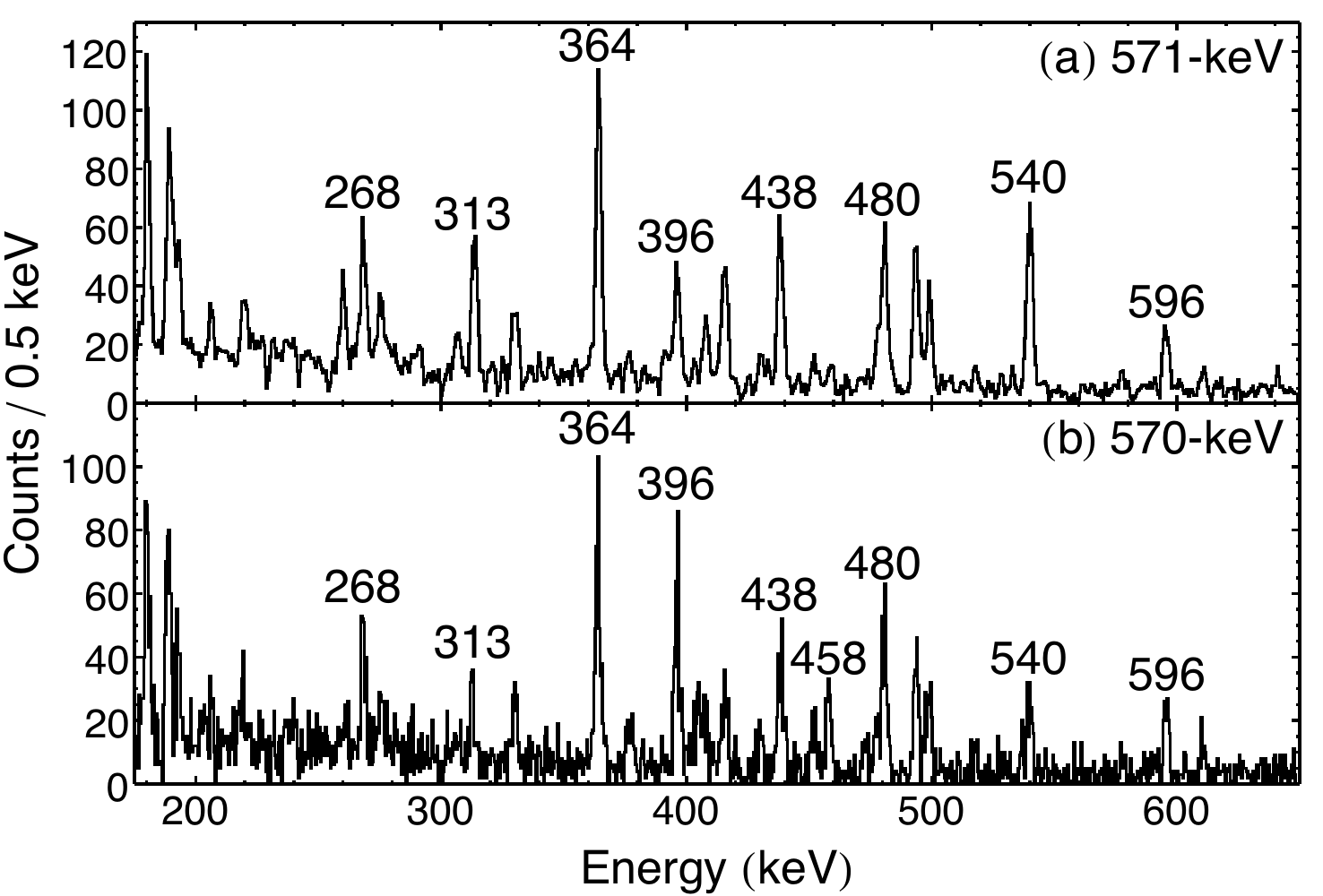}
\caption{Coincidence spectra showing transitions in coincidence with the (a)~571- and (b)~570-keV transitions.  Note the different intensities of the 396- and 540-keV peaks in each gate.}
\label{fig:145Sm-Gates_570,571}
\end{figure}

Other $\gamma$-ray gates in this region indicate the presence of a 714-keV transition, separate to the 713-keV, $8785 \rightarrow 8072$~keV transition already placed in the level scheme.  This $\gamma$~ray was observed in coincidence with both the 396- and 313-keV transitions, but not with the 570- or 571-keV transitions.  This observation, along with the energy sums for the 396-714- and 571-540-keV cascades, fixes the ordering of the 396- and 570-keV transitions, resulting in the addition of a new state at 6931~keV.  This new arrangement of $\gamma$~rays is shown in Figure~\hyperref[{fig:145Sm-levelscheme_396}]{\ref*{fig:145Sm-levelscheme_396}b}.  With the new level scheme, neither the 571- nor the 1047-keV transition is in coincidence with the 396-keV $\gamma$~ray, accounting for the lack of intensity in these $\gamma$-ray gates.  Furthermore, the population of both the 6361-keV and 6216-keV states through the 6931-keV state accounts for the comparable intensity of the 396-keV transition in gates on both the 458- and 313-keV $\gamma$~rays.

With the addition of the 6931-keV state, several other $\gamma$~rays can be assigned to the $^{145}$Sm level scheme.  The 518-keV transition appears in coincidence with the 885-keV $\gamma$~ray but not the 729-keV $\gamma$~ray, suggesting it depopulates the state at 7449~keV.  The agreement between the energy sums for the 885-518-570- and 885-729-358-keV cascades places this transition as feeding the 6931-keV state.  Similar logic was used to place the 474-keV, ${7404 \rightarrow 6931}$~keV transition.  These newly placed transitions are shown in red in Figure~\ref{fig:145Sm-levelscheme}.

Some comment is required for the assignment of the 714-keV transition.  In the level scheme of Odahara~\textit{et~al.}~\cite{Odahara1997}, a transition of similar energy was assigned as a part of a 714-394-keV, ${5029 \rightarrow 4315 \rightarrow 3921}$~keV cascade.  In the Odahara~\textit{et~al.}~data set, however, the relative intensity of the 714-keV transition is given as 14.3 (relative to the 364-keV $\gamma$~ray).  Comparatively, no intensity is reported for their subsequent 394-keV transition.  Similarly, no indication of a 394-keV transition is found in the present work.  This makes the presence of this transition in the $^{145}$Sm level scheme unlikely.  Moreover, the intensity of the 714-keV $\gamma$~ray in gates on both the 438- and 499-keV $\gamma$~rays agree within uncertainty (11.4(15) and 13(2), respectively).  If the 714-394-keV cascade fed the state at 3921~keV, bypassing the state at 4420~keV, greater intensity would be expected in the gate on the 438-keV transition.  As such, instead of introducing a third transition of $\approx$714~keV to the level scheme, it is proposed that the 714-keV transition be moved to depopulate the new 6931-keV state and that the 394-keV transition and 4315-keV state be removed entirely.

Also of note is the addition of a 15-keV transition between the states at 2979 and 2964~keV.  While not observed directly, the existence of this transition was inferred based on the observed coincidence between the 140-keV, $3119 \rightarrow 2979$~keV and 735-keV, $2964 \rightarrow 2230$~keV transitions.

\begin{figure}[htb]
\centering
\includegraphics[width=\columnwidth]{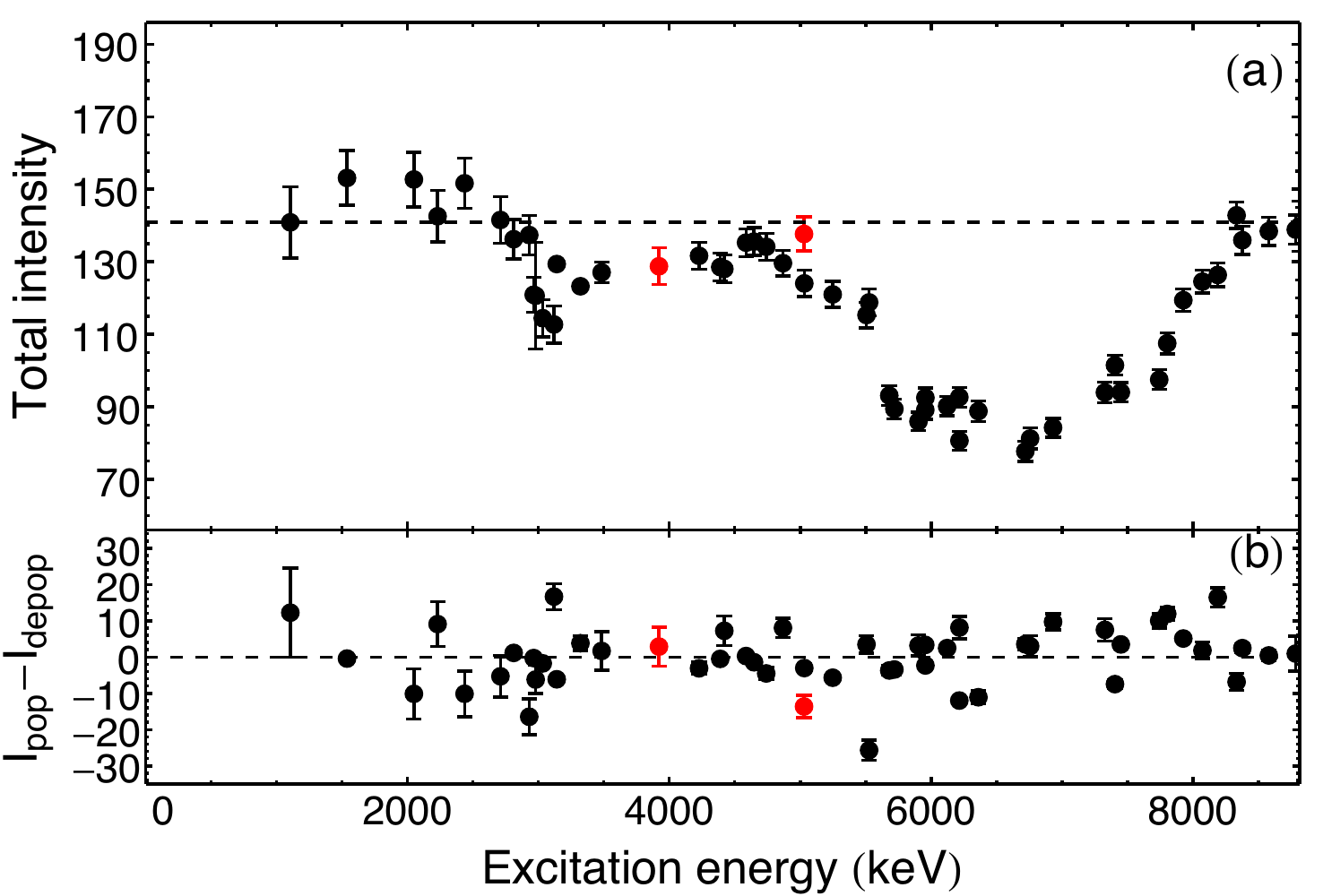}
\caption{Transition intensity in the $^{145}$Sm level scheme.  (a) The total transition intensity below each state (see text) (b) the difference in the total intensity populating and depopulating each state.  The points for the 3921-keV, 27/2$^{-}$ and 5029-keV, 31/2$^{+}$ states are highlighted in red.  In both parts the data points should fall on the dotted line if all transitions and their intensities were identified correctly.}
\label{fig:145Sm-intensities}
\end{figure}

Figure~\hyperref[{fig:145Sm-intensities}]{\ref*{fig:145Sm-intensities}a} shows the total intensity of assigned transitions in the $^{145}$Sm level scheme as a function of excitation energy.  These values were obtained by drawing a `line' across the level scheme at a given energy and summing the total intensity of all transitions that cross the line.  This was evaluated just below each state in the level scheme.  As all of the decay intensity flows between the 49/2$^{+}$ isomer and the 7/2$^{-}$ ground state, the total transition intensity should be constant throughout the level scheme.  However, approximately 40\% of the transition intensity is missing above the first 31/2$^{+}$ state at 5029~keV (a similar pattern is seen in the level scheme proposed by Odahara \textit{et~al.}~\cite{Odahara1997}).  Figure~\hyperref[{fig:145Sm-intensities}]{\ref*{fig:145Sm-intensities}b} shows the intensity balance for each state.  Despite the significant intensity missing in the level scheme, determining where this intensity is lost is not trivial.  A similar problem was present in the decay scheme for the $^{147}$Gd isomer~\cite{Broda1982}; however, a recent study of $^{147}$Gd~\cite{Broda2020} largely resolved this, deducing an extraordinarily complex decay scheme with the missing intensity split across hundreds of low-intensity ($\num{\approx 10}^{-3}$) $\gamma$ rays within the $5-8$-MeV excitation energy range.  It is speculated that similar fragmentation occurs in the $^{145}$Sm level scheme but was below the detection limit of this experiment ($I_{\gamma} \approx 10^{-2}$).

\subsubsection{Spins and parities}
\label{sec:SpinsandParities}

In the previous work on $^{145}$Sm, transition multipolarities were assigned based on angular distributions~\cite{Odahara1997} and conversion coefficients~\cite{Piiparinen1991} up to the state at 4420~keV.  Above this, tentative assignments had been made based on DIPM calculations and systematic arguments.  In the present work, transition multipolarities have been assigned using the measured conversion coefficients up to the high-spin isomer near 9~MeV.  While these are largely consistent with the tentative assignments made in Refs.~\cite{Piiparinen1991,Odahara1997}, there are a few key exceptions.

\begin{figure}[htb]
\centering
\includegraphics[width=\columnwidth]{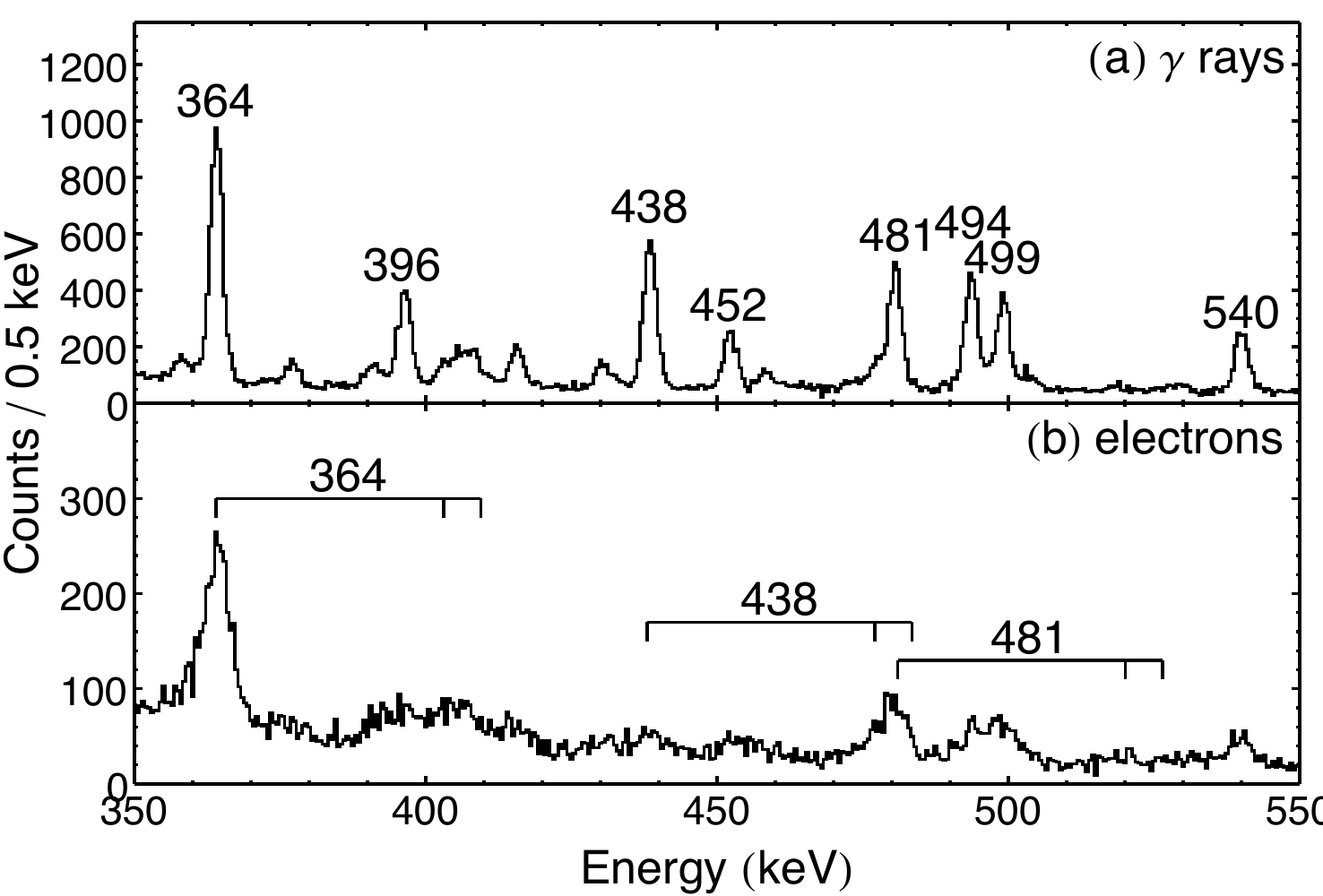}
\caption{Coincidence spectra showing the relative $K$-electron intensities for the 438-keV, $E1$, 364-keV, $M1$, and 481-keV, $M1$ transitions in a gate on the 1105-keV $\gamma$ ray. The electron energies have been shifted to align the $\gamma$-ray and $K$-conversion peaks.}
\label{fig:145Sm-Gates_1105}
\end{figure}

Piiparinen \textit{et~al.}~\cite{Piiparinen1991} assigned the 438-keV, $3921 \rightarrow 3483$~keV $\gamma$~ray as $M1$ based on angular distributions and conversion coefficients.  In the present work, however, the conversion coefficient clearly indicates an $E1$ multipolarity (note the low conversion-electron intensity in Figure~\ref{fig:145Sm-Gates_1105}).  This $E1$ assignment changes the spin and parity of the 3921-keV state to 27/2$^{-}$, making it consistent with the systematics of the other $N=83$ isotones that also feature a prominent 27/2$^{-}$ isomer, as shown in Figure~\ref{fig:145Sm-N=83_Isomers}.  In Ref.~\cite{Odahara1997}, Odahara \textit{et~al.} note that their DIPM calculations do not predict low lying 27/2$^{+}$ and 29/2$^{+}$ states and conclude that the lack of a 27/2$^{-}$ isomer remains an open problem.  Changing the multipolarity assignment of the 438-keV transition resolves both these issues.

Moreover, the transition strength for a 438-keV, $M1$ transition with a 1.1-ns half life is significantly hindered (see Table~\ref{tab:145Sm-BXL}) which would indicate a dramatic change in particle configuration between the 3921- and 3483-keV states.  Although no configuration assignment was made to the 3921-keV state in Ref.~\cite{Odahara1997}, some insight into the configuration changes can be gained by looking at other transitions in this part of the level scheme.  The presence of the 306-keV $M1/E2$ and the 745-keV $E1$ transitions between the non-isomeric state at 4228~keV and the states at 3921 and 3483~keV, suggests that a large difference in configuration between these states is unlikely.  Assuming that the configuration for the 4228-keV states is similar to one of the other two states (were it different to both, then the 4228-keV state should also be isomeric), this state should only decay to the state with which it has a similar configuration, the other branch being hindered.  The fact that both the 306- and 745-keV transitions are present (and with similar intensity) suggests that the particle configurations of all three states are similar, meaning a hindered $M1$ assignment for the 438-keV transition is difficult to justify.

On the other hand, the transition strength for an $E1$ assignment, also shown in Table~\ref{tab:145Sm-BXL} (labelled as `$^{145}$Sm (present work)'), is consistent with the expected strength for an $E1$ transition ($\approx$10$^{-6}$~W.u.) and does not require a significant change in particle configuration to explain the lifetime of the isomer.  Considering this, together with the new conversion-coefficient result and the systematic arguments presented above, an $E1$ assignment is made for the 438-keV transition leading to a 27/2$^{-}$ assignment for the isomer at 3921~keV.

\begin{table*}[htb]
\centering
\caption{Transition strengths in $^{143}$Nd, $^{145}$Sm, and $^{147}$Gd.  For $^{145}$Sm, transition strengths are presented for both the previous~\cite{Odahara1997} and current level scheme interpretations (see text).  The new lifetime for the high-spin isomer in $^{145}$Sm is used in both level scheme interpretations.  Transition strengths for electric transitions, $B(EL)$, are in units of $e^{2}$fm$^{2L}$ with those for magnetic transitions, $B(ML)$, in units of $\mu_{N}^{2}$fm$^{2L-2}$.  Limits are given at $1\sigma$.}
\label{tab:145Sm-BXL}
\resizebox{\textwidth}{!}{\begin{tabular}{c c c c c c c c c c c}
\toprule
\toprule
Nucleus & Ref. & $E_{x}$(keV) & $J^{\pi}$ & $t_{1/2}$ & $E_{\gamma}$(keV) & $I_{\gamma}$ & $XL$ & $B(XL)$ & $B/Bw$\\
\midrule
$^{143}$Nd & \cite{Zhou1999}     & 8987 & $49/2^{+}$ & 35(8)~ns         & 338.6    & 14.4(6) & $E1$ & $2.8(7)\times 10^{-7}$  & $1.6(4)\times 10^{-7}$  \\
           &                     &      &            &                  & 300.6    & 1.8(6)  &      &                      &                      \\
\midrule
$^{145}$Sm & \cite{Odahara1997}  & 3921 & $27/2^{+}$ & 1.1(2)~ns        & 438.4(3)  & 63.7(45)& $M1$ & $4.1(8)\times 10^{-4}$  & $2.3(4)\times 10^{-4}$ \\
           &                     & 8785 &($49/2^{+}$)& 3.52(16)~$\mu$s  & 206.2(4)  & 10.9(8) & $E1$ & $1.1(1)\times 10^{-9}$  & $6.3(5)\times 10^{-10}$ \\
           &                     &      &            &                  & 408.3(4)  & 11.6(8) & $M1$ & $1.4(1)\times 10^{-8}$  & $7.8(6)\times 10^{-9}$  \\
           &                     &      &            &                  & 452.4(5)  & 26.7(2) & $E2$ & $1.65(9)\times 10^{-3}$  & $3.6(2)\times 10^{-5}$  \\
           &                     &      &            &                  & 595.6(4)  & 29(2)   & $E2$ & $4.5(3)\times 10^{-4}$  & $1.0(7)\times 10^{-5}$  \\
           &                     &      &            &                  & 712.8(5)  & 23(2)   & $E1$ & $5.7(5)\times 10^{-11}$ & $3.2(3)\times 10^{-11}$ \\
           &                     &      &            &                  & 981.9(5)  & 20(1)   & $E2$ & $2.6(2)\times 10^{-5}$  & $5.7(4)\times 10^{-7}$  \\
           &                     &      &            &                  & 1042.4(2) & 4.1(3)  & $M2$ & $3.5(3)\times 10^{-4}$  & $7.7(7)\times 10^{-6}$  \\
           &                     &      &            &                  & 1457.9(8) & 10.3(8) & $E3$ & $1.85(16)$               & $1.5(1)\times 10^{-3}$  \\
\midrule
$^{145}$Sm & (present work)      & 3921 & $27/2^{-}$ & 1.1(2)~ns        & 438.4(3)  & 63.7(45)& $E1$ & $4.7(9)\times 10^{-6}$  & $2.6(5)\times 10^{-6}$ \\
           &                     & 8785 & $47/2^{-}$ & <5~ns            & 206.2(4)  & 10.9(8) & $E1$ & $\num{>7.3}\times 10^{-7}$ & $\num{>4.1}\times 10^{-7}$ \\
           &                     &      &            &                  & 408.3(4)  & 11.6(8) & $M1$ & $\num{>9.1}\times 10^{-6}$ & $\num{>5.1}\times 10^{-6}$ \\
           &                     &      &            &                  & 452.4(5)  & 26.7(2) & $E2$ & $\num{>1.1}$              & $\num{>2.5}\times 10^{-2}$             \\
           &                     &      &            &                  & 595.6(4)  & 29(2)   & $E2$ & $\num{>0.30}$              & $\num{>6.6}\times 10^{-3}$             \\
           &                     &      &            &                  & 712.8(5)  & 23(2)   & $E1$ & $\num{>3.7}\times 10^{-8}$ & $\num{>2.1}\times 10^{-8}$ \\
           &                     &      &            &                  & 981.9(5)  & 20(1)   & $E2$ & $\num{>0.017}$             & $\num{>3.8}\times 10^{-4}$            \\
           &                     &      &            &                  & 1042.4(2) & 4.1(3)  & $M2$ & $\num{>0.23}$              & $\num{>5.0}\times 10^{-3}$             \\
           &                     &      &            &                  & 1457.9(8) & 10.3(8) & $E3$ & $\num{>1200}$              & $\num{>0.96}$              \\
           &                     & 8815 & $49/2^{+}$ & 3.52(16)~$\mu$s  & 30.2(8)   & 61(3)   & $E1$ & $2.0(2)\times 10^{-6}$ & $1.1(1)\times 10^{-6}$ \\
           &                     &      &            &                  & (236)  & $\num{<0.96}$   &($E2$)& $\num{<1.5}\times 10^{-3}$ & $\num{<3.3}\times 10^{-5}$ \\
           &                     &      &            &                  & (743)  & $\num{<0.52}$   &($E2$)& $\num{<2.7}\times 10^{-6}$ & $\num{<5.9}\times 10^{-8}$ \\
\midrule
$^{147}$Gd & \cite{Broda2020}    & 8588 & $49/2^{+}$ & 510(20)~ns       & 254.4    & 828(25)  & $E2$ & $0.86(5)$             & $1.9(1)\times 10^{-2}$    \\
           &                     &      &            &                  & 434.5    & 57.8(42) & $E2$ & $4.1(4)\times 10^{3}$ & $9.0(8)\times 10^{-5}$                \\
           &                     &      &            &                  & 593.7    & 27.4(11) & $E3$ & $2500(200)$             & $2.0(1)$                \\
           &                     &      &            &                  & 623.8    & 1.3(3)   & $E3$ & $84(20)$                & $0.07(2)$  \\
\bottomrule
\bottomrule
\end{tabular}}
\end{table*}

Odahara \textit{et~al.}~\cite{Odahara1997} assigned the 571-keV, $7327 \rightarrow 6756$~keV transition as ${\Delta J=2}$ without explanation.  The conversion coefficient measured in the present work, however, indicates an $E1$ multipolarity.  Furthermore, an $E1$ assignment for this transition is consistent with the $E1$ assignments made to the newly added 570- and 714-keV transitions.  As a consequence of this proposed change, the 929-685-, 407-1207- and 885-729-keV cascades that depopulate the 8333-keV state are required to have $M1$-$M1$ assignments, despite the measured conversion coefficients for these transitions being more consistent with $E2$ assignments.  On average, an $E2$ admixture of $\approx$70\% is required for the measured conversion coefficients to agree with $M1/E2$ assignments for these transitions.  Note, however, that despite the $\Delta J=2$ assignment for the 929-keV transition in Ref.~\cite{Odahara1997}, the angular distribution measured in that work, $A_{2}/A_{0}=-0.25(4)$, is more consistent with a dipole assignment, in agreement with the $M1$ requirement.  Therefore, considering also that these are weak branches with little internal conversion, the disagreement observed for the conversion coefficients (closer to $E2$, but must be $M1$) is considered minor and an $E1$ assignment is made to the 571-keV transition, with tentative $M1/E2$ assignments to the 929-, 685-, 407-, 1207-, 885-, and 729-keV transitions.  Similar logic was used to assign the 982-keV, $8785 \rightarrow 7803$~keV and 1047-keV, $7803 \rightarrow 6756$~keV transitions as $E2$ and the 1042-keV, $8785 \rightarrow 7742$~keV transition as $M2$.

With these changes, the spin and parity of the isomeric state at 8785~keV becomes 47/2$^{-}$.  This is troubling, since the accepted spin and parity assignment for the high-spin isomers in the odd-mass, $N=83$ isotones is 49/2$^{+}$.  To resolve this puzzle, the neighbouring $N=83$ isotones were examined.  Comparing the $^{145}$Sm level scheme from Ref.~\cite{Odahara1997} with its neighbours, there is a qualitative difference between the decay of the 8785-keV state and the 49/2$^{+}$ isomers in $^{143}$Nd~\cite{Zhou1999} and $^{147}$Gd~\cite{Broda2020}.  In $^{147}$Gd, the isomer decays by four transitions: an enhanced $E3$, a hindered $E3$, and two hindered $E2$ transitions, while the isomer in $^{143}$Nd decays by two transitions: a hindered $E1$ and a second, weaker, unassigned transition.  In comparison, the decay of the 8785-keV state in $^{145}$Sm is highly fragmented, with a mix of eight $E1$, $M1$, $E2$, $M2$, and $E3$ transitions depopulating the state.  For this state to have a lifetime of 3.52~$\mu$s, all of these transitions would have to be severely hindered with transition strengths ranging from $10^{-5}$ to $10^{-11}$~W.u.~(details of the implied transition strengths are given in Table~\ref{tab:145Sm-BXL}).  As such hindrances are unlikely, irrespective of the spin and parity of the state, this suggests that the state at 8785~keV is not the isomer in $^{145}$Sm.

\subsubsection{High-spin isomer}
\label{sec:HSIsomer}

\begin{figure}[htb]
\centering
\includegraphics[width=\columnwidth]{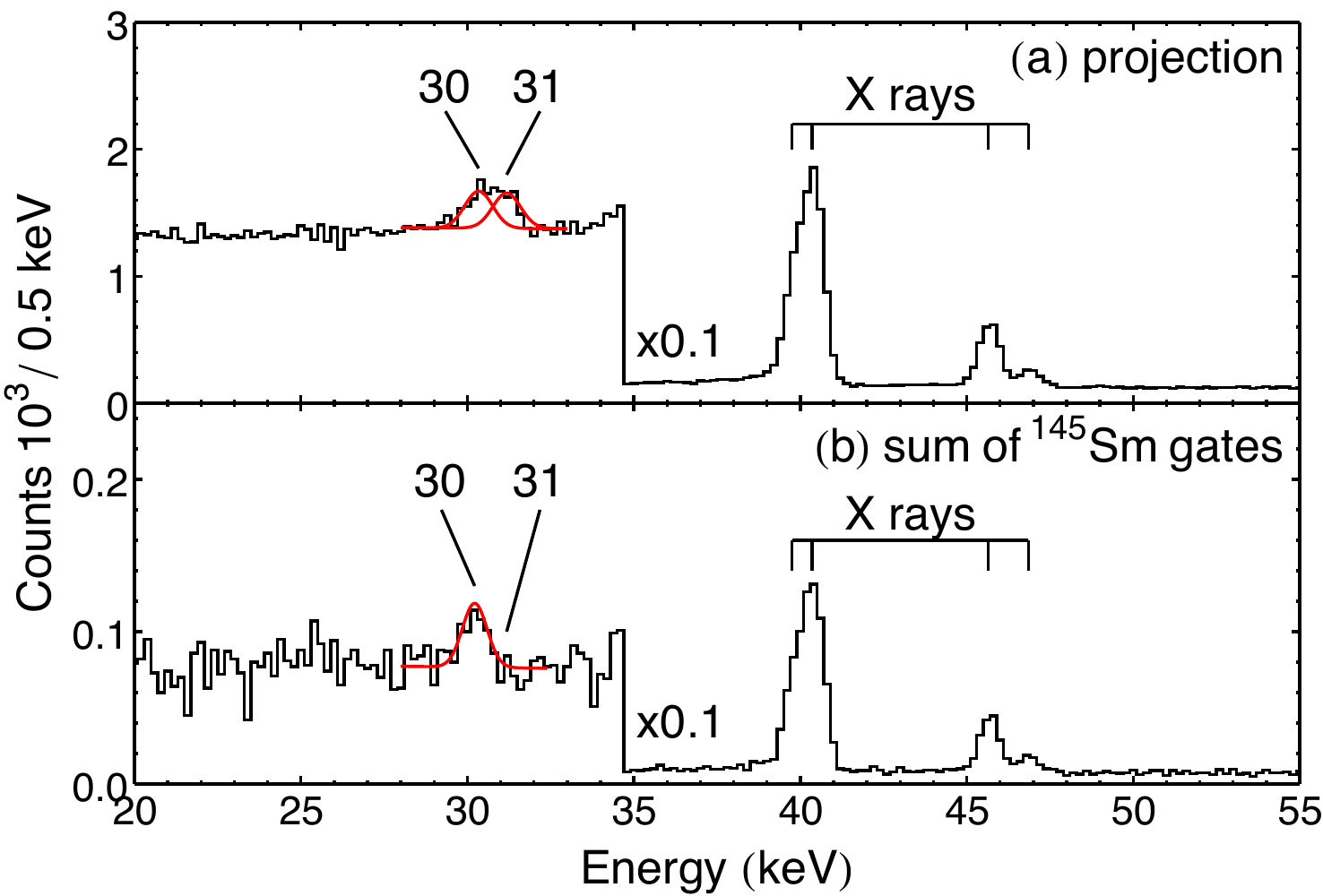}
\caption{Coincidence spectra showing the 30/31-keV doublet in (a) the full projection of the X-$\gamma$ coincidence matrix and (b) the sum of gates on strong $^{145}$Sm transitions (see text).  The large peaks at 39.5-, 40.6-, 45.9-, and 47.0-keV in both spectra are the $^{145}$Sm X rays.}
\label{fig:145Sm-xg}

\vspace{\floatsep}

\includegraphics[width=\columnwidth]{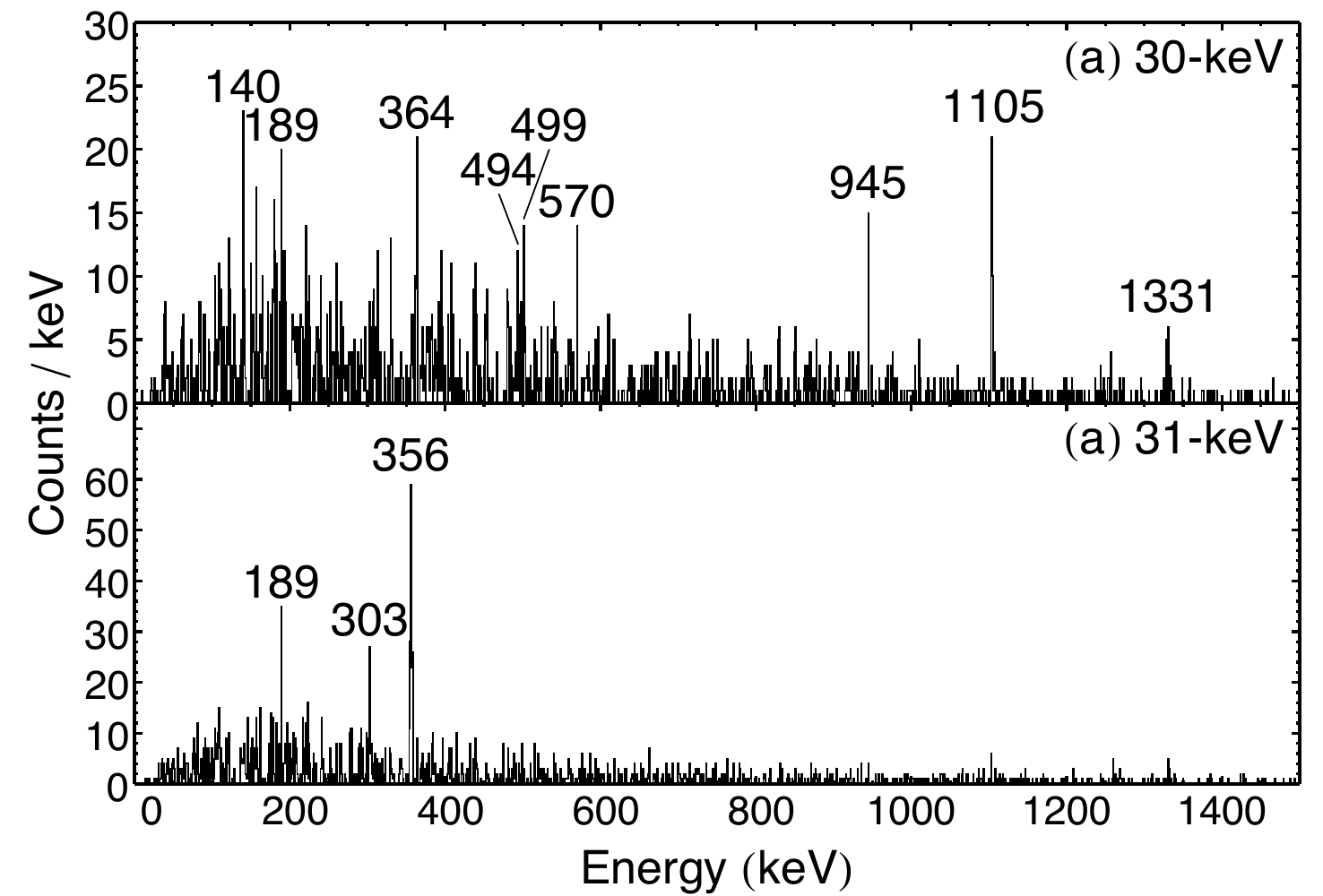}
\caption{Coincidence spectra showing $\gamma$ rays in the HPGe detectors observed in coincidence with gates on the (a) 30- and (b) 31-keV $\gamma$ rays in the LEPS detectors.  The 31-, 356-, and 303-keV transitions are associated with $\beta^{-}$ decay of $^{133}$Ba.  The single-channel spike at 189-keV is assigned as the $3119 \rightarrow 2930$-keV transition in $^{145}$Sm.}
\label{fig:145Sm-gx}
\end{figure}

Analysis of the X\footnote{Here `X' refers to $\gamma$~rays observed in the LEPS detector while `$\gamma$' refers to $\gamma$~rays observed in the HPGe detectors.}-$\gamma$ coincidence matrix shows a 30/31-keV doublet in the full X projection with only the 30-keV peak present in a sum of gates on $^{145}$Sm $\gamma$~rays (the 140-, 161-, 180-, 193-, 268-, 276-, 313-, 364-, 396-, 397-, 438-, 481-, 494-, 499-, 570-, 571-, 611-, 863-, 945-, 1105-, and 1331-keV $\gamma$~rays), as shown in Figure~\ref{fig:145Sm-xg}.  In the reverse $\gamma$-X matrix, known $^{145}$Sm $\gamma$~rays are seen in the gate on the 30-keV transition, while $\gamma$~rays associated with the $^{133}$Ba electron-capture decay~\cite{Khazov2011} are seen in coincidence with the 31-keV transition (see Figure~\ref{fig:145Sm-gx}).  Thus, the 30-keV $\gamma$~ray is assigned to $^{145}$Sm and the 31-keV transition assigned as the $^{133}$Cs X ray (it is suggested that the vacuum chamber has become weakly contaminated by the open $^{133}$Ba source used for energy and efficiency calibrations).

Firm placement of the 30-keV transition is difficult due to low statistics and the fragmentation of the level scheme; however, the fact that it is in coincidence with all of the intense transitions in the low-energy region of the $^{145}$Sm level scheme suggests that it is associated with a high-energy level.  Furthermore, the intensity of the 30-keV $\gamma$ ray suggests that most of the intensity in the level scheme flows through this transition.  While efficiency calibrations in this energy region are imprecise, a limit of $I_{\gamma} > 33$ has been placed.  Assuming that the transition has an $E1$ multipolarity (and therefore the least amount of internal conversion), the total intensity is $I_{\text{tot}} > 75$ (cf. $I_{\text{tot}}(1105) = 137.5$).  This could only be balanced if it were subsequently spread between the fragmented branches in the level scheme (the flow of intensity in the level scheme is discussed further in \S\ref{sec:States3921-5680}).  Considering this, along with the fact that there are no 30-keV gaps in the high-energy level scheme and that there are no new transitions in coincidence with the 30-keV $\gamma$ ray, this transition is assigned to populate the 8785-keV state, depopulating a newly proposed state at 8815~keV.

The conversion coefficient for this transition can be estimated using an intensity balance.  The total intensity (including internal conversion) of the 30-keV transition populating the 47/2$^{+}$ state at 8785~keV should balance the total intensity of transitions depopulating the state, $I_{\text{tot}} = 137.3(34)$.  The efficiency-corrected intensity of the 30-keV $\gamma$ ray was obtained from the X-projection of the gated X-$\gamma$ matrix (Figure~\hyperref[{fig:145Sm-xg}]{\ref*{fig:145Sm-xg}b}) and any difference between this value and the total intensity depopulating the state was attributed to internal conversion.  A conversion coefficient of $\alpha_{\text{tot}}(\text{expt})<3.2$ was measured using this method, consistent only with an $E1$ assignment (see Table~\ref{tab:transitions}).  This leads to a 49/2$^{+}$ assignment for the 8815-keV state which, considering the systematics of the high-spin isomers in the other odd-mass $N=83$ nuclei, suggests this is the equivalent isomer in $^{145}$Sm.  As the 30-keV transition now has a well defined multipolarity, a more accurate value for the $\gamma$-ray intensity can be calculated.  Reversing the above argument and using the theoretical conversion coefficient for an $E1$ transition ($\alpha_{\text{tot}}(E1) = 1.27(10)$~\cite{Kibedi2008}) gives a relative $\gamma$-ray intensity of $61(3)$, which is adopted in subsequent analysis and interpretation.

Transition strengths for the new level scheme are also presented in Table~\ref{tab:145Sm-BXL}.  For the 8785-keV, 47/2$^{-}$ state, transition strengths were calculated assuming a lifetime of $t_{1/2} < 5$~ns (chosen based on the experimental limit for lifetime measurements in the present work).  Without the long lifetime previously attributed to this state, the limits on the strengths of the depopulating transitions are greatly increased, resolving the issue of requiring so many transitions to be sufficiently hindered to result in a 3.52-$\mu$s half life.  For the proposed 49/2$^{+}$ isomer, the transition strength for the 30-keV $E1$ transition is $1.1 \times 10^{-6}$~W.u.  While this is approximately ten times greater than the strength of the equivalent transition in $^{143}$Nd, it is still reasonable for an $E1$ transition.  Upper limits have also been deduced for the unobserved $E2$ transitions that would cross over the 8785-keV state.  These imply strongly hindered $E2$ transitions whose strengths are discussed further in \S\ref{sec:StatesAbove5719}.

\section{Discussion}

The $^{145}$Sm nucleus is one neutron outside the $N=82$ shell closure and two proton holes outside the $Z=64$ sub-shell closure.  As such, the low-lying level scheme of $^{145}$Sm should be dominated by single-particle excitations.  Other $N=83$ isotones have been successfully described using a combination of deformed-independent-particle model (DIPM)~\cite{Neergard1981, Bakander1982} and semi-empirical shell-model calculations~\cite{Zhou1999, Bakander1982, Piiparinen1991} and a similar approach has been applied here.

The KShell program\footnote{The KShell program was chosen following its recent use describing the other $N=83$ isotone, $^{152}_{~\hfill69}$Tm$_{83}$~\cite{NaraSingh2018}.}~\cite{Shimizu2019} was used to carry out more detailed shell-model calculations using the basis space and single-particle energies given in Table~\ref{tab:spe} along with the `cw5082' residual interaction~\cite{Chou1992}.  To make the calculations more tractable, the problem space was truncated so that only relevant configurations were included in the basis space (for example, only configurations with 4-8 protons in the $0g_{7/2}$ orbital were considered).  These truncations are given in Table~\ref{tab:spe}.

\begin{table}[htb]
\centering
\caption{Single-particle energies~\cite{Chou1992} and truncations used in the present KShell calculations.  Single-particle energies are given relative to $^{132}_{~\hfill50}$Sn$_{82}$.}
\label{tab:spe}
\resizebox{\columnwidth}{!}{\begin{tabular}{c c c|c c c}
\toprule
\toprule
level & s.p.e (MeV) & truncation & level & s.p.e (MeV) & truncation \\
\midrule
$\nu 1f_{7/2}$  & -2.380 & 0-2 & $\pi 0g_{7/2}$  & -9.596 & 4-8 \\	
$\nu 2p_{3/2}$  & -1.625 & 0-2 & $\pi 1d_{5/2}$  & -8.675 & 2-6 \\
$\nu 2p_{1/2}$  & -1.160 & 0-2 & $\pi 1d_{3/2}$  & -6.955 & 0-2 \\
$\nu 0h_{9/2}$  & -0.895 & 0-2 & $\pi 2s_{1/2}$  & -6.927 & 0-2 \\
$\nu 1f_{5/2}$  & -0.890 & 0-2 & $\pi 0h_{11/2}$ & -6.835 & 0-4 \\
$\nu 0i_{13/2}$ & -0.290 & 0-2 & & \\
\bottomrule
\bottomrule
\end{tabular}}
\end{table}

\begin{turnpage}
\begin{figure*}
\centering
\includegraphics[width=\textheight]{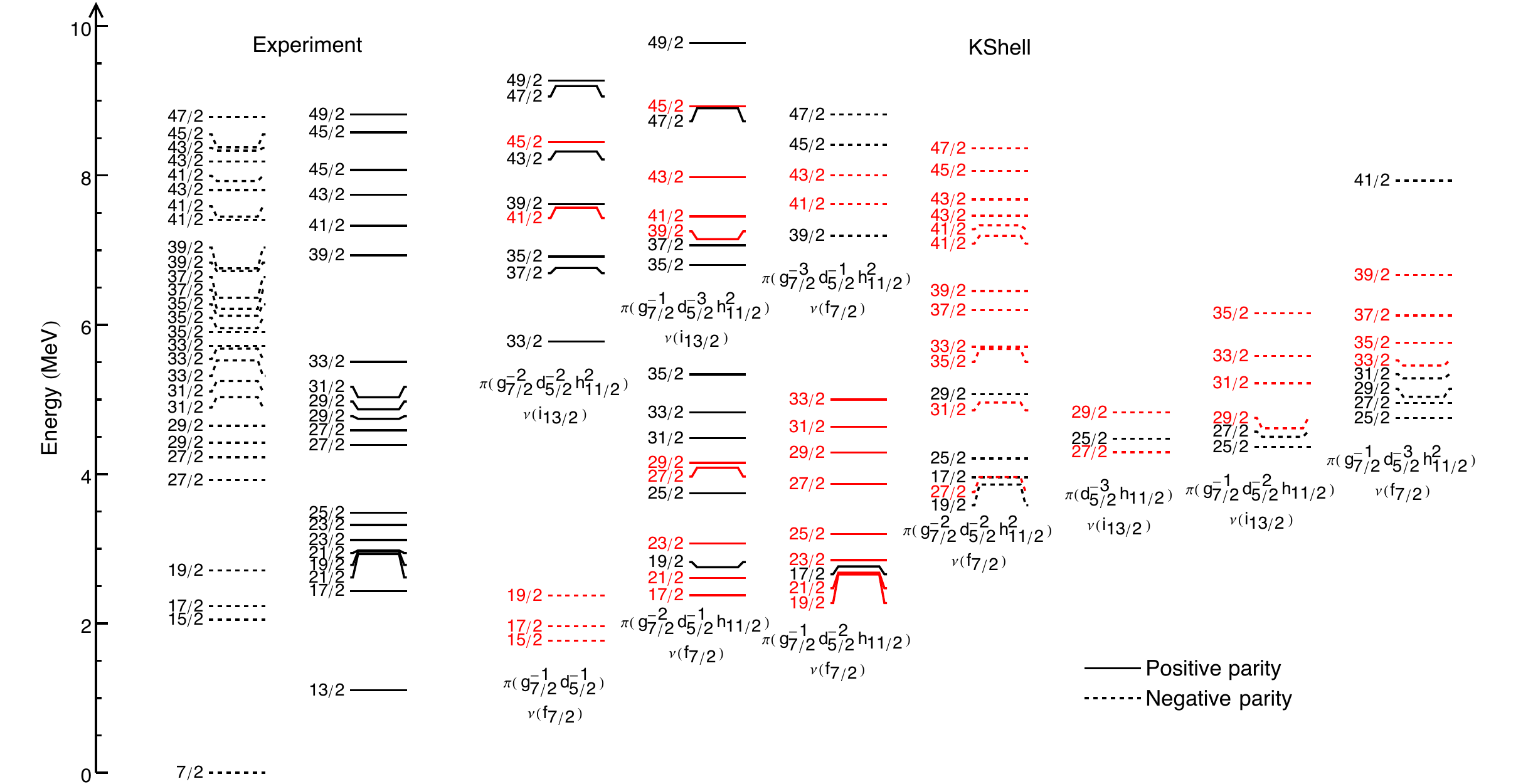}
\caption{Comparison between the experimental level scheme of $^{145}$Sm on the left and states predicted by the KShell calculations on the right.  Calculated states are grouped and labeled based on their dominant configuration.  Calculated states associated with the observed experimental states are highlighted in red (see also Table~\ref{tab:145Sm-KShell_assignments}).  The calculated 7/2$^{-}$ ground state from the $\pi(d_{5/2}^{-2})\nu(f_{7/2})$ configuration is not shown.}
\label{fig:145Sm-KShell_145Sm}
\end{figure*}
\end{turnpage}

Calculated and experimental levels for $^{145}$Sm are compared in Figure~\ref{fig:145Sm-KShell_145Sm}.  It should be noted that while these calculations were made with an inert $^{132}_{~\hfill50}$Sn$_{82}$ core, the configurations presented in the following discussion are given relative to the sub-shell closure at $^{146}_{~\hfill64}$Gd$_{82}$ (as if it were doubly magic).  Excepting core-neutron excitations for now, the states in $^{145}$Sm can be described using the following particle configurations: $\pi(l_{j}^{-2})\nu(f_{7/2})$, $\pi(l_{j}^{-3}h_{11/2})\nu(f_{7/2})$, $\pi(l_{j}^{-3}h_{11/2})\nu(i_{13/2})$, $\pi(l_{j}^{-4}h_{11/2}^{2})\nu(i_{13/2})$, and $\pi(l_{j}^{-4}h_{11/2}^{2})\nu(f_{7/2})$, where $l_{j}^{-n}$ represents some combination of $n$ $d_{5/2}$ and $g_{7/2}$ proton holes (e.g.~$\pi(l_{j}^{-4}) \equiv \pi(g_{7/2}^{-2}d_{5/2}^{-2})$).  Similar notation is used throughout the following discussion.

Due to the high degree of mixing in the states predicted by KShell (indicated by the low amplitude of the major configurations), precise configuration assignments are not realistic.  Nevertheless, suggested dominant configuration assignments for each state are given in Table~\ref{tab:145Sm-KShell_assignments}, and highlighted in Figure~\ref{fig:145Sm-KShell_145Sm}.  Good agreement is seen between the experimentally observed states in $^{145}$Sm and those predicted by KShell, with an overall root-mean-square deviation of 301~keV.

\begingroup

\begin{table*}[htb]
\centering
\caption{Configuration assignments for states below the 49/2$^{+}$ isomer in $^{145}$Sm.  Calculated energies are given relative to the 7/2$^{-}$ ground state.}
\label{tab:145Sm-KShell_assignments}
\begin{tabular}{c c c c c c}
\toprule
\toprule
$E_{x}$~(keV) & $J^{\pi}$ & $E_{\text{calc}}$~(keV) & Major configuration & $|\text{Amp.}|^{2}$ (\%) & $E_{\text{calc}} - E_{x}$~(keV)\\
\midrule
0    & $ 7/2^{-}$ & 0    & $\pi[            d_{5/2}^{-2}            ]\otimes \nu[f_{7/2}] $ & 22.6 & 0    \\
2049 & $15/2^{-}$ & 1771 & $\pi[g_{7/2}^{-1}d_{5/2}^{-1}h_{11/2}^{ }]\otimes \nu[f_{7/2}] $ & 42.8 & -278 \\
2230 & $17/2^{-}$ & 1963 & $\pi[g_{7/2}^{-1}d_{5/2}^{-1}h_{11/2}^{ }]\otimes \nu[f_{7/2}] $ & 46.1 & -267 \\
2436 & $17/2^{+}$ & 2378 & $\pi[g_{7/2}^{-2}d_{5/2}^{-1}h_{11/2}    ]\otimes \nu[f_{7/2}] $ & 28.1 & -58  \\
2710 & $19/2^{-}$ & 2375 & $\pi[g_{7/2}^{-1}d_{5/2}^{-1}h_{11/2}    ]\otimes \nu[f_{7/2}] $ & 46.3 & -335 \\
2930 & $21/2^{+}$ & 2610 & $\pi[g_{7/2}^{-2}d_{5/2}^{-1}h_{11/2}    ]\otimes \nu[f_{7/2}] $ & 29.8 & -320 \\
2964 & $19/2^{+}$ & 2656 & $\pi[g_{7/2}^{-1}d_{5/2}^{-2}h_{11/2}    ]\otimes \nu[f_{7/2}] $ & 36.9 & -308 \\
2979 & $21/2^{+}$ & 2680 & $\pi[g_{7/2}^{-1}d_{5/2}^{-2}h_{11/2}    ]\otimes \nu[f_{7/2}] $ & 40.3 & -299 \\
3119 & $23/2^{+}$ & 2849 & $\pi[g_{7/2}^{-1}d_{5/2}^{-2}h_{11/2}    ]\otimes \nu[f_{7/2}] $ & 45.1 & -270 \\
3322 & $23/2^{+}$ & 3070 & $\pi[g_{7/2}^{-2}d_{5/2}^{-1}h_{11/2}    ]\otimes \nu[f_{7/2}] $ & 24.9 & -252 \\
3483 & $25/2^{+}$ & 3196 & $\pi[g_{7/2}^{-1}d_{5/2}^{-2}h_{11/2}    ]\otimes \nu[f_{7/2}] $ & 45.2 & -287 \\
3921 & $27/2^{-}$ & 3961 & $\pi[g_{7/2}^{-2}d_{5/2}^{-2}h_{11/2}^{2}]\otimes \nu[f_{7/2}] $ & 36.9 &  40  \\
4228 & $27/2^{-}$ & 4293 & $\pi[g_{7/2}^{-1}d_{5/2}^{-2}h_{11/2}    ]\otimes \nu[i_{13/2}]$ & 20.5 &  65  \\
4389 & $27/2^{+}$ & 3870 & $\pi[g_{7/2}^{-1}d_{5/2}^{-2}h_{11/2}    ]\otimes \nu[f_{7/2}] $ & 37.0 & -519 \\
4420 & $29/2^{-}$ & 4613 & $\pi[g_{7/2}^{-1}d_{5/2}^{-2}h_{11/2}    ]\otimes \nu[i_{13/2}]$ & 42.9 &  193 \\
4586 & $27/2^{+}$ & 4085 & $\pi[g_{7/2}^{-2}d_{5/2}^{-1}h_{11/2}    ]\otimes \nu[f_{7/2}] $ & 32.6 & -501 \\
4647 & $29/2^{-}$ & 4827 & $\pi[            d_{5/2}^{-3}h_{11/2}    ]\otimes \nu[i_{13/2}]$ & 16.4 &  180 \\
4741 & $29/2^{+}$ & 4151 & $\pi[g_{7/2}^{-2}d_{5/2}^{-1}h_{11/2}    ]\otimes \nu[f_{7/2}] $ & 40.2 & -590 \\
4868 & $29/2^{+}$ & 4290 & $\pi[g_{7/2}^{-1}d_{5/2}^{-2}h_{11/2}    ]\otimes \nu[f_{7/2}] $ & 48.0 & -578 \\
5029 & $31/2^{+}$ & 4633 & $\pi[g_{7/2}^{-2}d_{5/2}^{-1}h_{11/2}    ]\otimes \nu[f_{7/2}] $ & 46.9 & -396 \\
5031 & $31/2^{-}$ & 4959 & $\pi[g_{7/2}^{-2}d_{5/2}^{-2}h_{11/2}^{2}]\otimes \nu[f_{7/2}] $ & 39.1 & -72  \\
5248 & $31/2^{-}$ & 5216 & $\pi[g_{7/2}^{-1}d_{5/2}^{-2}h_{11/2}^{1}]\otimes \nu[f_{7/2}] $ & 32.9 & -32  \\
5507 & $33/2^{+}$ & 5000 & $\pi[g_{7/2}^{-1}d_{5/2}^{-2}h_{11/2}    ]\otimes \nu[f_{7/2}] $ & 61.3 & -507 \\
5526 & $33/2^{-}$ & 5453 & $\pi[g_{7/2}^{-1}d_{5/2}^{-3}h_{11/2}^{ }]\otimes \nu[f_{7/2}] $ & 31.7 & -73  \\
5680 & $33/2^{-}$ & 5581 & $\pi[g_{7/2}^{-1}d_{5/2}^{-2}h_{11/2}    ]\otimes \nu[f_{7/2}] $ & 24.6 & -99  \\
5719 & $33/2^{-}$ & 5705 & $\pi[g_{7/2}^{-2}d_{5/2}^{-2}h_{11/2}^{2}]\otimes \nu[f_{7/2}] $ & 22.2 & -14  \\
5903 & $35/2^{-}$ & 5676 & $\pi[g_{7/2}^{-2}d_{5/2}^{-2}h_{11/2}^{2}]\otimes \nu[f_{7/2}] $ & 31.2 & -227 \\
5956 & $35/2^{-}$ & 5760 & $\pi[g_{7/2}^{-1}d_{5/2}^{-3}h_{11/2}^{2}]\otimes \nu[f_{7/2}] $ & 29.2 & -196 \\
6121 & $35/2^{-}$ & 6154 & $\pi[g_{7/2}^{-1}d_{5/2}^{-2}h_{11/2}    ]\otimes \nu[f_{7/2}] $ & 25.9 &  33  \\
6216 & $37/2^{-}$ & 6123 & $\pi[g_{7/2}^{-1}d_{5/2}^{-3}h_{11/2}^{2}]\otimes \nu[f_{7/2}] $ & 25.8 & -93  \\
6361 & $37/2^{-}$ & 6196 & $\pi[g_{7/2}^{-2}d_{5/2}^{-2}h_{11/2}^{2}]\otimes \nu[f_{7/2}] $ & 39.7 & -165 \\
6719 & $39/2^{-}$ & 6454 & $\pi[g_{7/2}^{-2}d_{5/2}^{-3}h_{11/2}^{2}]\otimes \nu[f_{7/2}] $ & 53.0 & -265 \\
6756 & $39/2^{-}$ & 6667 & $\pi[g_{7/2}^{-1}d_{5/2}^{-3}h_{11/2}^{2}]\otimes \nu[f_{7/2}] $ & 30.3 & -89  \\
6931 & $39/2^{+}$ & 7147 & $\pi[g_{7/2}^{-1}d_{5/2}^{-3}h_{11/2}^{2}]\otimes \nu[i_{13/2}]$ & 40.6 &  216 \\
7327 & $41/2^{+}$ & 7451 & $\pi[g_{7/2}^{-1}d_{5/2}^{-3}h_{11/2}^{2}]\otimes \nu[i_{13/2}]$ & 50.9 &  124 \\
7404 & $41/2^{-}$ & 7191 & $\pi[g_{7/2}^{-2}d_{5/2}^{-2}h_{11/2}^{2}]\otimes \nu[f_{7/2}] $ & 29.6 & -213 \\
7449 & $41/2^{-}$ & 7334 & $\pi[g_{7/2}^{-2}d_{5/2}^{-2}h_{11/2}^{2}]\otimes \nu[f_{7/2}] $ & 50.0 & -115 \\
7742 & $43/2^{+}$ & 7976 & $\pi[g_{7/2}^{-1}d_{5/2}^{-3}h_{11/2}^{2}]\otimes \nu[i_{13/2}]$ & 44.5 &  234 \\
7803 & $43/2^{-}$ & 7461 & $\pi[g_{7/2}^{-2}d_{5/2}^{-2}h_{11/2}^{2}]\otimes \nu[f_{7/2}] $ & 51.2 & -342 \\
7926 & $41/2^{-}$ & 7617 & $\pi[g_{7/2}^{-3}d_{5/2}^{-1}h_{11/2}^{2}]\otimes \nu[f_{7/2}] $ & 23.7 & -309 \\
8072 & $45/2^{+}$ & 8448 & $\pi[g_{7/2}^{-2}d_{5/2}^{-2}h_{11/2}^{2}]\otimes \nu[i_{13/2}]$ & 47.0 &  376 \\
8190 & $43/2^{-}$ & 8002 & $\pi[g_{7/2}^{-3}d_{5/2}^{-1}h_{11/2}^{2}]\otimes \nu[f_{7/2}] $ & 31.8 & -188 \\
8333 & $43/2^{-}$ & 7677 & $\pi[g_{7/2}^{-3}d_{5/2}^{-1}h_{11/2}^{2}]\otimes \nu[f_{7/2}] $ & 28.1 & -656 \\
8377 & $45/2^{-}$ & 8063 & $\pi[g_{7/2}^{-2}d_{5/2}^{-2}h_{11/2}^{2}]\otimes \nu[f_{7/2}] $ & 57.7 & -314 \\
8579 & $45/2^{+}$ & 8924 & $\pi[g_{7/2}^{-1}d_{5/2}^{-3}h_{11/2}^{2}]\otimes \nu[i_{13/2}]$ & 22.3 &  345 \\
8785 & $47/2^{-}$ & 8363 & $\pi[g_{7/2}^{-2}d_{5/2}^{-2}h_{11/2}^{2}]\otimes \nu[f_{7/2}] $ & 62.5 & -422 \\

\bottomrule
\bottomrule
\end{tabular}
\end{table*}

\subsection{States up to 3483~keV}

The ground state and lowest-energy, negative-parity states in $^{145}$Sm can be readily explained by the $\pi(l_{j}^{-2})\nu(f_{7/2})$ configurations.  Previous studies of the $^{145}$Sm level scheme have attributed the $13/2^{+}$ state at 1105-keV to a $3^{-}$ octupole vibration coupled to the ground state~\cite{Trache1989,Kader1989}.  Due to the collective nature of this excitation, accurately describing this state was outside the scope of the KShell calculations.  Other positive-parity states below the 25/2$^{+}$ state at 3483~keV are produced by promoting a proton from either of the $d_{5/2}$ or $g_{7/2}$ orbitals into the $h_{11/2}$ orbital.

\subsection{States between 3921~keV and 5507~keV}
\label{sec:States3921-5680}

Above the 3483-keV state the level scheme splits into two branches.

The first branch feeds the 25/2$^{+}$ state at 3483~keV and the 23/2$^{+}$ state at 3119~keV by a number of relatively high-energy $E2$ transitions.  The relevant particle configurations show that $J=23/2$ and $J=25/2$ are the maximum angular momenta that can be produced with one pair of proton-holes coupled to zero and one proton in the $h_{11/2}$ single-particle state.  These configurations are:
\begin{equation}
[\pi(g_{7/2}^{-1}[d_{5/2}^{-2}]_{0^{+}}h_{11/2})_{9^{+}}\nu(f_{7/2})]_{25/2^{+}}
\end{equation}
and
\begin{equation}
[\pi([g_{7/2}^{-2}]_{0^{+}}d_{5/2}^{-1}h_{11/2})_{8^{+}}\nu(f_{7/2})]_{23/2^{+}},
\end{equation}
respectively.  As such, it appears that these high-energy $E2$ transitions are associated with the energy that comes from breaking the proton-hole pairs coupled to spin zero.

Once these pairs are broken, the maximum angular momentum available for the $\pi(g_{7/2}^{-2}d_{5/2}^{-1}h_{11/2})\nu(f_{7/2})$ configuration is $J^{\pi}=35/2^{+}$; however, no such state has been observed in the $^{145}$Sm level scheme.  Similarly, the $J^{\pi}=33/2^{+}$, 35/2$^{+}$, and 37/2$^{+}$ states expected from the $\pi(l_{j}^{-4}h_{11/2}^{2})\nu(i_{13/2})$ configurations are also absent.  Several $\gamma$~rays observed in the present work (344-, 350-, 591-, 721-, 974-, 1052-, 1057-, 1128-, 1176-, 1244-keV transitions) are in coincidence with the known $^{145}$Sm $\gamma$~rays but could not be placed in the level scheme.  It is speculated that some of these or other unobserved transitions may connect to the missing positive-parity states.

The second branch involves exciting a second proton into the $h_{11/2}$ orbital, resulting in the 27/2$^{-}$ isomer at 3921~keV with the $\pi(g_{7/2}^{-2}d_{5/2}^{-2}h_{11/2}^{2})\nu(f_{7/2})$ configuration.  This state is favored due to the residual interaction energy from the
\begin{equation}
[\pi([g_{7/2}^{-2}]_{6^{+}}[d_{5/2}^{-2}]_{4^{+}}[h_{11/2}^{2}]_{0^{+}})_{10^{+}}\nu(f_{7/2})]_{27/2^{-}}
\end{equation}
and 
\begin{equation}
[\pi([g_{7/2}^{-2}]_{0^{+}}[d_{5/2}^{-2}]_{0^{+}}[h_{11/2}^{2}]_{10^{+}})_{10^{+}}\nu(f_{7/2})]_{27/2^{-}}
\end{equation}
configurations, where the $h_{11/2}^{2}$, $g_{7/2}^{-2}$, and $d_{5/2}^{-2}$ protons and proton holes are variously and favourably coupled to either spin zero or their maximum angular momentum.  It should be noted that there are no candidates in the KShell calculations for a 27/2$^{+}$ isomer near 3.9~MeV, as required by the level scheme of Odahara \textit{et~al.}~\cite{Odahara1997}.  This adds confidence in the new 27/2$^{-}$ assignment for the 3921-keV state.  Other negative-parity states in this energy region are produced by the $\pi(l_{j}^{-3}h_{11/2})\nu(i_{13/2})$ and $\pi(l_{j}^{-4}h_{11/2}^{2})\nu(f_{7/2})$ configurations.

As an aside, the reassignment of the 3921-keV isomer in $^{145}$Sm as a 27/2$^{-}$ state calls into question the assignment of the similar state at 4076~keV in $^{143}$Nd.  In their work, Zhou~\textit{et~al.}~\cite{Zhou1999} tentatively assign the spin and parity of this state as 27/2$^{+}$, noting that this state was not predicted by calculations and the assignment necessitates an unlikely $M3$ assignment to the 1165-keV transition feeding the 21/2$^{+}$ state at 2911~keV.  Based on the reassignment of the 3921-keV state in $^{145}$Sm and the systematic appearance of a 27/2$^{-}$ isomer in the $N=83$ isotones (see Figure~\ref{fig:145Sm-N=83_Isomers}), this suggests that the 4076-keV state in $^{143}$Nd also has a spin and parity of 27/2$^{-}$.  This reassignment would also change the multipolarity of the 1165-keV, $4076 \rightarrow 2911$~keV transition to $E3$, similar to the enhanced 822-keV, $3582 \rightarrow 2760$~keV transition in $^{147}$Gd\footnote{The equivalent transition has not yet been observed in $^{145}$Sm; however, one candidate is a 943-keV transition from $3921 \rightarrow 2979$~keV.  It is possible that this transition is present but is overwhelmed by the intense 945-keV, $2049 \rightarrow 1105$~keV transition in the singles data, and is too weak to see in gated coincidence data.}~\cite{Broda2020}, removing the requirement for an $M3$ transition.

\subsection{States above 5719~keV}
\label{sec:StatesAbove5719}

The deformed, neutron-core excitations that have played an important role in explaining the high-spin states in the $N=83$ isotones are outside the spherical basis space of the KShell calculations.  As such, to help understand the remaining states leading up to the isomer, $^{145}$Sm was again compared with the neighbouring nuclei, $^{143}$Nd and $^{147}$Gd.

In $^{147}$Gd, the states directly connected to the isomer have been associated with core neutron excitations (see Ref.~\cite{Bakander1982}).  However, with the newly assigned isomer at 8815~keV, the decay pattern in $^{145}$Sm is closer to that of $^{143}$Nd (see Ref.~\cite{Zhou1999}).  DIPM calculations performed for $^{143}$Nd~\cite{Zhou1999} associate only the 49/2$^{+}$ isomer with the neutron-core excitation, and use spherical $\pi(l_{j}^{-4}h_{11/2}^{2})\nu(f_{7/2})$ configurations to explain the other high-energy states.  Taking a similar approach here, the high-energy, negative-parity states in $^{145}$Sm can be explained by $\pi(l_{j}^{-4}h_{11/2}^{2})\nu(f_{7/2})$ configurations up to a spin of 47/2$^{-}$, the maximum allowed for such configurations.  Positive-parity states in this region can be similarly produced by promoting the $f_{7/2}$ neutron into the $i_{13/2}$ orbital with the same proton configurations.

The isomer itself has been associated with the $\pi(d_{5/2}^{-2}h_{11/2}^{2})\nu(d_{3/2}^{-2}f_{7/2}h_{9/2}i_{13/2})$ configuration~\cite{Odahara1997}.  This assignment is based on the good agreement between DIPM calculations and experimental energy systematics of the isomers in the isotone chain, as well as the measured quadrupole moment and $g$ factor of the isomer in $^{147}$Gd~\cite{Neergard1981}.  While the KShell calculations predicted a number of 49/2$^{+}$ states without invoking core-neutron excitations, these are predicted at higher energies than the experimental isomer and none of these states have configurations that stand out as a candidate for an isomer.  More specifically, each of the predicted states has available decay transitions of a moderate energy and low multipolarity that can feed states within the same particle configuration, providing no clear reason for a long lifetime.  Furthermore, if the 49/2$^{+}$ isomer were produced by a $\pi(h^{2}_{11/2})$ or even a $\pi(h^{3}_{11/2})$ configuration, then unhindered $E2$ transitions between the isomer and the 45/2$^{+}$ states at 8072- and 8579-keV would be expected due to the similarity of the associated configurations.  The lack of $E2$ transitions crossing over the 8785-keV state (upper limits on the $\gamma$-ray intensities and transition strengths for the 743- and 236-keV transitions are given in Table~\ref{tab:145Sm-BXL}) implies a complex configuration change, which can be explained by the excitation of core neutrons.  These observations, together with the new level scheme and the success of the DIPM calculations in describing the $N=83$ high-spin isomers, affirms the previous deformed core-excited configuration assigned to the 49/2$^{+}$ isomer in $^{145}$Sm.

\section{Conclusion}

The decay of the high-spin isomer in $^{145}$Sm was studied with the aim of making firm spin and parity assignments for states up to and including the long-lived isomer.  The half life of the isomer was measured as 3.52(16)~$\mu$s, significantly longer than the previous value of 0.96~$\mu$s~\cite{Ferragut1993}.  Measured transition multipolarities and analysis of $\gamma$-$\gamma$ coincidences has led to a revised level scheme.  Most significantly, the spin and parity assignment of the 3921-keV state (a low-lying isomer in the level scheme) changed from 27/2$^{+}$ to 27/2$^{-}$ and the 8785-keV state (historically considered to be the isomer~\cite{Odahara1997}) changed from 49/2$^{+}$ to 47/2$^{-}$.  A new state at 8815~keV has been proposed as the long-lived 49/2$^{+}$ isomer, decaying to the 8785-keV state by a 30-keV $E1$ transition.

The new level scheme was explained using shell-model calculations alongside comparisons with, and analysis of, the $N=83$ systematics.  The new assignment for the 3921-keV isomeric state is consistent with the systematics of $^{147}$Gd, $^{149}$Dy, and $^{151}$Er~(see Figure~\ref{fig:145Sm-N=83_Isomers}), calling into question the (27/2$^{+}$) assignment for the equivalent isomer in $^{143}$Nd.  The interpretation of the 49/2$^{+}$ isomers as a deformed excitation of the core neutrons remains unchanged.

\begin{acknowledgments}
The authors are grateful to the academic and technical staff of the Department of Nuclear Physics (Australian National University) and the Heavy Ion Accelerator Facility for their continued support.  In particular, thanks go to M.~Dasgupta, D.~J.~Hinde, L.~Bezzina, and S.~Battisson for their continued assistance in the development and operation of SOLITAIRE.  This research was supported by the Australian Research Council through grant numbers FT100100991, DP120101417, DP14102986, DP140103317, and DP170101675.  M.~S.~M.~G., A.~A., B.~J.~C., J.~T.~H.~D., T.~J.~G., and B.~P.~M. acknowledge the support of the Australian Government Research Training Program.  Support for the ANU Heavy Ion Accelerator Facility operations through the Australian National Collaborative Research Infrastructure Strategy (NCRIS) program is also acknowledged.
\end{acknowledgments}

\bibliography{mybibfile}

\begin{thebibliography}{28}%
\makeatletter
\providecommand \@ifxundefined [1]{%
 \@ifx{#1\undefined}
}%
\providecommand \@ifnum [1]{%
 \ifnum #1\expandafter \@firstoftwo
 \else \expandafter \@secondoftwo
 \fi
}%
\providecommand \@ifx [1]{%
 \ifx #1\expandafter \@firstoftwo
 \else \expandafter \@secondoftwo
 \fi
}%
\providecommand \natexlab [1]{#1}%
\providecommand \enquote  [1]{``#1''}%
\providecommand \bibnamefont  [1]{#1}%
\providecommand \bibfnamefont [1]{#1}%
\providecommand \citenamefont [1]{#1}%
\providecommand \href@noop [0]{\@secondoftwo}%
\providecommand \href [0]{\begingroup \@sanitize@url \@href}%
\providecommand \@href[1]{\@@startlink{#1}\@@href}%
\providecommand \@@href[1]{\endgroup#1\@@endlink}%
\providecommand \@sanitize@url [0]{\catcode `\\12\catcode `\$12\catcode
  `\&12\catcode `\#12\catcode `\^12\catcode `\_12\catcode `\%12\relax}%
\providecommand \@@startlink[1]{}%
\providecommand \@@endlink[0]{}%
\providecommand \url  [0]{\begingroup\@sanitize@url \@url }%
\providecommand \@url [1]{\endgroup\@href {#1}{\urlprefix }}%
\providecommand \urlprefix  [0]{URL }%
\providecommand \Eprint [0]{\href }%
\providecommand \doibase [0]{http://dx.doi.org/}%
\providecommand \selectlanguage [0]{\@gobble}%
\providecommand \bibinfo  [0]{\@secondoftwo}%
\providecommand \bibfield  [0]{\@secondoftwo}%
\providecommand \translation [1]{[#1]}%
\providecommand \BibitemOpen [0]{}%
\providecommand \bibitemStop [0]{}%
\providecommand \bibitemNoStop [0]{.\EOS\space}%
\providecommand \EOS [0]{\spacefactor3000\relax}%
\providecommand \BibitemShut  [1]{\csname bibitem#1\endcsname}%
\let\auto@bib@innerbib\@empty
\bibitem [{\citenamefont {Zhou}\ \emph {et~al.}(1999)\citenamefont {Zhou},
  \citenamefont {Tsuchida}, \citenamefont {Gono}, \citenamefont {Odahara},
  \citenamefont {Ideguchi}, \citenamefont {Morikawa}, \citenamefont {Shibata},
  \citenamefont {Watanabe}, \citenamefont {Miyake}, \citenamefont {Tsutsumi},
  \citenamefont {Motomura}, \citenamefont {Kishida}, \citenamefont {Mitarai},\
  and\ \citenamefont {Ishihara}}]{Zhou1999}%
  \BibitemOpen
  \bibfield  {author} {\bibinfo {author} {\bibfnamefont {X.~H.}\ \bibnamefont
  {Zhou}}, \bibinfo {author} {\bibfnamefont {H.}~\bibnamefont {Tsuchida}},
  \bibinfo {author} {\bibfnamefont {Y.}~\bibnamefont {Gono}}, \bibinfo {author}
  {\bibfnamefont {A.}~\bibnamefont {Odahara}}, \bibinfo {author} {\bibfnamefont
  {E.}~\bibnamefont {Ideguchi}}, \bibinfo {author} {\bibfnamefont
  {T.}~\bibnamefont {Morikawa}}, \bibinfo {author} {\bibfnamefont
  {M.}~\bibnamefont {Shibata}}, \bibinfo {author} {\bibfnamefont
  {H.}~\bibnamefont {Watanabe}}, \bibinfo {author} {\bibfnamefont
  {M.}~\bibnamefont {Miyake}}, \bibinfo {author} {\bibfnamefont
  {T.}~\bibnamefont {Tsutsumi}}, \bibinfo {author} {\bibfnamefont
  {S.}~\bibnamefont {Motomura}}, \bibinfo {author} {\bibfnamefont
  {T.}~\bibnamefont {Kishida}}, \bibinfo {author} {\bibfnamefont
  {S.}~\bibnamefont {Mitarai}}, \ and\ \bibinfo {author} {\bibfnamefont
  {M.}~\bibnamefont {Ishihara}},\ }\href {\doibase 10.1103/PhysRevC.61.014303}
  {\bibfield  {journal} {\bibinfo  {journal} {Phys. Rev. C}\ }\textbf {\bibinfo
  {volume} {61}},\ \bibinfo {pages} {014303} (\bibinfo {year}
  {1999})}\BibitemShut {NoStop}%
\bibitem [{\citenamefont {Murakami}\ \emph {et~al.}(1993)\citenamefont
  {Murakami}, \citenamefont {Gono}, \citenamefont {Ferragut}, \citenamefont
  {Zhang}, \citenamefont {Morita}, \citenamefont {Yoshida}, \citenamefont
  {Ogawa}, \citenamefont {Nakajima}, \citenamefont {Min}, \citenamefont
  {Kumagai}, \citenamefont {Oshima}, \citenamefont {Morikawa}, \citenamefont
  {Sugawara},\ and\ \citenamefont {Kusakari}}]{Murakami1993}%
  \BibitemOpen
  \bibfield  {author} {\bibinfo {author} {\bibfnamefont {T.}~\bibnamefont
  {Murakami}}, \bibinfo {author} {\bibfnamefont {Y.}~\bibnamefont {Gono}},
  \bibinfo {author} {\bibfnamefont {A.}~\bibnamefont {Ferragut}}, \bibinfo
  {author} {\bibfnamefont {Y.~H.}\ \bibnamefont {Zhang}}, \bibinfo {author}
  {\bibfnamefont {K.}~\bibnamefont {Morita}}, \bibinfo {author} {\bibfnamefont
  {A.}~\bibnamefont {Yoshida}}, \bibinfo {author} {\bibfnamefont
  {M.}~\bibnamefont {Ogawa}}, \bibinfo {author} {\bibfnamefont
  {M.}~\bibnamefont {Nakajima}}, \bibinfo {author} {\bibfnamefont {B.~J.}\
  \bibnamefont {Min}}, \bibinfo {author} {\bibfnamefont {H.}~\bibnamefont
  {Kumagai}}, \bibinfo {author} {\bibfnamefont {M.}~\bibnamefont {Oshima}},
  \bibinfo {author} {\bibfnamefont {T.}~\bibnamefont {Morikawa}}, \bibinfo
  {author} {\bibfnamefont {M.}~\bibnamefont {Sugawara}}, \ and\ \bibinfo
  {author} {\bibfnamefont {H.}~\bibnamefont {Kusakari}},\ }\href {\doibase
  10.1007/BF01290348} {\bibfield  {journal} {\bibinfo  {journal} {Z. Phys. A}\
  }\textbf {\bibinfo {volume} {345}},\ \bibinfo {pages} {123} (\bibinfo {year}
  {1993})}\BibitemShut {NoStop}%
\bibitem [{\citenamefont {Odahara}\ \emph {et~al.}(1997)\citenamefont
  {Odahara}, \citenamefont {Gono}, \citenamefont {Mitarai}, \citenamefont
  {Morikawa}, \citenamefont {Shizuma}, \citenamefont {Kidera}, \citenamefont
  {Shibata}, \citenamefont {Kishida}, \citenamefont {Ideguchi}, \citenamefont
  {Morita}, \citenamefont {Yoshida}, \citenamefont {Kumagai}, \citenamefont
  {Zhang}, \citenamefont {Ferragut}, \citenamefont {Murakami}, \citenamefont
  {Oshima}, \citenamefont {Iimura}, \citenamefont {Hamada}, \citenamefont
  {Kusakari}, \citenamefont {Sugawara}, \citenamefont {Ogawa}, \citenamefont
  {Nakajima}, \citenamefont {Min}, \citenamefont {Kim}, \citenamefont {Chae},\
  and\ \citenamefont {Sagawa}}]{Odahara1997}%
  \BibitemOpen
  \bibfield  {author} {\bibinfo {author} {\bibfnamefont {A.}~\bibnamefont
  {Odahara}}, \bibinfo {author} {\bibfnamefont {Y.}~\bibnamefont {Gono}},
  \bibinfo {author} {\bibfnamefont {S.}~\bibnamefont {Mitarai}}, \bibinfo
  {author} {\bibfnamefont {T.}~\bibnamefont {Morikawa}}, \bibinfo {author}
  {\bibfnamefont {T.}~\bibnamefont {Shizuma}}, \bibinfo {author} {\bibfnamefont
  {M.}~\bibnamefont {Kidera}}, \bibinfo {author} {\bibfnamefont
  {M.}~\bibnamefont {Shibata}}, \bibinfo {author} {\bibfnamefont
  {T.}~\bibnamefont {Kishida}}, \bibinfo {author} {\bibfnamefont
  {E.}~\bibnamefont {Ideguchi}}, \bibinfo {author} {\bibfnamefont
  {K.}~\bibnamefont {Morita}}, \bibinfo {author} {\bibfnamefont
  {A.}~\bibnamefont {Yoshida}}, \bibinfo {author} {\bibfnamefont
  {H.}~\bibnamefont {Kumagai}}, \bibinfo {author} {\bibfnamefont {Y.~H.}\
  \bibnamefont {Zhang}}, \bibinfo {author} {\bibfnamefont {A.}~\bibnamefont
  {Ferragut}}, \bibinfo {author} {\bibfnamefont {T.}~\bibnamefont {Murakami}},
  \bibinfo {author} {\bibfnamefont {M.}~\bibnamefont {Oshima}}, \bibinfo
  {author} {\bibfnamefont {H.}~\bibnamefont {Iimura}}, \bibinfo {author}
  {\bibfnamefont {S.}~\bibnamefont {Hamada}}, \bibinfo {author} {\bibfnamefont
  {H.}~\bibnamefont {Kusakari}}, \bibinfo {author} {\bibfnamefont
  {M.}~\bibnamefont {Sugawara}}, \bibinfo {author} {\bibfnamefont
  {M.}~\bibnamefont {Ogawa}}, \bibinfo {author} {\bibfnamefont
  {M.}~\bibnamefont {Nakajima}}, \bibinfo {author} {\bibfnamefont {B.~J.}\
  \bibnamefont {Min}}, \bibinfo {author} {\bibfnamefont {J.~C.}\ \bibnamefont
  {Kim}}, \bibinfo {author} {\bibfnamefont {S.~J.}\ \bibnamefont {Chae}}, \
  and\ \bibinfo {author} {\bibfnamefont {H.}~\bibnamefont {Sagawa}},\ }\href
  {\doibase 10.1016/S0375-9474(97)00189-9} {\bibfield  {journal} {\bibinfo
  {journal} {Nucl. Phys. A}\ }\textbf {\bibinfo {volume} {620}},\ \bibinfo
  {pages} {363} (\bibinfo {year} {1997})}\BibitemShut {NoStop}%
\bibitem [{\citenamefont {Ideguchi}\ \emph {et~al.}(1995)\citenamefont
  {Ideguchi}, \citenamefont {Gono}, \citenamefont {Mitarai}, \citenamefont
  {Morikawa}, \citenamefont {Odahara}, \citenamefont {Kidera}, \citenamefont
  {Sibata}, \citenamefont {Tsuchida}, \citenamefont {Miyazaki}, \citenamefont
  {Oshima}, \citenamefont {Hatsukawa}, \citenamefont {Hamada}, \citenamefont
  {Iimura}, \citenamefont {Shibata}, \citenamefont {Ishii}, \citenamefont
  {Murakami}, \citenamefont {Kusakari}, \citenamefont {Sugawara}, \citenamefont
  {Kishida}, \citenamefont {Morita}, \citenamefont {Kumagai},\ and\
  \citenamefont {Ishihara}}]{Ideguchi1995}%
  \BibitemOpen
  \bibfield  {author} {\bibinfo {author} {\bibfnamefont {E.}~\bibnamefont
  {Ideguchi}}, \bibinfo {author} {\bibfnamefont {Y.}~\bibnamefont {Gono}},
  \bibinfo {author} {\bibfnamefont {S.}~\bibnamefont {Mitarai}}, \bibinfo
  {author} {\bibfnamefont {T.}~\bibnamefont {Morikawa}}, \bibinfo {author}
  {\bibfnamefont {A.}~\bibnamefont {Odahara}}, \bibinfo {author} {\bibfnamefont
  {M.}~\bibnamefont {Kidera}}, \bibinfo {author} {\bibfnamefont
  {M.}~\bibnamefont {Sibata}}, \bibinfo {author} {\bibfnamefont
  {H.}~\bibnamefont {Tsuchida}}, \bibinfo {author} {\bibfnamefont
  {K.}~\bibnamefont {Miyazaki}}, \bibinfo {author} {\bibfnamefont
  {M.}~\bibnamefont {Oshima}}, \bibinfo {author} {\bibfnamefont
  {Y.}~\bibnamefont {Hatsukawa}}, \bibinfo {author} {\bibfnamefont
  {S.}~\bibnamefont {Hamada}}, \bibinfo {author} {\bibfnamefont
  {H.}~\bibnamefont {Iimura}}, \bibinfo {author} {\bibfnamefont
  {M.}~\bibnamefont {Shibata}}, \bibinfo {author} {\bibfnamefont
  {T.}~\bibnamefont {Ishii}}, \bibinfo {author} {\bibfnamefont
  {T.}~\bibnamefont {Murakami}}, \bibinfo {author} {\bibfnamefont
  {H.}~\bibnamefont {Kusakari}}, \bibinfo {author} {\bibfnamefont
  {M.}~\bibnamefont {Sugawara}}, \bibinfo {author} {\bibfnamefont
  {T.}~\bibnamefont {Kishida}}, \bibinfo {author} {\bibfnamefont
  {K.}~\bibnamefont {Morita}}, \bibinfo {author} {\bibfnamefont
  {H.}~\bibnamefont {Kumagai}}, \ and\ \bibinfo {author} {\bibfnamefont
  {M.}~\bibnamefont {Ishihara}},\ }\href {\doibase 10.1007/BF01299753}
  {\bibfield  {journal} {\bibinfo  {journal} {Zeitschrift f{\"{u}}r Phys. A}\
  }\textbf {\bibinfo {volume} {352}},\ \bibinfo {pages} {363} (\bibinfo {year}
  {1995})}\BibitemShut {NoStop}%
\bibitem [{\citenamefont {Broda}\ \emph {et~al.}(2020)\citenamefont {Broda},
  \citenamefont {Wrzesinski}, \citenamefont {Michelagnoli}, \citenamefont
  {Lunardi}, \citenamefont {Ur}, \citenamefont {Bazzacco}, \citenamefont
  {Menegazzo}, \citenamefont {Mengoni},\ and\ \citenamefont
  {Recchia}}]{Broda2020}%
  \BibitemOpen
  \bibfield  {author} {\bibinfo {author} {\bibfnamefont {R.}~\bibnamefont
  {Broda}}, \bibinfo {author} {\bibfnamefont {J.}~\bibnamefont {Wrzesinski}},
  \bibinfo {author} {\bibfnamefont {C.}~\bibnamefont {Michelagnoli}}, \bibinfo
  {author} {\bibfnamefont {S.}~\bibnamefont {Lunardi}}, \bibinfo {author}
  {\bibfnamefont {C.}~\bibnamefont {Ur}}, \bibinfo {author} {\bibfnamefont
  {D.}~\bibnamefont {Bazzacco}}, \bibinfo {author} {\bibfnamefont
  {R.}~\bibnamefont {Menegazzo}}, \bibinfo {author} {\bibfnamefont
  {D.}~\bibnamefont {Mengoni}}, \ and\ \bibinfo {author} {\bibfnamefont
  {F.}~\bibnamefont {Recchia}},\ }\href {\doibase 10.1103/PhysRevC.101.064320}
  {\bibfield  {journal} {\bibinfo  {journal} {Phys. Rev. C}\ }\textbf {\bibinfo
  {volume} {101}},\ \bibinfo {pages} {064320} (\bibinfo {year}
  {2020})}\BibitemShut {NoStop}%
\bibitem [{\citenamefont {Ideguchi}\ \emph {et~al.}(1999)\citenamefont
  {Ideguchi}, \citenamefont {Zhou}, \citenamefont {Gono}, \citenamefont
  {Mitarai}, \citenamefont {Morikawa}, \citenamefont {Kidera}, \citenamefont
  {Tsuchida}, \citenamefont {Shibata}, \citenamefont {Watanabe}, \citenamefont
  {Miyake}, \citenamefont {Odahara}, \citenamefont {Oshima}, \citenamefont
  {Hatsukaa}, \citenamefont {Hamada}, \citenamefont {Iimura}, \citenamefont
  {Shibata}, \citenamefont {Ishi}, \citenamefont {Kishida},\ and\ \citenamefont
  {Ishihara}}]{Ideguchi1999}%
  \BibitemOpen
  \bibfield  {author} {\bibinfo {author} {\bibfnamefont {E.}~\bibnamefont
  {Ideguchi}}, \bibinfo {author} {\bibfnamefont {X.~H.}\ \bibnamefont {Zhou}},
  \bibinfo {author} {\bibfnamefont {Y.}~\bibnamefont {Gono}}, \bibinfo {author}
  {\bibfnamefont {S.}~\bibnamefont {Mitarai}}, \bibinfo {author} {\bibfnamefont
  {T.}~\bibnamefont {Morikawa}}, \bibinfo {author} {\bibfnamefont
  {M.}~\bibnamefont {Kidera}}, \bibinfo {author} {\bibfnamefont
  {H.}~\bibnamefont {Tsuchida}}, \bibinfo {author} {\bibfnamefont
  {M.}~\bibnamefont {Shibata}}, \bibinfo {author} {\bibfnamefont
  {H.}~\bibnamefont {Watanabe}}, \bibinfo {author} {\bibfnamefont
  {M.}~\bibnamefont {Miyake}}, \bibinfo {author} {\bibfnamefont
  {A.}~\bibnamefont {Odahara}}, \bibinfo {author} {\bibfnamefont
  {M.}~\bibnamefont {Oshima}}, \bibinfo {author} {\bibfnamefont
  {Y.}~\bibnamefont {Hatsukaa}}, \bibinfo {author} {\bibfnamefont
  {S.}~\bibnamefont {Hamada}}, \bibinfo {author} {\bibfnamefont
  {H.}~\bibnamefont {Iimura}}, \bibinfo {author} {\bibfnamefont
  {M.}~\bibnamefont {Shibata}}, \bibinfo {author} {\bibfnamefont
  {T.}~\bibnamefont {Ishi}}, \bibinfo {author} {\bibfnamefont {T.}~\bibnamefont
  {Kishida}}, \ and\ \bibinfo {author} {\bibfnamefont {M.}~\bibnamefont
  {Ishihara}},\ }\href {\doibase 10.1007/s100500050360} {\bibfield  {journal}
  {\bibinfo  {journal} {Eur. Phys. J. A}\ }\textbf {\bibinfo {volume} {6}},\
  \bibinfo {pages} {387} (\bibinfo {year} {1999})}\BibitemShut {NoStop}%
\bibitem [{\citenamefont {Gono}\ \emph {et~al.}(2002)\citenamefont {Gono},
  \citenamefont {Odahara}, \citenamefont {Fukuchi}, \citenamefont {Ideguchi},
  \citenamefont {Kishida}, \citenamefont {Kubo}, \citenamefont {Watanabe},
  \citenamefont {Motomura}, \citenamefont {Saito}, \citenamefont {Kashiyama},
  \citenamefont {Morikawa}, \citenamefont {Cederwall}, \citenamefont {Zhang},
  \citenamefont {Zhou}, \citenamefont {Ishihara},\ and\ \citenamefont
  {Sagawa}}]{Gono2002}%
  \BibitemOpen
  \bibfield  {author} {\bibinfo {author} {\bibfnamefont {Y.}~\bibnamefont
  {Gono}}, \bibinfo {author} {\bibfnamefont {A.}~\bibnamefont {Odahara}},
  \bibinfo {author} {\bibfnamefont {T.}~\bibnamefont {Fukuchi}}, \bibinfo
  {author} {\bibfnamefont {E.}~\bibnamefont {Ideguchi}}, \bibinfo {author}
  {\bibfnamefont {T.}~\bibnamefont {Kishida}}, \bibinfo {author} {\bibfnamefont
  {T.}~\bibnamefont {Kubo}}, \bibinfo {author} {\bibfnamefont {H.}~\bibnamefont
  {Watanabe}}, \bibinfo {author} {\bibfnamefont {S.}~\bibnamefont {Motomura}},
  \bibinfo {author} {\bibfnamefont {K.}~\bibnamefont {Saito}}, \bibinfo
  {author} {\bibfnamefont {O.}~\bibnamefont {Kashiyama}}, \bibinfo {author}
  {\bibfnamefont {T.}~\bibnamefont {Morikawa}}, \bibinfo {author}
  {\bibfnamefont {B.}~\bibnamefont {Cederwall}}, \bibinfo {author}
  {\bibfnamefont {Y.~H.}\ \bibnamefont {Zhang}}, \bibinfo {author}
  {\bibfnamefont {X.~H.}\ \bibnamefont {Zhou}}, \bibinfo {author}
  {\bibfnamefont {M.}~\bibnamefont {Ishihara}}, \ and\ \bibinfo {author}
  {\bibfnamefont {H.}~\bibnamefont {Sagawa}},\ }\href {\doibase
  10.1140/epja1339-02} {\bibfield  {journal} {\bibinfo  {journal} {Eur. Phys.
  J. A}\ }\textbf {\bibinfo {volume} {13}},\ \bibinfo {pages} {5} (\bibinfo
  {year} {2002})}\BibitemShut {NoStop}%
\bibitem [{\citenamefont {Fukuchi}\ \emph {et~al.}(2006)\citenamefont
  {Fukuchi}, \citenamefont {Tanaka}, \citenamefont {Sasaki}, \citenamefont
  {Gono}, \citenamefont {Odahara}, \citenamefont {Morikawa}, \citenamefont
  {Shibata}, \citenamefont {Watanabe}, \citenamefont {Motomura}, \citenamefont
  {Tsutsumi}, \citenamefont {Kashiyama}, \citenamefont {Saitoh}, \citenamefont
  {Wakabayashi}, \citenamefont {Kishida}, \citenamefont {Kubono},\ and\
  \citenamefont {Ishihara}}]{Fukuchi2006}%
  \BibitemOpen
  \bibfield  {author} {\bibinfo {author} {\bibfnamefont {T.}~\bibnamefont
  {Fukuchi}}, \bibinfo {author} {\bibfnamefont {S.}~\bibnamefont {Tanaka}},
  \bibinfo {author} {\bibfnamefont {T.}~\bibnamefont {Sasaki}}, \bibinfo
  {author} {\bibfnamefont {Y.}~\bibnamefont {Gono}}, \bibinfo {author}
  {\bibfnamefont {A.}~\bibnamefont {Odahara}}, \bibinfo {author} {\bibfnamefont
  {T.}~\bibnamefont {Morikawa}}, \bibinfo {author} {\bibfnamefont
  {M.}~\bibnamefont {Shibata}}, \bibinfo {author} {\bibfnamefont
  {H.}~\bibnamefont {Watanabe}}, \bibinfo {author} {\bibfnamefont
  {S.}~\bibnamefont {Motomura}}, \bibinfo {author} {\bibfnamefont
  {T.}~\bibnamefont {Tsutsumi}}, \bibinfo {author} {\bibfnamefont
  {O.}~\bibnamefont {Kashiyama}}, \bibinfo {author} {\bibfnamefont
  {K.}~\bibnamefont {Saitoh}}, \bibinfo {author} {\bibfnamefont
  {Y.}~\bibnamefont {Wakabayashi}}, \bibinfo {author} {\bibfnamefont
  {T.}~\bibnamefont {Kishida}}, \bibinfo {author} {\bibfnamefont
  {S.}~\bibnamefont {Kubono}}, \ and\ \bibinfo {author} {\bibfnamefont
  {M.}~\bibnamefont {Ishihara}},\ }\href {\doibase 10.1103/PhysRevC.73.067303}
  {\bibfield  {journal} {\bibinfo  {journal} {Phys. Rev. C}\ }\textbf {\bibinfo
  {volume} {73}},\ \bibinfo {pages} {067303} (\bibinfo {year}
  {2006})}\BibitemShut {NoStop}%
\bibitem [{\citenamefont {Foin}\ \emph {et~al.}(2000)\citenamefont {Foin},
  \citenamefont {Gizon}, \citenamefont {Barn{\'{e}}oud}, \citenamefont
  {Genevey}, \citenamefont {Gizon}, \citenamefont {Santos}, \citenamefont
  {Farget}, \citenamefont {Paris}, \citenamefont {Liang}, \citenamefont
  {Bucurescu},\ and\ \citenamefont {P{\l}ochocki}}]{Foin2000}%
  \BibitemOpen
  \bibfield  {author} {\bibinfo {author} {\bibfnamefont {C.}~\bibnamefont
  {Foin}}, \bibinfo {author} {\bibfnamefont {A.}~\bibnamefont {Gizon}},
  \bibinfo {author} {\bibfnamefont {D.}~\bibnamefont {Barn{\'{e}}oud}},
  \bibinfo {author} {\bibfnamefont {J.}~\bibnamefont {Genevey}}, \bibinfo
  {author} {\bibfnamefont {J.}~\bibnamefont {Gizon}}, \bibinfo {author}
  {\bibfnamefont {D.}~\bibnamefont {Santos}}, \bibinfo {author} {\bibfnamefont
  {F.}~\bibnamefont {Farget}}, \bibinfo {author} {\bibfnamefont
  {P.}~\bibnamefont {Paris}}, \bibinfo {author} {\bibfnamefont {C.~F.}\
  \bibnamefont {Liang}}, \bibinfo {author} {\bibfnamefont {D.}~\bibnamefont
  {Bucurescu}}, \ and\ \bibinfo {author} {\bibfnamefont {A.}~\bibnamefont
  {P{\l}ochocki}},\ }\href {\doibase 10.1007/s10050-000-4512-z} {\bibfield
  {journal} {\bibinfo  {journal} {Eur. Phys. J. A}\ }\textbf {\bibinfo {volume}
  {8}},\ \bibinfo {pages} {451} (\bibinfo {year} {2000})}\BibitemShut {NoStop}%
\bibitem [{\citenamefont {Neerg{\aa}rd}\ \emph {et~al.}(1981)\citenamefont
  {Neerg{\aa}rd}, \citenamefont {Dossing},\ and\ \citenamefont
  {Sagawa}}]{Neergard1981}%
  \BibitemOpen
  \bibfield  {author} {\bibinfo {author} {\bibfnamefont {K.}~\bibnamefont
  {Neerg{\aa}rd}}, \bibinfo {author} {\bibfnamefont {T.}~\bibnamefont
  {Dossing}}, \ and\ \bibinfo {author} {\bibfnamefont {H.}~\bibnamefont
  {Sagawa}},\ }\href {\doibase 10.1016/0370-2693(81)91105-9} {\bibfield
  {journal} {\bibinfo  {journal} {Phys. Lett. B}\ }\textbf {\bibinfo {volume}
  {99}},\ \bibinfo {pages} {191} (\bibinfo {year} {1981})}\BibitemShut
  {NoStop}%
\bibitem [{\citenamefont {H{\"{a}}usser}\ \emph {et~al.}(1979)\citenamefont
  {H{\"{a}}usser}, \citenamefont {Taras}, \citenamefont {Trautmann},
  \citenamefont {Ward}, \citenamefont {Alexander}, \citenamefont {Andrews},
  \citenamefont {Haas},\ and\ \citenamefont {Horn}}]{Hausser1979}%
  \BibitemOpen
  \bibfield  {author} {\bibinfo {author} {\bibfnamefont {O.}~\bibnamefont
  {H{\"{a}}usser}}, \bibinfo {author} {\bibfnamefont {P.}~\bibnamefont
  {Taras}}, \bibinfo {author} {\bibfnamefont {W.}~\bibnamefont {Trautmann}},
  \bibinfo {author} {\bibfnamefont {D.}~\bibnamefont {Ward}}, \bibinfo {author}
  {\bibfnamefont {T.~K.}\ \bibnamefont {Alexander}}, \bibinfo {author}
  {\bibfnamefont {H.~R.}\ \bibnamefont {Andrews}}, \bibinfo {author}
  {\bibfnamefont {B.}~\bibnamefont {Haas}}, \ and\ \bibinfo {author}
  {\bibfnamefont {D.}~\bibnamefont {Horn}},\ }\href {\doibase
  10.1103/PhysRevLett.42.1451} {\bibfield  {journal} {\bibinfo  {journal}
  {Phys. Rev. Lett.}\ }\textbf {\bibinfo {volume} {42}},\ \bibinfo {pages}
  {1451} (\bibinfo {year} {1979})}\BibitemShut {NoStop}%
\bibitem [{\citenamefont {H{\"{a}}usser}\ \emph {et~al.}(1982)\citenamefont
  {H{\"{a}}usser}, \citenamefont {Mahnke}, \citenamefont {Alexander},
  \citenamefont {Andrews}, \citenamefont {Sharpey-Schafer}, \citenamefont
  {Swanson}, \citenamefont {Ward}, \citenamefont {Taras},\ and\ \citenamefont
  {Keinonen}}]{Hausser1982}%
  \BibitemOpen
  \bibfield  {author} {\bibinfo {author} {\bibfnamefont {O.}~\bibnamefont
  {H{\"{a}}usser}}, \bibinfo {author} {\bibfnamefont {H.~E.}\ \bibnamefont
  {Mahnke}}, \bibinfo {author} {\bibfnamefont {T.~K.}\ \bibnamefont
  {Alexander}}, \bibinfo {author} {\bibfnamefont {H.~R.}\ \bibnamefont
  {Andrews}}, \bibinfo {author} {\bibfnamefont {J.~F.}\ \bibnamefont
  {Sharpey-Schafer}}, \bibinfo {author} {\bibfnamefont {M.~L.}\ \bibnamefont
  {Swanson}}, \bibinfo {author} {\bibfnamefont {D.}~\bibnamefont {Ward}},
  \bibinfo {author} {\bibfnamefont {P.}~\bibnamefont {Taras}}, \ and\ \bibinfo
  {author} {\bibfnamefont {J.}~\bibnamefont {Keinonen}},\ }\href {\doibase
  10.1016/0375-9474(82)90394-3} {\bibfield  {journal} {\bibinfo  {journal}
  {Nucl. Phys. A}\ }\textbf {\bibinfo {volume} {379}},\ \bibinfo {pages} {287}
  (\bibinfo {year} {1982})}\BibitemShut {NoStop}%
\bibitem [{\citenamefont {Ferragut}\ \emph {et~al.}(1993)\citenamefont
  {Ferragut}, \citenamefont {Gono}, \citenamefont {Murakami}, \citenamefont
  {Morikawa}, \citenamefont {Zhang}, \citenamefont {Morita}, \citenamefont
  {Yoshica}, \citenamefont {Oshima}, \citenamefont {Kusakari}, \citenamefont
  {Sugawara}, \citenamefont {Ogama}, \citenamefont {Makajima}, \citenamefont
  {Mitarai}, \citenamefont {Odahara}, \citenamefont {Ideguchi}, \citenamefont
  {Shizuma}, \citenamefont {Kidera}, \citenamefont {Kim}, \citenamefont {Chae},
  \citenamefont {Min},\ and\ \citenamefont {Kumagai}}]{Ferragut1993}%
  \BibitemOpen
  \bibfield  {author} {\bibinfo {author} {\bibfnamefont {A.}~\bibnamefont
  {Ferragut}}, \bibinfo {author} {\bibfnamefont {Y.}~\bibnamefont {Gono}},
  \bibinfo {author} {\bibfnamefont {T.}~\bibnamefont {Murakami}}, \bibinfo
  {author} {\bibfnamefont {T.}~\bibnamefont {Morikawa}}, \bibinfo {author}
  {\bibfnamefont {Y.~H.}\ \bibnamefont {Zhang}}, \bibinfo {author}
  {\bibfnamefont {K.}~\bibnamefont {Morita}}, \bibinfo {author} {\bibfnamefont
  {A.}~\bibnamefont {Yoshica}}, \bibinfo {author} {\bibfnamefont
  {M.}~\bibnamefont {Oshima}}, \bibinfo {author} {\bibfnamefont
  {H.}~\bibnamefont {Kusakari}}, \bibinfo {author} {\bibfnamefont
  {M.}~\bibnamefont {Sugawara}}, \bibinfo {author} {\bibfnamefont
  {M.}~\bibnamefont {Ogama}}, \bibinfo {author} {\bibfnamefont
  {M.}~\bibnamefont {Makajima}}, \bibinfo {author} {\bibfnamefont
  {S.}~\bibnamefont {Mitarai}}, \bibinfo {author} {\bibfnamefont
  {A.}~\bibnamefont {Odahara}}, \bibinfo {author} {\bibfnamefont
  {E.}~\bibnamefont {Ideguchi}}, \bibinfo {author} {\bibfnamefont
  {T.}~\bibnamefont {Shizuma}}, \bibinfo {author} {\bibfnamefont
  {M.}~\bibnamefont {Kidera}}, \bibinfo {author} {\bibfnamefont {J.~C.}\
  \bibnamefont {Kim}}, \bibinfo {author} {\bibfnamefont {S.~J.}\ \bibnamefont
  {Chae}}, \bibinfo {author} {\bibfnamefont {B.~J.}\ \bibnamefont {Min}}, \
  and\ \bibinfo {author} {\bibfnamefont {H.}~\bibnamefont {Kumagai}},\ }\href
  {\doibase 10.1143/JPSJ.62.3343} {\bibfield  {journal} {\bibinfo  {journal}
  {J. Phys. Soc. Japan}\ }\textbf {\bibinfo {volume} {62}},\ \bibinfo {pages}
  {3343} (\bibinfo {year} {1993})}\BibitemShut {NoStop}%
\bibitem [{\citenamefont {Gerathy}\ \emph {et~al.}(2020)\citenamefont
  {Gerathy}, \citenamefont {Lane}, \citenamefont {Dracoulis}, \citenamefont
  {Nieminen}, \citenamefont {Kib{\'{e}}di}, \citenamefont {Reed}, \citenamefont
  {Akber}, \citenamefont {Coombes}, \citenamefont {Dasgupta}, \citenamefont
  {Dowie}, \citenamefont {Gray}, \citenamefont {Hinde}, \citenamefont {Lee},
  \citenamefont {Mitchell}, \citenamefont {Palazzo}, \citenamefont {Stuchbery},
  \citenamefont {Whichello},\ and\ \citenamefont {Wright}}]{Gerathy2020}%
  \BibitemOpen
  \bibfield  {author} {\bibinfo {author} {\bibfnamefont {M.~S.~M.}\
  \bibnamefont {Gerathy}}, \bibinfo {author} {\bibfnamefont {G.~J.}\
  \bibnamefont {Lane}}, \bibinfo {author} {\bibfnamefont {G.~D.}\ \bibnamefont
  {Dracoulis}}, \bibinfo {author} {\bibfnamefont {P.}~\bibnamefont {Nieminen}},
  \bibinfo {author} {\bibfnamefont {T.}~\bibnamefont {Kib{\'{e}}di}}, \bibinfo
  {author} {\bibfnamefont {M.~W.}\ \bibnamefont {Reed}}, \bibinfo {author}
  {\bibfnamefont {A.}~\bibnamefont {Akber}}, \bibinfo {author} {\bibfnamefont
  {B.~J.}\ \bibnamefont {Coombes}}, \bibinfo {author} {\bibfnamefont
  {M.}~\bibnamefont {Dasgupta}}, \bibinfo {author} {\bibfnamefont {J.~T.~H.}\
  \bibnamefont {Dowie}}, \bibinfo {author} {\bibfnamefont {T.~J.}\ \bibnamefont
  {Gray}}, \bibinfo {author} {\bibfnamefont {D.~J.}\ \bibnamefont {Hinde}},
  \bibinfo {author} {\bibfnamefont {B.~Q.}\ \bibnamefont {Lee}}, \bibinfo
  {author} {\bibfnamefont {A.~J.}\ \bibnamefont {Mitchell}}, \bibinfo {author}
  {\bibfnamefont {T.}~\bibnamefont {Palazzo}}, \bibinfo {author} {\bibfnamefont
  {A.~E.}\ \bibnamefont {Stuchbery}}, \bibinfo {author} {\bibfnamefont
  {L.}~\bibnamefont {Whichello}}, \ and\ \bibinfo {author} {\bibfnamefont
  {A.}~\bibnamefont {Wright}},\ }\href {\doibase 10.1016/j.nima.2019.163136}
  {\bibfield  {journal} {\bibinfo  {journal} {Nucl. Instrum. Methods A}\
  }\textbf {\bibinfo {volume} {953}},\ \bibinfo {pages} {163136} (\bibinfo
  {year} {2020})}\BibitemShut {NoStop}%
\bibitem [{\citenamefont {Rodr{\'{i}}guez}\ \emph {et~al.}(2010)\citenamefont
  {Rodr{\'{i}}guez}, \citenamefont {Brown}, \citenamefont {Dasgupta},
  \citenamefont {Hinde}, \citenamefont {Weisser}, \citenamefont {Kib{\'{e}}di},
  \citenamefont {Lane}, \citenamefont {Cherry}, \citenamefont {Muirhead},
  \citenamefont {Turkentine}, \citenamefont {Lobanov}, \citenamefont {Cooper},
  \citenamefont {Harding}, \citenamefont {Blacksell},\ and\ \citenamefont
  {Davidson}}]{Rodriguez2010}%
  \BibitemOpen
  \bibfield  {author} {\bibinfo {author} {\bibfnamefont {M.~D.}\ \bibnamefont
  {Rodr{\'{i}}guez}}, \bibinfo {author} {\bibfnamefont {M.~L.}\ \bibnamefont
  {Brown}}, \bibinfo {author} {\bibfnamefont {M.}~\bibnamefont {Dasgupta}},
  \bibinfo {author} {\bibfnamefont {D.~J.}\ \bibnamefont {Hinde}}, \bibinfo
  {author} {\bibfnamefont {D.~C.}\ \bibnamefont {Weisser}}, \bibinfo {author}
  {\bibfnamefont {T.}~\bibnamefont {Kib{\'{e}}di}}, \bibinfo {author}
  {\bibfnamefont {M.~A.}\ \bibnamefont {Lane}}, \bibinfo {author}
  {\bibfnamefont {P.~J.}\ \bibnamefont {Cherry}}, \bibinfo {author}
  {\bibfnamefont {A.~G.}\ \bibnamefont {Muirhead}}, \bibinfo {author}
  {\bibfnamefont {R.~B.}\ \bibnamefont {Turkentine}}, \bibinfo {author}
  {\bibfnamefont {N.}~\bibnamefont {Lobanov}}, \bibinfo {author} {\bibfnamefont
  {A.~K.}\ \bibnamefont {Cooper}}, \bibinfo {author} {\bibfnamefont {A.~B.}\
  \bibnamefont {Harding}}, \bibinfo {author} {\bibfnamefont {M.}~\bibnamefont
  {Blacksell}}, \ and\ \bibinfo {author} {\bibfnamefont {P.~M.}\ \bibnamefont
  {Davidson}},\ }\href {\doibase 10.1016/j.nima.2009.12.039} {\bibfield
  {journal} {\bibinfo  {journal} {Nucl. Instrum. Methods A}\ }\textbf {\bibinfo
  {volume} {614}},\ \bibinfo {pages} {119} (\bibinfo {year}
  {2010})}\BibitemShut {NoStop}%
\bibitem [{\citenamefont {Gavron}(1980)}]{Gavron1980}%
  \BibitemOpen
  \bibfield  {author} {\bibinfo {author} {\bibfnamefont {A.}~\bibnamefont
  {Gavron}},\ }\href {\doibase 10.1103/PhysRevC.21.230} {\bibfield  {journal}
  {\bibinfo  {journal} {Phys. Rev. C}\ }\textbf {\bibinfo {volume} {21}},\
  \bibinfo {pages} {230} (\bibinfo {year} {1980})}\BibitemShut {NoStop}%
\bibitem [{\citenamefont {XIA}(2018)}]{XIA2018}%
  \BibitemOpen
  \bibfield  {author} {\bibinfo {author} {\bibnamefont {XIA}},\ }\href
  {https://www.xia.com/dgf\_pixie-16.html} {\enquote {\bibinfo {title}
  {{Pixie-16|XIA LLC}},}\ } (\bibinfo {year} {2018})\BibitemShut {NoStop}%
\bibitem [{\citenamefont {Kib{\'{e}}di}\ \emph {et~al.}(2008)\citenamefont
  {Kib{\'{e}}di}, \citenamefont {Burrows}, \citenamefont {Trzhaskovskaya},
  \citenamefont {Davidson},\ and\ \citenamefont {Nestor}}]{Kibedi2008}%
  \BibitemOpen
  \bibfield  {author} {\bibinfo {author} {\bibfnamefont {T.}~\bibnamefont
  {Kib{\'{e}}di}}, \bibinfo {author} {\bibfnamefont {T.~W.}\ \bibnamefont
  {Burrows}}, \bibinfo {author} {\bibfnamefont {M.~B.}\ \bibnamefont
  {Trzhaskovskaya}}, \bibinfo {author} {\bibfnamefont {P.~M.}\ \bibnamefont
  {Davidson}}, \ and\ \bibinfo {author} {\bibfnamefont {C.~W.}\ \bibnamefont
  {Nestor}},\ }\href {\doibase 10.1016/j.nima.2008.02.051} {\bibfield
  {journal} {\bibinfo  {journal} {Nucl. Instrum. Methods A}\ }\textbf {\bibinfo
  {volume} {589}},\ \bibinfo {pages} {202} (\bibinfo {year}
  {2008})}\BibitemShut {NoStop}%
\bibitem [{\citenamefont {Browne}\ and\ \citenamefont
  {Tuli}(2009)}]{Browne2001}%
  \BibitemOpen
  \bibfield  {author} {\bibinfo {author} {\bibfnamefont {E.}~\bibnamefont
  {Browne}}\ and\ \bibinfo {author} {\bibfnamefont {J.~K.}\ \bibnamefont
  {Tuli}},\ }\href {\doibase 10.1016/j.nds.2009.02.001} {\bibfield  {journal}
  {\bibinfo  {journal} {Nucl. Data Sheets}\ }\textbf {\bibinfo {volume}
  {110}},\ \bibinfo {pages} {507} (\bibinfo {year} {2009})}\BibitemShut
  {NoStop}%
\bibitem [{\citenamefont {Broda}\ \emph {et~al.}(1982)\citenamefont {Broda},
  \citenamefont {Kleinheinz}, \citenamefont {Lunardi}, \citenamefont
  {Styczen},\ and\ \citenamefont {Blomqvist}}]{Broda1982}%
  \BibitemOpen
  \bibfield  {author} {\bibinfo {author} {\bibfnamefont {R.}~\bibnamefont
  {Broda}}, \bibinfo {author} {\bibfnamefont {P.}~\bibnamefont {Kleinheinz}},
  \bibinfo {author} {\bibfnamefont {S.}~\bibnamefont {Lunardi}}, \bibinfo
  {author} {\bibfnamefont {J.}~\bibnamefont {Styczen}}, \ and\ \bibinfo
  {author} {\bibfnamefont {J.}~\bibnamefont {Blomqvist}},\ }\href {\doibase
  10.1007/BF01419074} {\bibfield  {journal} {\bibinfo  {journal} {Z. Phys. A}\
  }\textbf {\bibinfo {volume} {305}},\ \bibinfo {pages} {281} (\bibinfo {year}
  {1982})}\BibitemShut {NoStop}%
\bibitem [{\citenamefont {Piiparinen}\ \emph {et~al.}(1991)\citenamefont
  {Piiparinen}, \citenamefont {Nagai}, \citenamefont {Kleinheinz},
  \citenamefont {Bosca}, \citenamefont {Rubio}, \citenamefont {Lach},\ and\
  \citenamefont {Blomqvist}}]{Piiparinen1991}%
  \BibitemOpen
  \bibfield  {author} {\bibinfo {author} {\bibfnamefont {M.}~\bibnamefont
  {Piiparinen}}, \bibinfo {author} {\bibfnamefont {Y.}~\bibnamefont {Nagai}},
  \bibinfo {author} {\bibfnamefont {P.}~\bibnamefont {Kleinheinz}}, \bibinfo
  {author} {\bibfnamefont {M.~C.}\ \bibnamefont {Bosca}}, \bibinfo {author}
  {\bibfnamefont {B.}~\bibnamefont {Rubio}}, \bibinfo {author} {\bibfnamefont
  {M.}~\bibnamefont {Lach}}, \ and\ \bibinfo {author} {\bibfnamefont
  {J.}~\bibnamefont {Blomqvist}},\ }\href {\doibase 10.1007/BF01295769}
  {\bibfield  {journal} {\bibinfo  {journal} {Z. Phys. A}\ }\textbf {\bibinfo
  {volume} {338}},\ \bibinfo {pages} {417} (\bibinfo {year}
  {1991})}\BibitemShut {NoStop}%
\bibitem [{\citenamefont {Khazov}\ \emph {et~al.}(2011)\citenamefont {Khazov},
  \citenamefont {Rodionov},\ and\ \citenamefont {Kondev}}]{Khazov2011}%
  \BibitemOpen
  \bibfield  {author} {\bibinfo {author} {\bibfnamefont {Y.}~\bibnamefont
  {Khazov}}, \bibinfo {author} {\bibfnamefont {A.}~\bibnamefont {Rodionov}}, \
  and\ \bibinfo {author} {\bibfnamefont {F.~G.}\ \bibnamefont {Kondev}},\
  }\href {\doibase 10.1016/j.nds.2011.03.001} {\bibfield  {journal} {\bibinfo
  {journal} {Nucl. Data Sheets}\ }\textbf {\bibinfo {volume} {112}},\ \bibinfo
  {pages} {855} (\bibinfo {year} {2011})}\BibitemShut {NoStop}%
\bibitem [{\citenamefont {Bakander}\ \emph {et~al.}(1982)\citenamefont
  {Bakander}, \citenamefont {Baktash}, \citenamefont {Borggreen}, \citenamefont
  {Jensen}, \citenamefont {Kownacki}, \citenamefont {Pedersen}, \citenamefont
  {Sletten}, \citenamefont {Ward}, \citenamefont {Andrews}, \citenamefont
  {H{\"{a}}usser}, \citenamefont {Skensved},\ and\ \citenamefont
  {Taras}}]{Bakander1982}%
  \BibitemOpen
  \bibfield  {author} {\bibinfo {author} {\bibfnamefont {O.}~\bibnamefont
  {Bakander}}, \bibinfo {author} {\bibfnamefont {C.}~\bibnamefont {Baktash}},
  \bibinfo {author} {\bibfnamefont {J.}~\bibnamefont {Borggreen}}, \bibinfo
  {author} {\bibfnamefont {J.~B.}\ \bibnamefont {Jensen}}, \bibinfo {author}
  {\bibfnamefont {K.}~\bibnamefont {Kownacki}}, \bibinfo {author}
  {\bibfnamefont {J.}~\bibnamefont {Pedersen}}, \bibinfo {author}
  {\bibfnamefont {G.}~\bibnamefont {Sletten}}, \bibinfo {author} {\bibfnamefont
  {D.}~\bibnamefont {Ward}}, \bibinfo {author} {\bibfnamefont {H.~R.}\
  \bibnamefont {Andrews}}, \bibinfo {author} {\bibfnamefont {O.}~\bibnamefont
  {H{\"{a}}usser}}, \bibinfo {author} {\bibfnamefont {P.}~\bibnamefont
  {Skensved}}, \ and\ \bibinfo {author} {\bibfnamefont {P.}~\bibnamefont
  {Taras}},\ }\href {\doibase 10.1016/0375-9474(82)90293-7} {\bibfield
  {journal} {\bibinfo  {journal} {Nucl. Phys. A}\ }\textbf {\bibinfo {volume}
  {389}},\ \bibinfo {pages} {93} (\bibinfo {year} {1982})}\BibitemShut
  {NoStop}%
\bibitem [{\citenamefont {{Nara Singh}}\ \emph {et~al.}(2018)\citenamefont
  {{Nara Singh}}, \citenamefont {Cullen}, \citenamefont {Taylor}, \citenamefont
  {Srivastava}, \citenamefont {{Van Isacker}}, \citenamefont {Beeke},
  \citenamefont {Dodson}, \citenamefont {Scholey}, \citenamefont {O'Donell},
  \citenamefont {Jakobson}, \citenamefont {Grahn}, \citenamefont {Greenlees},
  \citenamefont {Jones}, \citenamefont {Julin}, \citenamefont {Khan},
  \citenamefont {Leino}, \citenamefont {Lepp{\"{a}}nen}, \citenamefont
  {Eeckhaudt}, \citenamefont {M{\"{a}}ntyniemi}, \citenamefont {Pakarinen},
  \citenamefont {Peura}, \citenamefont {Rahkila}, \citenamefont {Sar{\'{e}}n},
  \citenamefont {Sorri}, \citenamefont {Uusitalo},\ and\ \citenamefont
  {Venhart}}]{NaraSingh2018}%
  \BibitemOpen
  \bibfield  {author} {\bibinfo {author} {\bibfnamefont {B.~S.}\ \bibnamefont
  {{Nara Singh}}}, \bibinfo {author} {\bibfnamefont {D.~M.}\ \bibnamefont
  {Cullen}}, \bibinfo {author} {\bibfnamefont {M.~J.}\ \bibnamefont {Taylor}},
  \bibinfo {author} {\bibfnamefont {P.~C.}\ \bibnamefont {Srivastava}},
  \bibinfo {author} {\bibfnamefont {P.}~\bibnamefont {{Van Isacker}}}, \bibinfo
  {author} {\bibfnamefont {O.}~\bibnamefont {Beeke}}, \bibinfo {author}
  {\bibfnamefont {B.}~\bibnamefont {Dodson}}, \bibinfo {author} {\bibfnamefont
  {C.}~\bibnamefont {Scholey}}, \bibinfo {author} {\bibfnamefont
  {D.}~\bibnamefont {O'Donell}}, \bibinfo {author} {\bibfnamefont
  {U.}~\bibnamefont {Jakobson}}, \bibinfo {author} {\bibfnamefont
  {T.}~\bibnamefont {Grahn}}, \bibinfo {author} {\bibfnamefont {P.~T.}\
  \bibnamefont {Greenlees}}, \bibinfo {author} {\bibfnamefont {P.~M.}\
  \bibnamefont {Jones}}, \bibinfo {author} {\bibfnamefont {R.}~\bibnamefont
  {Julin}}, \bibinfo {author} {\bibfnamefont {S.}~\bibnamefont {Khan}},
  \bibinfo {author} {\bibfnamefont {M.}~\bibnamefont {Leino}}, \bibinfo
  {author} {\bibfnamefont {A.~P.}\ \bibnamefont {Lepp{\"{a}}nen}}, \bibinfo
  {author} {\bibfnamefont {S.}~\bibnamefont {Eeckhaudt}}, \bibinfo {author}
  {\bibfnamefont {K.}~\bibnamefont {M{\"{a}}ntyniemi}}, \bibinfo {author}
  {\bibfnamefont {J.}~\bibnamefont {Pakarinen}}, \bibinfo {author}
  {\bibfnamefont {P.}~\bibnamefont {Peura}}, \bibinfo {author} {\bibfnamefont
  {P.}~\bibnamefont {Rahkila}}, \bibinfo {author} {\bibfnamefont
  {J.}~\bibnamefont {Sar{\'{e}}n}}, \bibinfo {author} {\bibfnamefont
  {J.}~\bibnamefont {Sorri}}, \bibinfo {author} {\bibfnamefont
  {J.}~\bibnamefont {Uusitalo}}, \ and\ \bibinfo {author} {\bibfnamefont
  {M.}~\bibnamefont {Venhart}},\ }\href {\doibase 10.1103/PhysRevC.98.024319}
  {\bibfield  {journal} {\bibinfo  {journal} {Phys. Rev. C}\ }\textbf {\bibinfo
  {volume} {98}},\ \bibinfo {pages} {024319} (\bibinfo {year}
  {2018})}\BibitemShut {NoStop}%
\bibitem [{\citenamefont {Shimizu}\ \emph {et~al.}(2019)\citenamefont
  {Shimizu}, \citenamefont {Mizusaki}, \citenamefont {Utsuno},\ and\
  \citenamefont {Tsunoda}}]{Shimizu2019}%
  \BibitemOpen
  \bibfield  {author} {\bibinfo {author} {\bibfnamefont {N.}~\bibnamefont
  {Shimizu}}, \bibinfo {author} {\bibfnamefont {T.}~\bibnamefont {Mizusaki}},
  \bibinfo {author} {\bibfnamefont {Y.}~\bibnamefont {Utsuno}}, \ and\ \bibinfo
  {author} {\bibfnamefont {Y.}~\bibnamefont {Tsunoda}},\ }\href {\doibase
  10.1016/j.cpc.2019.06.011} {\bibfield  {journal} {\bibinfo  {journal}
  {Comput. Phys. Commun.}\ }\textbf {\bibinfo {volume} {244}},\ \bibinfo
  {pages} {372} (\bibinfo {year} {2019})}\BibitemShut {NoStop}%
\bibitem [{\citenamefont {Chou}\ and\ \citenamefont
  {Warburton}(1992)}]{Chou1992}%
  \BibitemOpen
  \bibfield  {author} {\bibinfo {author} {\bibfnamefont {W.~T.}\ \bibnamefont
  {Chou}}\ and\ \bibinfo {author} {\bibfnamefont {E.~K.}\ \bibnamefont
  {Warburton}},\ }\href {\doibase 10.1103/PhysRevC.45.1720} {\bibfield
  {journal} {\bibinfo  {journal} {Phys. Rev. C}\ }\textbf {\bibinfo {volume}
  {45}},\ \bibinfo {pages} {1720} (\bibinfo {year} {1992})}\BibitemShut
  {NoStop}%
\bibitem [{\citenamefont {Trache}\ \emph {et~al.}(1989)\citenamefont {Trache},
  \citenamefont {Clauberg}, \citenamefont {Wesselborg}, \citenamefont {von
  Brentano}, \citenamefont {Wrzesinski}, \citenamefont {Broda}, \citenamefont
  {Berinde},\ and\ \citenamefont {Iacob}}]{Trache1989}%
  \BibitemOpen
  \bibfield  {author} {\bibinfo {author} {\bibfnamefont {L.}~\bibnamefont
  {Trache}}, \bibinfo {author} {\bibfnamefont {A.}~\bibnamefont {Clauberg}},
  \bibinfo {author} {\bibfnamefont {C.}~\bibnamefont {Wesselborg}}, \bibinfo
  {author} {\bibfnamefont {P.}~\bibnamefont {von Brentano}}, \bibinfo {author}
  {\bibfnamefont {J.}~\bibnamefont {Wrzesinski}}, \bibinfo {author}
  {\bibfnamefont {R.}~\bibnamefont {Broda}}, \bibinfo {author} {\bibfnamefont
  {A.}~\bibnamefont {Berinde}}, \ and\ \bibinfo {author} {\bibfnamefont
  {V.~E.}\ \bibnamefont {Iacob}},\ }\href {\doibase 10.1103/PhysRevC.40.1006}
  {\bibfield  {journal} {\bibinfo  {journal} {Phys. Rev. C}\ }\textbf {\bibinfo
  {volume} {40}},\ \bibinfo {pages} {1006} (\bibinfo {year}
  {1989})}\BibitemShut {NoStop}%
\bibitem [{\citenamefont {Kader}\ \emph {et~al.}(1989)\citenamefont {Kader},
  \citenamefont {Graw}, \citenamefont {Eckle}, \citenamefont {Eckle},
  \citenamefont {Schiemenz}, \citenamefont {Kleinheinz}, \citenamefont {Rubio},
  \citenamefont {{De Angelis}}, \citenamefont {Massey}, \citenamefont {Mann},\
  and\ \citenamefont {Blomqvist}}]{Kader1989}%
  \BibitemOpen
  \bibfield  {author} {\bibinfo {author} {\bibfnamefont {H.}~\bibnamefont
  {Kader}}, \bibinfo {author} {\bibfnamefont {G.}~\bibnamefont {Graw}},
  \bibinfo {author} {\bibfnamefont {F.~J.}\ \bibnamefont {Eckle}}, \bibinfo
  {author} {\bibfnamefont {G.}~\bibnamefont {Eckle}}, \bibinfo {author}
  {\bibfnamefont {P.}~\bibnamefont {Schiemenz}}, \bibinfo {author}
  {\bibfnamefont {P.}~\bibnamefont {Kleinheinz}}, \bibinfo {author}
  {\bibfnamefont {B.}~\bibnamefont {Rubio}}, \bibinfo {author} {\bibfnamefont
  {G.}~\bibnamefont {{De Angelis}}}, \bibinfo {author} {\bibfnamefont {T.~N.}\
  \bibnamefont {Massey}}, \bibinfo {author} {\bibfnamefont {L.~G.}\
  \bibnamefont {Mann}}, \ and\ \bibinfo {author} {\bibfnamefont
  {J.}~\bibnamefont {Blomqvist}},\ }\href {\doibase
  10.1016/0370-2693(89)90938-6} {\bibfield  {journal} {\bibinfo  {journal}
  {Phys. Lett. B}\ }\textbf {\bibinfo {volume} {227}},\ \bibinfo {pages} {325}
  (\bibinfo {year} {1989})}\BibitemShut {NoStop}%
\end{thebibliography}%

\end{document}